\documentclass[12pt]{article}

\usepackage{amsfonts,amsbsy,bm,euscript,mathrsfs}
\usepackage{amssymb,stmaryrd,faktor}
\usepackage[tbtags]{amsmath}
\usepackage[title,titletoc]{appendix}
\usepackage[bookmarks=true,colorlinks=true,linkcolor=blue,citecolor=blue,urlcolor=blue,bookmarksnumbered]{hyperref}
\usepackage{graphics}
\usepackage{tikz}
\usetikzlibrary{matrix,arrows}

\newcommand{\mathsym}[1]{{}}
\makeatletter
\renewcommand\section{\@startsection {section}{1}{\z@}
{-3.5ex \@plus -1ex \@minus -.2ex}
{2.3ex \@plus.2ex}
{\normalfont\large\bfseries}}
\renewcommand\subsection{\@startsection{subsection}{2}{\z@}
{-3.25ex\@plus -1ex \@minus -.2ex}
{1.5ex \@plus.2ex}
{\normalfont\normalsize\bfseries}}
\renewcommand\subsubsection{\@startsection{subsubsection}{2}{\z@}%V4
{-3.25ex\@plus -1ex \@minus -.2ex}
{1.5ex \@plus.2ex}
{\normalfont\small\bfseries}}
\makeatother

\def\id{\protect{{1 \kern-.28em {\rm l}}}}
\def\be{\begin{eqnarray}}
\def\ee{\end{eqnarray}}
\def\p{{\partial}}

\def\Tr{{\rm Tr}}
\def\foot{\footnote}
\def\bi{\bibitem}
\def\tr{{\rm tr}}
\def\ha{{1 \over 2}}
\def\td{\tilde}
\def\ci{\cite}

\def\t{\tau}

 \def \J {{\mathcal J}}

\def\z{\zeta}

\def\a{\alpha}

\def\e{\varepsilon}
\def\p{\phi}

\def\del{\partial}
\def\a{\alpha}

\def\g{\gamma}
\def\s{\sigma}
\def\z{\zeta}

\def\ov{\over}

\def\l{\lambda}

\def\k{\kappa}
\def\foot{\footnote}

\def\third{\textstyle {1\ov 3}}

\def\ci{\cite}

\def\Tr{{\rm Tr}}

\def\l{\lambda}

\def\td{\tilde}

\def\m{\mu}

\def\e{\epsilon}
\def\bi{\bibitem}
\def\la{\label}
\def\l{\lambda}
\def\foot{\footnote}

\def\sql{{\sqrt \l}}
\def\adss{$AdS_5 \times S^5~$ }
\newcommand{\rf}[1]{(\ref{#1})}
\def\ov{\over}

\def\ha{{1\ov 2}}

\def\no{\nonumber}
\def\J{\mathcal{J}}
\def\del{\partial}

\def\J{{\cal J}}

\def\om{\omega}

\def\bi{\bibitem}
\def\la{\label}
\def\l{\lambda}
\def\foot{\footnote}

\def\sql{{\sqrt \l}}

\def\adss{$AdS_5 \times S^5$\ }
\def\t{\tau}
\def\p{\phi}

\def\ov{\over}

\def\varpi{{\rm w}}

\def\ep{\epsilon}

\def\t{\tau}

\def\Tr{{\rm Tr}}

\def\s{\sigma}

 \def \sql {\sqrt{\lambda}}
\def\vp{\varphi}
\def\hh{{\rm h}}

\def\tv{ \td \vp}
\def\fo{{{\textstyle {1 \ov 4}}}}
\def\SS{{\rm S}} 

\def\ifo{\iffalse}
\def\edd{\end{document}}
\def\ha{{{\textstyle{1 \ov2}}}}
\def\fo{{\textstyle{1 \ov4}}}

\def\sql{{\sqrt{\l}}}

\def\ket{\rangle}

\def\sql{\sqrt{\l}}

\def\bea{\be}
\def\eea{\ee}
\def\eqref{\rf}
\def\da{\partial_+}
\def\db{\partial_-}
\def\t{\theta}
\def\p{\phi}  \def \J {{\cal J}}\def \sql {{\sqrt\lambda}}
\def\ty{\hat y}
\def\vc{{v^*}}
\def\text{ {\textstyle} }
\def\vo{ {\textstyle{ 1 \ov 8}} }
\def\adst{$AdS_3 \times S^3 \times T^4$ }
\def\rWZW{{\rm WZW}}
\def\rS{{\rm S}}
\def \bc {{\rm c}}

\begin{document}
%%%%%%%%%%%%%%%%%%%%%%%%%%%%%%%%%%%%%%

\overfullrule=0pt
\parskip=2pt
\parindent=12pt
\headheight=0in \headsep=0in \topmargin=0in \oddsidemargin=0in

\vspace{ -3cm}
\thispagestyle{empty}
\vspace{-1cm}

\rightline{ Imperial-TP-AT-2013-01}
\rightline{ HU-EP-13/10}

\begin{center}
\vspace{1cm}
{\Large\bf
On string theory on $AdS_3 \times S^3 \times T^4$ \\
\vspace{0.2cm}
with mixed 3-form flux: tree-level S-matrix \\
\vspace{0.2cm}
}
\vspace{1.5cm}

{B. Hoare$^{a,}$\footnote{ben.hoare@physik.hu-berlin.de} and A.A. Tseytlin$^{b,}$\footnote{Also at Lebedev Institute, Moscow. tseytlin@imperial.ac.uk }}\\

\vskip 0.6cm

{\em $^{a}$ Institut f\"ur Physik, Humboldt-Universit\"at zu Berlin, \\ Newtonstra\ss e 15, D-12489 Berlin, Germany}

\vskip 0.3cm

{\em
$^{b}$ The Blackett Laboratory, Imperial College,
London SW7 2AZ, U.K.
}

\vspace{.2cm}

\end{center}

\begin{abstract}
We consider superstring theory on $AdS_3 \times S^3 \times T^4$ supported
by a combination of RR and NSNS 3-form fluxes
(with parameter of the NSNS 3-form $q$).
This theory interpolates between the
pure RR flux model ($q=0$) whose spectrum is expected to be described by a (thermodynamic) Bethe ansatz
and the pure NSNS flux model ($q=1$) which is described by the supersymmetric extension of the $SL(2,R) \times SU(2)$
WZW model. As a first step towards the solution of this integrable theory for generic value of
$q$ we compute the corresponding
tree-level S-matrix for massive BMN-type excitations. We find that this S-matrix has a surprisingly
simple dependence on $q$: the diagonal amplitudes
have exactly the same structure as in the $q=0$ case
but with the BMN dispersion relation $e^2 = p^2 + 1$ replaced
by the one with shifted momentum and mass, $e^2 = ( p \pm q)^2 + 1 - q^2 $.
The off-diagonal amplitudes are then determined from the classical Yang-Baxter equation.
We also construct the Pohlmeyer reduced model corresponding to this superstring
theory and find that it depends on $q$ only through the rescaled mass parameter, $\mu \to \sqrt{1-q^2}\, \mu$,
implying that its relativistic S-matrix is $q$-independent.

\end{abstract}

\newpage
\tableofcontents
\setcounter{equation}{0}
\setcounter{footnote}{0}
\setcounter{section}{0}

\newpage
%%%%%%%%%%%%%%%%%%%%%%%%%%%%%%%%%%%%%%
\renewcommand{\theequation}{1.\arabic{equation}}
\setcounter{equation}{0}
\section{Introduction}
%%%%%%%%%%%%%%%%%%%%%%%%%%%%%%%%%%%%%%

The subject of this paper is the superstring theory on $AdS_3 \times S^3 \times T^4$ space-time supported
by a combination of RR and NSNS 3-form fluxes.
The corresponding type IIB supergravity background
is the near-horizon limit of the mixed NS5-NS1\ + \ D5-D1 solution that is the basis for an interesting example
of $AdS$/CFT duality \ci{mal,gks}.
In the near-horizon limit the dilaton is constant, the 3-form fluxes
through $AdS_3$ and $S^3$ have related coefficients and the radii of the two spaces are equal.

The S-duality symmetry of type IIB supergravity transforms the NSNS 3-form into the RR 3-form,
so that if the coefficients of the NSNS and RR fluxes are chosen as $q$ and $q'$,
respectively, then they enter symmetrically into the supergravity equations, e.g., as
$q^2 + q'^2 =1$ (we set the curvature radius to 1).
The perturbative fundamental superstring theory is { not} invariant under the S-duality and should thus depend
non-trivially on the parameter $q$.\foot{The (leading-order) conformal invariance (or $\kappa$-symmetry)
conditions of the superstring theory, being equivalent to the supergravity field equations,
will still depend on the coefficients of the two types of fluxes through the S-duality invariant combination
$q^2 + q'^2$, relating it to the square of the radius of the $AdS_3$ and $S^3$ (here set to 1).}

We shall assume that
$0 \leq q \leq 1$, with $q=0$ corresponding to the \adst theory with pure RR flux and
$q=1$ -- to the \adst theory with pure NSNS flux.
The NSNS flux theory is given by the
superstring generalization of the $SL(2) \times SU(2)$ WZW model
while the RR flux theory has a Green-Schwarz (GS) formulation \ci{pes,rahm}
similar to the one \ci{mt} in the \adss case.
The ``mixed''  theory for generic value of $q$ (first discussed in  GS formulation in \ci{pes})
 has a nice  $PSU(1,1|2) \times PSU(1,1|2) \over SU(1,1) \times SU(2)$
supercoset formulation \ci{cz}, exposing its classical integrability  and UV finiteness.\foot{In our notation
the parameters of the GS action in \ci{cz} are $\chi= q, \ \k= \sqrt{ 1- q^2}$.}

The free string spectrum of the NSNS ($q=1$) theory can be found using the chiral decomposition
property of the WZW model
\ci{ogm} while the apparently more complicated spectral problem of the
RR ($q=0$) theory
is expected to be solved, as in the \adss case \ci{rev}, by a thermodynamic Bethe ansatz
(for recent progress towards its construction see \ci{bsz,david,sax,wulf,ahn,bor,ab,bec,wu}).

Solving the ``interpolating'' theory with $0 < q < 1$ is thus a very interesting problem as
that may help to
understand the relation between the more standard CFT approach in the $q=1$ case
and the integrability-based TBA approach in the $q\not=1$ case.\foot{It should be noted that the
$SU(2)$ principal chiral model with a WZ term was studied in the past
using the Bethe ansatz approach \ci{pw,fr,zz,ber} but this quantum solution is not directly relevant
for the \adst superstring case where the presence of fermions makes the world-sheet theory UV finite:
there is thus no RG flow (or dynamical mass generation in the $q=0$ case)
while a fiducial (BMN-type \ci{bmn}) mass scale is introduced by a ``vacuum'' choice or
a gauge fixing. }

\

The first step towards constructing the $q\not=0$ generalization of the Bethe ansatz for the string spectrum
is to find the corresponding S-matrix for the elementary (BMN-like) massive excitations.
Our aim here will be to determine the string tree-level term in this S-matrix following the approach used in the \adss case in
\ci{fpz,rtt,roi,km,afr}, i.e. computing it directly from the gauge-fixed string action.

We shall start in section \ref{sec2} with an explicit description of the bosonic part of the string action in the sector
corresponding to the string moving on $R \times S^3$. In conformal gauge it is described
by the $SU(2)$ principal chiral model with a WZ term
(with coefficient proportional to $q$).
Fixing a gauge corresponding the BMN vacuum in which the center of mass of the string moves along a circle in $S^3$
we expand the action to quartic order in the fields, sufficient to compute the two-particle tree-level S-matrix.

The computation of this S-matrix for the bosonic string on $S^3$ with $B$-field flux is the subject of
section \ref{sec3}. This S-matrix has a direct generalization to the full bosonic $AdS_3 \times S^3$ sector
found by using the expression for the relevant gauge-fixed action given in Appendix \ref{A_1}.

The simplicity of the bosonic result and the requirements of integrability (symmetry factorization and the Yang-Baxter equation)
suggest a natural generalization to the fermionic sector, i.e. leading to the
full tree-level S-matrix of the
GS superstring theory on \adst with mixed RR-NSNS flux, which we present in the section \ref{sec4}.

In Appendix \ref{A_2}  we  explain how  the quadratic fermionic  action that  reproduces the non-trivial $BBFF$ 
   part of this S-matrix should follow 
 from the \adst   superstring action upon light-cone gauge fixing and expansion in powers of bosons.  
A candidate for the   symmetry algebra of this S-matrix  is discussed in Appendix \ref{appb} 
where we also comment on  the symmetry of the S-matrix in the $q=0$ case. 
Section \ref{sec5} contains some concluding remarks.

Imposing the conformal gauge, one may solve the Virasoro conditions explicitly and reformulate
the classical string
theory in terms of current field variables.
One way to do this
is the Pohlmeyer reduction and another is the Faddeev-Reshetikhin construction (that applies in the bosonic $SU(2)$ case).
In Appendix \ref{appc} we construct the Faddeev-Reshetikhin model corresponding
to the bosonic string on $R \times S^3$ with $B$-flux, generalizing the discussion in \ci{kz},
and present the $q\not=0$ expression for the corresponding tree-level S-matrix (which is different from the
bosonic string sigma model one).

In Appendix \ref{appd} we give a detailed construction of the Pohlmeyer-reduced model corresponding to the
superstring theory on \adst with mixed 3-form flux.
Somewhat unexpectedly, we find that the reduced action is the same as in the $q=0$ case \ci{gt1,gt2} but has
the rescaled mass scale parameter, $\mu \to \sqrt{1-q^2}\, \mu$. As a result, the corresponding relativistic
$S$-matrix does not depend on $q$.

%%%%%%%%%%%%%%%%%%%%%%%%%%%%%%%%%%%%%%
\renewcommand{\theequation}{2.\arabic{equation}}
\setcounter{equation}{0}
\section{Bosonic string action on $R\times S^3$ with $B$-flux\label{sec2}}
%%%%%%%%%%%%%%%%%%%%%%%%%%%%%%%%%%%%%%

Let us start with some basic definitions.
In general, the bosonic string sigma model action is\foot{We use $(-,+)$ world-sheet signature and $\ep^{01}=1$.}
\be
\rS = { 1 \ov 2 \pi\a'} \int d \tau d\s\ L \ , \ \ \ \ \ \ \
L= - \ha \big[ \sqrt g g^{ab} G_{mn}(x) + \ep^{ab} B_{mn} (x) \big] \del_a x^m \del_b x^n \ . \la{01} \ee
The action of the \adst superstring theory with mixed 3-form flux has the following structure:
\be
\rS_{\rm tot}= \rS_{AdS} + \rS_{S} + {\rm fermionic\ terms} \ , \la{011} \ee
where the first two terms are given by the principal chiral models with an extra
WZ term for the groups
$SL(2,R)$ and $SU(2)$ respectively. The WZ term represents the NSNS 3-form flux
(proportional to $q$).\foot{As already mentioned, we shall always assume that
$0 \leq q\leq 1$.}
Both the NSNS 3-form coupling and the RR 3-form coupling (proportional to
$\sqrt{1-q^2}$) appear in the fermionic terms.

In general,
the string moving on a group space with a $B$-flux is described, in the conformal gauge, by the
the action of a principal chiral model with a WZ term
\be \la{4.1}
&&\rS = \ha \hh \Big[ \int d^2 \s\ \ha \tr ( J^a J_a )
+ { q} \int d^3 \s \ {\textstyle {1\ov 3}} \epsilon^{abc} \tr ( J_a J_b J_c)\Big] \ , \\
&&
J_a = g^{-1} \del_a g\ , \ \ \ \ \ \ \ \ \ \ \ \hh = {R^2\ov 2\pi \a'} = {\sql \ov 2 \pi} \ . \la{041}
\ee
Here $\hh$ is the effective coupling (string tension). The
quantized coefficient of the WZ term is $k=\sql \, q$.
%$k=2\sql \, q$.
$q=1$ corresponds to the case of the WZW model.
The corresponding classical equations of motion can be written as ($\del_\pm = \del_0 \pm \del_1$)
\be
(1 - q) \da J_- + (1 +q) \db J_+ = 0 \ , \ \ \ \ \ \ \ \ \ \
\da J_- - \db J_+ + [J_+, J_-] =0 \ , \la{4.2} \ee
or as
\be
\da J_- + \ha ( 1 + q) [J_+, J_-] =0 \ , \ \ \ \ \ \ \ \ \ \
\db J_+ - \ha ( 1 - q) [J_+, J_-] =0 \ ,
\la{4.21} \ee
implying the existence of the Lax pair, i.e. the classical integrability of this model.\foot{Let us note
that these equations may be written also as $\partial^a L_a =0, \ \ L_a \equiv J_a + q \e_{ab} J^b$, where
$J_a =g^{-1} \del_a g$
or as $\partial^a R_a =0, \ \ R_a \equiv K_a - q \e_{ab} K^b$, where $K_a = \del_a g \, g^{-1}$.
}
Explicitly, the Lax pair is given by
\begin{equation}
\mathcal{L}_\pm = \tfrac12(1 \pm q + z^{\pm 1}\sqrt{1-q^2})\, J_\pm \ , \la{laxx}
\end{equation}
where $z$ is the spectral parameter.\foot{Local and non-local conserved charges in such model were discussed in \ci{tah,ev}.}

In this section we shall concentrate on the sector of the full superstring theory when the string is
moving on $R\times S^3$, i.e. when the
Lagrangian is $L = - \ha \da t \db t + L_S $ where $L_S$ is given by
\rf{4.1} with $g \in SU(2)$.

%%%%%%%%%%%%%%%%%%%%%%%%%%%%%%%%%%%%%%
\subsection{Explicit form of the Lagrangian}
%%%%%%%%%%%%%%%%%%%%%%%%%%%%%%%%%%%%%%

Using the familiar parametrization of $g \in SU(2)$ the Lagrangian $L_S$
may be written as
\be
&&
L_S = \ha \Big[ \da \t \db \t + \sin^2 \t\ \da \p_1 \db \p_1+ \cos^2 \t\ \da \p_2 \db \p_2 \no\\
&& \ \ \ \ \ \ \ \ \ \ \ \ \ \ \ \ \ \ \ \ \ \ \ \ \
+ \ q \, \sin^2 \t\ ( \da \p_1 \db \p_2 - \da \p_2 \db \p_1)\Big]
\ . \la{1} \ee
Below we shall use also an alternative parametrization of $S^3$ in terms of an angle $\vp$ and two ``cartesian''
coordinates $y_s$ ($s=1,2$)\foot{A similar parametrization of $S^5$ (and $AdS_5$) was used, e.g., in \ci{km}.
It is related to the one in \rf{1} as follows:
if we set $y_1 + i y_2 = y e^{i\p_1} $, $y = 2 \tan { \t \ov 2} $ and $\vp= \p_2$ then %v5
$G d \vp^2 + F dy_s dy_s = d\t^2 + \sin^2 \t\ d\p_1^2 + \cos^2 \t\ d\vp^2$ and %v5
$ B_s dy_s = q F \epsilon_{rs} y_{r} dy_s = q \sin^2 \t\ d\p_1$. } %v5
%if we set $y_1 + i y_2 = y e^{i\t} $, $y = 2 \tan { \p_1 \ov 2} $ and $\vp= \p_2$ then
%$G d \vp^2 + F dy_s dy_s = d\t^2 + \sin^2 \t\ d\p_1^2 + \cos^2 \t\ d\vp^2$ and
%$ B_s dy_s = q F \epsilon_{rs} y_{r} dy_s = q \sin^2 \t\ d\p_1$. }
\be \la{5.22}
&& L_S= - \ha\Big[ G(y) \del^a \vp \del_a \vp + F(y) \del^a y_s \del_a y_s
+ 2 B_s(y) \ep^{ab} \del_a y_{s} \del_b \vp \Big] \ , \\
&& G= { ( 1 - \fo y^2 )^2 \ov ( 1 + \fo y^2 )^2 }= 1 - y^2 F \ , \ \ \ \ \ \ \ \ \ \ \ \ \ \ \ F = { 1 \ov ( 1 + \fo y^2 )^2 } \ , \la{225} \\
&& B_s = { q F (y) } \ \ty_s \ , \ \ \ \ \ \ \ \ \ \ty_s\equiv \epsilon_{rs} y_{r} \ . \la{522} \ee
Note that \rf{5.22} can be written also as
\be
&& L_S= - \ha \big[ 1 - (1-q^2) y^2 F(y)\big] \del^a \vp \del_a \vp - \ha
F(y) (\del^a y_s + q \ty_s \ep^{ab} \del_b \vp )^2 \ . \la{rrr}
\ee
In particular, for $q=1$ (i.e. in the WZW case) eq.~\rf{rrr} becomes\foot{This form reveals the chiral structure of the model
as after a local rotation of $y_s$ it is proportional to
$ \del_+ \vp \del_-\vp + F(\td y) (\del_+ \td y_s ) (\del_- \td y_s - 2 \epsilon_{rs} \td y_{r} \del_- \vp)$ and, e.g., the equation for
$\vp$ is readily integrated.}
\be
L_S(q=1) \equiv L_{\rm WZW} =\ha \del_+ \vp \del_-\vp +
\ha F(y) (\del_+ y_s + \ty_s \del_+ \vp) (\del_- y_s - \ty_s \del_- \vp) \ . \la{wz} \ee
Applying T-duality in $\vp$ to \rf{wz} gives the T-dual sigma model which has no WZ term
\be
&& \td L_{\rWZW} = \ha G^{-1} ( \del_+\td \vp - F \ty_{s} \del_+ y_s) ( \del_-\td \vp - F \ty_{s} \del_- y_s)
+ \ha F \del_+ y_s \del_- y_s\no \\
&&\ \ \ \ = \ha \del_+ \td \vp \del_-\td \vp +
\ha \td F(y) (\del_+ y_s - \ty_s \del_+ \td \vp) (\del_- y_s - \ty_s \del_-\td \vp)
- \vo \td F(y) F(y) \del_+ y^2 \del_- y^2 \la{twz} \\
&& \ \ \ \ \td F\equiv G^{-1} F = { 1 \ov ( 1 - \fo y^2 )^2 } \ .
\la{twzw}
\ee
The angle $\td \vp$ can be completely decoupled (by a local rotation of $y_s$ with phase $\td \vp$),
resulting in the familiar relation \ci{kiri} to a free field plus a 2d sigma model representing the $SU(2)/U(1)$ coset\foot{Setting $y_1 + i y_2 = y e^{i \phi_1}$ we find the corresponding metric to be
%\foot{Setting $y_1 + i y_2 = y e^{i \t}$ we find the corresponding metric to be
$ ds^2 = F(y) dy^2 + \td F (y)y^2 ( d \phi_1 + d \td \vp)^2 + d \td \vp^2 = dx^2 + \tan^2 {x} \ d\td \phi_1^2 + d \td \vp^2 , $
where $\td \phi_1=\phi_1 + \td \vp$, etc. }
\be \la{roo}
\td L_{\rWZW} =\ha \del_+ \td \vp \del_-\td \vp +
\ha \td F(\td y) \big[ \del_+ \td y_s \del_- \td y_s - \fo F(\td y) \del_+ \td y^2 \del_- \td y^2 \big] \ . \ee
This observation explains why we will find a relativistic expression
when expanding the corresponding Nambu action
near $\td \vp= \s$ as discussed below.
Note also that this T-dual of the WZW action is parity-invariant, which will also be reflected in the corresponding S-matrix.

%%%%%%%%%%%%%%%%%%%%%%%%%%%%%%%%%%%%%%
\def\J{{\cal J}}
\def\k{\J}
\def\hp{\hat p}
\subsection{Dispersion relation}
%%%%%%%%%%%%%%%%%%%%%%%%%%%%%%%%%%%%%%

The presence of the WZ term does not change the simplest BMN geodesic solution
which in the conformal gauge is given by
\be t= \J \tau\ ,\ \ \ \ \ \ \ \ \ \vp=\J \tau \ , \ \ \ \ \ \ \ \ y_s=0 \ . \la{5.1} \ee
Here $\J$ is proportional to the $S^3$ angular momentum of the string.
Our aim will be to study the 2-particle scattering of small $y_s$ excitations around this solution.
To find the spectrum of quadratic fluctuations it is sufficient to
set $ \vp=\k \tau$ in \rf{5.22} and expand in $y_s$:
\be
&& L= \ha ( \dot y^2_r - y'^2 _r - \k^2 y_r^2 ) + q \k \ep_{sr} y_s y'_r + O(y^4) \no \\
&& \ \ \ = \ha \Big[ \dot y^2_r - ( y' _r - \k q \ep_{sr} y_s)^2 - \k^2 (1 - q^2) y_r^2 \Big] + O(y^4) \ . \la{5.3} \ee
After a local $\s$-dependent rotation of $y_s$ (which shifts the spatial momentum by $\k q$)
one finds two massive modes with $m^2 = \k^2 (1 - q^2) $.
If $\s$ is $2\pi$-periodic as required for the closed-string world sheet, this local rotation of $y_s$
is not possible in general.
Indeed, the momentum-space
dispersion relation that follows directly from \rf{5.3} is
\be \la{dii}
&& e^2 - ( p \pm \k q)^2 = (1-q^2) \k^2 \ , \ \ \ \ \ \ \ \ \ \ \ e\equiv p_0 \ , \ \ \ \ \ \ p\equiv p_1 \ , \\
&& e =\pm \sqrt{ p^2 \pm 2 \k q p + \k^2} \ , \la{did}
\ee
where $p$ takes integer values $p=n=0,1,2,...\ $. One can therefore shift $p$ and get the standard massive relativistic dispersion relation only
if $\k q$ is integer.
Note that in the WZW case we get a massless dispersion relation :
\be \la{wes}
q=1: \ \ \ \ \ \ \ \ \ \ \ \ \ \ \ \ \ \ \ e= \pm p \pm \k \ . \ee
The formal redefinition of $y_s$ or shift of the spatial momentum is allowed, however,
in the discussion of the S-matrix we are interested here, for which we
consider the limit $\J \gg 1, \ \ p=n \gg 1$, and thus effectively decompactify the $\s$ direction.
Explicitly, we may rescale $ e \to \J e , \ p \to \J p$ so that the new momentum $ p = { n \ov \J}$
takes continuous values. We then find (for the particle $e > 0$ branch)
\be
e= \sqrt{ \hp^2 + 1-q^2 } \ , \ \ \ \ \ \ \ \ \ \ \ \ \ \ \hp = p \pm q \ . \la{hap} \ee
The minimal energy states correspond to $\hp =0$ and fluctuations near this vacuum have the
non-relativistic massive dispersion relation,
\be \la{mss} e= m + {\hp^2 \ov 2 m} +... \ , \ \ \ \ \ \ \ \ \ \ \ \ \ \ \ m= \sqrt{1-q^2} \ . \ee
Therefore, as long as all finite size effects are ignored and $p$ is treated is continuous,
it can be shifted by $\pm q$ and we end up with the standard massive magnon dispersion relation
with $q$-dependent mass $m= \sqrt{1-q^2} $.

This suggests that for $0< q < 1$ the corresponding spin chain interpretation
(e.g., via the connection to the Landau-Lifshitz model \ci{kr1,kr2}) of this near-BMN
expansion should be based again on a picture of magnon scattering
near a ferromagnetic vacuum just as for the $q=0$ case.
As we shall see below (in Appendix D), this conclusion is also supported by the analysis of the corresponding Pohlmeyer-reduced theory
which is given again by the complex sine-Gordon model but with the rescaled mass parameter
$ \mu \to \mu \sqrt{1-q^2} $.

A similar analysis applies to other massive \adst superstring modes, which turn out to
have the same mass $\sqrt{1-q^2} $, in agreement with what was found directly in the BMN limit in \ci{bmn,rut,cz}
(see also Appendix \ref{A_2}). 

%%%%%%%%%%%%%%%%%%%%%%%%%%%%%%%%%%%%%%
\subsection{Gauge fixing and expansion of the action to quartic order}\label{2_3}
%%%%%%%%%%%%%%%%%%%%%%%%%%%%%%%%%%%%%%

A systematic way to compute the S-matrix for the above elementary massive excitations
is to fix a gauge where in addition to $t\sim \tau$ one sets
the momentum density corresponding to the angle $\vp$ of $S^3$ to be constant.\foot{Note that in general
one cannot fix the conformal gauge and
in addition fix the fluctuation of $\vp$ to be zero:
the residual conformal transformations are parametrized only by the two functions $f(\s^+)$ and $\td f (\s^-)$ but $\vp$
will not satisfy a free massless equation. This is still possible at quadratic
order in the expansion in $y_s$ as discussed above.}
This can be done as in \ci{rtt,km} (following \ci{kr2,kt})  by first
applying  T-duality
in the $\vp$ direction and then fixing the static gauge $t=\k \tau, \ \ \td \vp = \k \s$.\foot{Use of 
T-duality   here is a formal trick  to  choose   a  gauge    where the momentum  conjugate to 
$\vp$ is fixed; we are not interested in  T-dual theory as such. 
Alternative ways of doing near-BMN expansion were discussed in \ci{ft2,cal,AF}.}
 
To compare to the $AdS_5 \times S^5$ case \ci{km} it is useful to consider a
one-parameter family of gauges which includes also the uniform light-cone gauge \ci{AF,fpz}. % a more general gauge
We shall thus start with the Polyakov action for string in $R_t \times S^3$, set
\be
t= u   -    b \, \vp \ , \ \ \ \ \ \ \ \ \ \ \ \ b\equiv { a \ov 1-a}
\ , \la{b} \ee
where $u$ is a new coordinate and $a$ is a gauge parameter,
and then perform the
T-duality transformation in the $\vp$ direction.
The resulting dual Lagrangian is (cf. \rf{5.22},\rf{225},\rf{522})
\be
&&\td L = - \sqrt g g^{cd} h_{cd} - b P \ep^{cd} \del_d u ( \del_c \tv - B_s \del_c y_s) \la{le} \ , \\
&&h_{cd}= - Q \del_c u \del_d u + P ( \del_c \tv - B_s \del_c y_s) ( \del_d \tv - B_s \del_d y_s)
+ F\del_c y_s \del_d y_s \ , \la{523} \\
&& Q = 1 + b^2 P \ , \ \ \ \ \ \ \ \ \ \ P = ( G - b^2 )^{-1} \ . \la{p} \ee
Finally, we shall fix the following gauge:
\be
u=
\bc \ \tau \ , \ \ \ \ \ \ \ \ \ \ \ \ \ \ \td \vp = \bc \, \J \, \s \ , \ \ \ \ \ \ \ \ \ \ \ \ \ \bc \equiv { 1\ov 1-a} \ . \la{gai} \ee
Here $a=0$ is the analog of the ``static'' gauge and $a=\ha$ is the uniform light-cone gauge
(when $u=t + \vp =2 \tau$, cf. \rf{5.1}). 
We shall also rescale $\s$ to absorb the $\J= { J \ov \sql}$ factor so that the cylinder is
decompactified in the $\J \gg 1$ limit.\foot{As already mentioned,
all finite size effects will be ignored as we will be interested only in the S-matrix.
}
The resulting dispersion relation then takes the form \rf{hap}.

Solving for the 2d metric $g_{cd}$
(and setting from now on $\J=1$ in \rf{gai} as discussed above)
the corresponding Nambu Lagrangian in the gauge \rf{b} takes the form
\be
&&\td L = - \sqrt h + { b }\, \bc P ( \bc - B_s y'_s) \ , \ \ \ \ \ \ \ \ \ \ \ \la{223} \\
&&h = \big[ \bc^2 Q - P (B_r \dot y_r)^2 - F \dot y^2_r \big]\big[ P (\bc - B_s y'_s)^2 + F y'^2_s\big] +
\big[ P B_s \dot y_s (\bc - B_r y'_r) - F \dot y_r y'_r \big]^2. \no
\ee
Expanding $\td L$ in powers of $y_s$ we get the following Lagrangian
\be
&&\td L= L_2 + L_4 +... \ , \no \\
&&L_2 = \ha ( \dot y^2_s - y'^2 _s - y_s^2 ) + q \ep_{sp} y_s y'_p \ , \la{222} \\
&& L_4=
\ha y^2_s y'^2_r + \ha q \big[ \dot y_r y'_r \ep_{sp} y_s \dot y_p
- \ha ( \dot y^2_r + y'^2_r + y^2_r ) \ep_{sp} y_s y'_p \big] \no \\
&& \ \ \ \ \ \ \ \ \ + ( a- \ha) \Big\{ \fo ( y^2_s)^2 - \fo (\dot y^2_s + y'^2_s)^2 + ( y' _s \dot y_s )^2 \no \\
&& \ \ \ \ \ \ \ \ \ \ \ \ \ \qquad \qquad + \ q \big[ - \dot y_r y'_r \ep_{sp} y_s \dot y_p
+ \ha ( \dot y^2_r + y'^2_r - y^2_r )\ep_{sp} y_s y'_p \big] \Big\} \ . \la{434}
\ee
The quadratic part $L_2$ here is the same as in \rf{5.3} leading to the massive
dispersion relation \rf{hap}.
For $q=0$ the quartic part $L_4$ agrees with eq.~(4.16) in \ci{km}.\foot{As was observed there,
the quartic part simplifies in the ``light-cone'' gauge $a= \ha$.
This does not appear to apply to the $q$-dependent terms.
Note also that to quartic order the Lagrangian is linear in $q$.
However, the tree S-matrix will not be linear in $q$ as $q$ appears also in the dispersion relation.}
Let us quote also the explicit expressions for $a=0, \ha, 1$:
\be
&& L_4(a=0) = \ha\big[ y^2_s y'^2_r - \fo (y^2_s)^2 + \fo (\dot y^2_s + y'^2_s)^2
- ( y' _s \dot y_s )^2\big]
\no\\
&&\ \ \ \ \ \ \ \ \ \ \ \ \ \ \ \ \ \ \ \ \ \ \ \ \ + q \big[ \dot y_r y'_r \ep_{sp} y_s \dot y_p - \ha ( \dot y^2_r
+ y'^2_r ) \ep_{sp} y_s y'_p \big] \ \la{333}\ , \\
&& L_4(a=\ha ) =
\ha y^2_s y'^2_r + \ha q \big[ \dot y_r y'_r \ep_{sp} y_s \dot y_p
- \ha ( \dot y^2_r + y'^2_r + y^2_r ) \ep_{sp} y_s y'_p \big] \ , \la{haha} \\
&& L_4(a=1) = \ha\big[ y^2_s y'^2_r + \fo (y^2_s)^2 - \fo (\dot y^2_s + y'^2_s)^2
+ ( y' _s \dot y_s )^2 \big] \ - \ \ha q\, y^2_r \, \ep_{sp} y_s y'_p \la{3ha}\ .
\ee
The gauge dependence of the resulting S-matrix implies that it is not a directly observable quantity;
in the corresponding Bethe ansatz it reflects the ambiguity in the choice of the corresponding spin chain length \ci{fpz,km,afr}.

The above discussion admits a straightforward generalization to the case of
bosonic string theory on $AdS_3 \times S^3$: the Lagrangian
\rf{5.22} picks up three extra terms representing the $AdS_3$ part, which can be obtained from the $S^3$ terms
by a formal analytic continuation: $\vp \to t$, $y_s \to i z_s$,
and then reversing the overall sign.
We present the combined action in Appendix \ref{appa}   where we also include the massless $T^4$  directions. 

\

The quadratic Lagrangian $L_2$ in \rf{222} can be ``diagonalized'' by a $\s$-dependent rotation of $y_s$, i.e.
by setting
\be
y= y_1 + i y_2 = e^{ i q \s} v \ , \ \ \ \ \ \ \ \ \ \ \ \ y^*= y_1 - i y_2 = e^{- i q \s} v^* \ . \la{ze} \ee
We then find the standard massive Lagrangian for the complex scalar $v$
\be \la{vev}
&&L_2 =\ha \big[ \dot y \dot y^* - ( y' - i q y) ( y^*{}' + i q y^*) - (1-q^2) yy^* \big] \no \\
&& \ \ \ \ =
\ha \big[ \dot v \dot v^* - v' v'^* - (1-q^2) v v^* \big] \ , \ee
corresponding once again to the dispersion relation \rf{hap} with ``shifted'' momentum ($\hat p \to p$).

Applying the field redefinition \rf{ze} to $L_4$ in \rf{434} we get\foot{All $\s$-dependence cancels out of course due to the global $U(1)$ invariance. Therefore the transformation simply amounts to
shifts of the spatial derivatives $y' = v' + i q v, \ y'^* = v'^* - i q v^*$, i.e. shifts of the corresponding spatial momenta.} %V4 %vv4
\be && L_4 = \vo \Big\{ 2 q^2 (1 - q^2) v^2 \vc^2 + 3 i q (1 - q^2) v \vc (v \vc' - \vc v') + 4 (1-q^2) v \vc v' v'^* \no \\
&&
\ \ \ \ \ \ \ +\, q^2 \big[v^2 (v'^*{}^2 - \dot v^*{}^2) + \vc^2 (v'^2 - \dot v^2) \big]
- i q \big[ vv' (v'^*{}^2 - \dot v^*{}^2) - v^* v'^* (v'^2 - \dot v^2) \big] \Big\} \no \\
&&\ \ \ \ \ \ \ \ \ \ \ \ +\, \fo ( a- \ha) \Big\{
(1 - q^2)^2 v^2 \vc^2 - i q (1 - q^2) v \vc (v \vc' - \vc v') \no \\
&& \ \ \ \ \ \ \ \ \ \ \ \ \ \ \ \ - \, i q \big[ vv' (v'^*{}^2 - \dot v^*{}^2) - v^* v'^* (v'^2 - \dot v^2)\big]
- (v'^*{}^2 - \dot v^*{}^2) (v'^2 -\dot v^2)\Big\} \ . \la{fee} \ee
For $q=0$ and $q=1$ we have
\be && L_4(q=0) = \ha v \vc v' v'^* + \fo ( a- \ha) \Big[
v^2 \vc^2
- (v'^*{}^2 - \dot v^*{}^2) (v'^2 -\dot v^2)\Big] \ , \la{fee1} \\
&& L_4(q=1) = \vo \Big[ v^2 (v'^*{}^2 - \dot v^*{}^2) + \vc^2 (v'^2 - \dot v^2)
- i \big[ vv' (v'^*{}^2 - \dot v^*{}^2) - v^* v'^* (v'^2 - \dot v^2) \big] \Big] \no \\
&&\ \ \ \ \ +\, \fo ( a- \ha) \Big[
- i \big[ vv' (v'^*{}^2 - \dot v^*{}^2) - v^* v'^* (v'^2 - \dot v^2) \big]
- (v'^*{}^2 - \dot v^*{}^2) (v'^2 -\dot v^2)\Big]
\ . \la{fee2} \ee
Note that for $q=1$, i.e. at the WZW point,
the Lorentz invariance and also the parity invariance
of the quartic Lagrangian is restored in the $a=0$ gauge:
\be && L_4(q=1, a=0) = \vo \Big[ v^2 (v'^*{}^2 - \dot v^*{}^2) +
\vc^2 (v'^2 - \dot v^2) + (v'^*{}^2 - \dot v^*{}^2) (v'^2 -\dot v^2)\Big] \no\\
&& \ \ \ \ \ \ \ \ \ \ \qquad \qquad =\vo \Big[ - (v^2 \del_+ v^* \del_- v^* + v^*{}^2 \del_+ v \del_- v )
+ \del_+ v \del_- v \del_+ v^* \del_- v^* \Big] \ . \la{ff} \ee
The reason for this was previously mentioned below \rf{roo}.\foot{Note that
for $q=1$ we have massless excitations, i.e. $v= u_+ + u_-$ where $\del_- u_+ =0, \ \del_+ u_-=0$
are formal ``in''-fields. Then integrating by parts we get
\be
L_4(q=1,a=0) = \fo ( u^*_+ u^*_- \del_+ u_+ \del_- u_- + c.c. )
+ \vo \del_+ u_+ \del_+ u_+^* \del_- u_- \del_- u_-^* \ . \no \ee
While the meaning of the corresponding S-matrix is not immediately clear, there are no LL or RR, only LR scattering processes
(see also section 3.2 below). }

%%%%%%%%%%%%%%%%%%%%%%%%%%%%%%%%%%%%%%
\def\TT{{\rm T}}
\def\SS{{\rm S}}
\def\de{\delta}
\newcommand{\yp}{{y_\pm}}
\newcommand{\ym}{{y_\mp}}
\newcommand{\zp}{{z_\pm}}
\newcommand{\zm}{{z_\mp}}
\newcommand{\ip}{{\zeta_\pm}}
\newcommand{\im}{{\zeta_\mp}}
\newcommand{\up}{{\chi_\pm}}
\newcommand{\um}{{\chi_\mp}}
\def\tT{\text{T}}
\def\tTv{\text{T}^v{}}
\def\htTv{{\text{\hat T}^v}}
\renewcommand{\theequation}{3.\arabic{equation}}
\setcounter{equation}{0}
\section{Tree-level S-matrix of bosonic string on $R\times S^3$ with $B$-flux\label{sec3}}
%%%%%%%%%%%%%%%%%%%%%%%%%%%%%%%%%%%%%%

Starting with the Lagrangian \rf{222},\rf{434} or \rf{vev},\rf{fee} it is straightforward
(following \ci{rtt,roi,km}) to compute
the corresponding tree-level 2-particle S-matrix for the elementary massive excitations of the bosonic string
on $S^3$ with non-zero $B$-field flux.

Despite the apparently complicated dependence of $L_4$ on $q$ (cf. eq.~\rf{fee})
we shall find that
the resulting S-matrix has a very simple structure: its expression for $q\not=0$
can be found from its $q=0$ limit by replacing $e(p) = \sqrt{ p^2 + 1}$ with
the modified dispersion relation $e(p) = \sqrt{ (p\pm q)^2 + 1-q^2}$.
%V4
Here  the plus sign corresponds  to,  e.g., boson $y$  and the minus sign to $y^*$ 
in \rf{vev}:  we assume that $y \sim e^{- i e \tau - i p \sigma}$ 
as appropriate for  ingoing fields  in a  scattering amplitude.

The S-matrix may be written as
\begin{equation}\la{fffi}
\mathbb{S} = \mathbb{I} + {i}{\hh^{-1}}\, \mathbb{T} \ , \qquad \qquad \hh^{-1} \equiv \frac{2\pi}{\sqrt{\lambda}} \ ,
\end{equation}
where
$\mathbb{I}$ is the identity operator, and $\mathbb{T}$ is the interaction part of the scattering operator.
As the theory under consideration is
an integrable one we should have particle-number conservation and thus we can
represent $\mathbb{T}$ as a map from two-particle states to two-particle states ($p$ and $p'$ are spatial momenta)
\begin{equation}\la{smatdef}
\mathbb{T} \left|y_m(p) y_n(p')\right> = \tT_{mn}^{rs}(p,p')\left|y_r(p) y_s(p')\right> \ .
\end{equation}

The expression for the S-matrix of the $R \times S^3$ theory in the case of the vanishing NSNS flux (for $q=0$)
is a truncation \ci{roi} of the $S^5$ part of the $AdS_5 \times S^5$ result in \cite{km}.
Let us start with summarizing this expression and then present our result for non-zero $q$.

%%%%%%%%%%%%%%%%%%%%%%%%%%%%%%%%%%%%%%
\def\hhi{\kappa}
\subsection{Vanishing $B$-flux} % ($q=0$)}
%%%%%%%%%%%%%%%%%%%%%%%%%%%%%%%%%%%%%%

Considering only the $S^3$ sector (i.e. two massive states) we
may simply take the $S^5$ expression for the S-matrix in \ci{km} and restrict the $SO(4)$ index to an $SO(2)$ index
(the same result follows of course directly from \rf{434} with $q=0$)
\begin{equation}\begin{split}
\tT_{mn}^{rs}(p,p') = & \Big[\frac{p^2+p'^2}{2(e'p-ep')} +
\big(a-\ha\big)(ep'-e'p)\Big] \delta_{m}^{r}\delta_{n}^{s} - \frac{pp'}{e'p-ep'} \epsilon_{m}^{\ r} \epsilon_{n}^{\ s} \ ,
\\ & e =\sqrt{p^2 + 1} \ , \qquad e' = \sqrt{p'{}^2 + 1} \ . \la{311}
\end{split}\end{equation}
Here $\ep_{rs}=- \ep_{sr} $ and $ \epsilon_{m}^{\ r} \epsilon_{n}^{\ s}= \delta_{mn}\delta^{rs}- \delta_{m}^{s}\delta_{n}^{r} $.
It is useful to change the basis of fields from the real $y_s=(y_1,y_2)$ to the complex $(y,y^*)=y_1 \pm iy_2$ one.
We shall use labels $+$ for $y$ and $-$ for $y^*$.
In this basis one finds that
$ \tT_{++}^{++}= \tT_{--}^{--}\not=0, \ \ \tT_{+-}^{+-}= \tT_{-+}^{-+}\not=0$,
with all other amplitudes vanishing. Explicitly,
\begin{equation}\begin{split}\label{bhreduce}
& \tT_{\pm\pm}^{\pm\pm}(p,p') =
\frac{(p+p')^2}{2(e'p-ep')} + \big(a-\ha\big)(ep'-e'p) = \ha \big(\frac{p+p'}{p-p'}\big)(ep'+e'p) + \big(a-\ha\big)(ep'-e'p) \ ,
\\ & \tT_{\pm\mp}^{\pm\mp}(p,p') = \frac{(p-p')^2}{2(e'p-ep')} + \big(a-\ha\big)(ep'-e'p) = \ha \big(\frac{p-p'}{p+p'}\big)(ep'+e'p) + \big(a-\ha\big)(ep'-e'p) \ ,
\\ & \tT_{\pm\mp}^{\mp\pm}(p,p') = 0 \ , \qquad \qquad e =\sqrt{p^2 + 1} \ , \qquad e' = \sqrt{p'{}^2 + 1} \ ,
\end{split}\end{equation}
where we have simplified the expressions using the identity
\begin{equation}\label{bhidentity}
(e'p+ep')(e'p-ep') = (p+p')(p-p') \ .
\end{equation}
The above tree amplitude $\tT\equiv \tT_{++}^{++}$ matches \ci{rtt,roi,km} the expansion
of the AFS \ci{afs} spin chain S-matrix in the $SU(2)$ sector.

Let us recall for completeness
the expressions for leading-order (tree-level) S-matrices in several similar models.
In the Landau-Lifshitz (LL) model \ci{kz} (here we use $\hhi= \hh^{-1} $ as an effective coupling)
\be \label{LLe}
\SS_{\rm LL} (p,p') = \frac{1 + { i\hhi \, p p' \ov p - p' } }{1 - {i \hhi\, p p' \ov p - p' } }= 1
+ {2i \hhi \, p p' \ov p - p' } +...
\ , \ee
where $ \tT= {2 ip p' \ov p - p' } $ matches the expansion of the magnon S-matrix in
the XXX spin chain Bethe ansatz.
In the Faddeev-Reshetikhin (FR) model \ci{kz}
\be \la{FRe}
&& \SS_{\rm FR} (p,p') = \frac{1 + {2i\hhi\, \ov x'-x } }{1 - {2i\hhi\, \ov x'-x } }= 1 + { 4i\hhi\, \ov x' - x} +... \ , \ \ \ \ \ \\
&&
x (p) = { 1 \ov p } \big[ e(p) +1\big] \ , \ \ \ \ \ \ \ \ \ \ \ \ \ \ e(p) = \sqrt{ p^2 + 1} \ , \ee
i.e.
\be \SS_{\rm FR} (p,p') = 1 + { 4i \hhi\, p p' \ov p (e' +1) - p' ( e + 1) } +...
= 1 + { 2i \hhi\, [p p' - (e-1) (e'-1)] \ov p e' - p' e } +...\ .
\la{ffr} \ee
This agrees with \rf{LLe} for small momenta.
The FR S-matrix \eqref{ffr} is somewhat
similar to the expansion of the BDS \ci{bds} spin chain S-matrix given by
\ci{rtt}
\be \la{bdsa}
&& \SS_{\rm BDS} (p,p') = \frac{1 + { i \hhi\ov u'-u } }{1 - { i\hhi \ov u'-u } }= 1 + { 2i \hhi \ov u' - u} +...
= 1 + { 2i \hhi\, pp' \ov pe' - p'e } +... \ , \ \ \ \ \ \ \ \ \ \\
&& {\textstyle u(p) = \ha \cot { p\ov 2} \sqrt{ 1 + {\textstyle { \l\ov \pi^2} \sin^2 { p \ov 2}}} \ , \ \ \ \ \ \ \ \ u_{p\to 0, \ \l p^2 ={\rm fixed}} \ \to { 1 \ov p }
e( \bar p) \ , \ \ \bar p= {\sql \ov 4 \pi} p\ . }
\ee
The asymptotic Bethe ansatz S-matrix \ci{rev} contains, in addition to the BDS factor, the dressing phase, given
at strong coupling by the AFS expression. The limit that allows to compare to the string world-sheet sigma model S-matrix is
$p\to 0$, $\l \gg 1$ with $\l p^2$=fixed.
In terms of the fixed (rescaled or string sigma model) momenta one then finds for
the corresponding leading (string tree-level) part of the spin chain S-matrix \ci{rtt,roi,km}
\be
&& \SS_{\rm AFS} (p,p') = 1 + { 2i\hhi\, F(p,p') \ov pe' - p'e } +... \ , \ \ \ \ \ \ \ \ \ \ \ \ \ e = \sqrt{ p^2 + 1} \ , \la{aff}\\
&& { 2F(p,p') \ov pe' - p'e } = { 2pp' \ov pe' - p'e } + \theta_{\rm AFS} (p,p')\ , \la{faf} \\
&& \theta_{\rm AFS} (p,p') = { ( p-p' )^2 \ov 2(pe' - p'e) } + \ha ( pe' - p'e) - ( p-p') \ , \la{fai} \\
&&
F(p,p') =
pp' - (e-1) (e'-1) + \fo \big[pp' - (e-1) (e'-1) \big]^2 \ . \la{ifa}
\ee
A useful equivalent form is
\be
{ 2F(p,p') \ov pe' - p'e } =
{ ( p+p' )^2 \ov 2(e' p - e p') } + \ha ( e'p - e p') - (p - p' ) \ . \la{vot}
\ee
The expression \rf{vot} should be compared to
the one for $T^{++}_{++}$ in \rf{bhreduce} taken in the $a=0$ gauge
(where the BMN charge $J$ plays the role of the spin chain length)
\be
\tT^{++}_{++}(p,p'){}_{_{a=0}}
= { ( p+p' )^2 \ov 2(e' p - e p') } + \ha ( e'p - e p')= { e\, p'^2 + e'\, p^2 \ov p-p'} \ . \la{tat}
\ee
They match up to the last term in \rf{vot}, which is linear in momentum and hence cancels in the Bethe ansatz equations \ci{km}.

%%%%%%%%%%%%%%%%%%%%%%%%%%%%%%%%%%%%%%
\subsection{Non-vanishing $B$-flux} %($ 0 \leq q \leq 1$)}
%%%%%%%%%%%%%%%%%%%%%%%%%%%%%%%%%%%%%%

To find the $q\not=0$ generalization of \rf{bhreduce} let us start with
the action \rf{vev}, \rf{fee} written in terms of the complex scalar $v$ \rf{ze}. The corresponding
particle dispersion relation is the standard massive one, $ e(p) = \sqrt{ p^2 + 1-q^2} $.
Labelling again the excitations associated to $v$ and $ v^*$ with the index $+$ and $-$
the tree-level S-matrix that follows directly
from the quartic action \eqref{fee} is found to be\footnote{As usual, the factor $\frac1{4(e'p-ep')}$
arises when solving the two-dimensional delta-function constraint imposing the energy conservation. The superscript $v$ on $T$
indicates that
we are scattering excitations associated to $v$ and $v^*$. Later we will return to the S-matrix for $y$ and $y^*$.
}
\begin{equation}\begin{split}\label{bhscatter1}
\tTv_{\pm\pm}^{\pm\pm} (p,p') = &
\frac{p+p'\mp 2 q}{2 (e' p-e p')} \Big[(1-q^2) (p+p'\mp q)\mp q (e e'-p p')\Big]
\\ & \ \ \ \ \ + \big(a-\ha\big) \Big[(e p'-e' p)\mp \frac{q (p+p') \left(e e'-p p'-\left(1-q^2\right)\right)}{e' p-e p'}\Big]\ ,
\\\tTv_{\pm\mp}^{\pm\mp}(p,p') = &
\frac{p-p'\mp 2 q}{2 (e' p-e p')} \Big[(1-q^2) (p-p'\mp q)\pm q (e e'-p p')\Big]
\\ & \ \ \ \ \ + \big(a-\ha\big) \Big[(e p'-e' p)\pm \frac{q (p-p') \left(e e'-p p'+\left(1-q^2\right)\right)}{e' p-e p'}\Big] \ , %v5 corrected signs
\\\tTv_{\pm\mp}^{\mp\pm}(p,p') = & 0 \ , \qquad \qquad e = \sqrt{p^2 + 1-q^2} \ , \qquad e' = \sqrt{p'{}^2 + 1 - q^2} \ .
\end{split}
\end{equation}
The dependence on $q$ is explicit as well as implicit through the energies $e,e'$.
These expressions of course agree with \eqref{bhreduce} for $q= 0$.

Using the generalization of the identity \eqref{bhidentity} to non-zero $q$\foot{Note also that \
$ { ee' - pp' + (1-q^2) \ov e'p - ep'} = { e + e' \ov p-p'} $ \ and \
$ { ee' - pp' - ( 1-q^2 ) \ov e'p - ep'} = { e - e' \ov p+ p'} \ .$ }
\begin{equation}\label{bhidentity2}
(e'p+ep')(e'p-ep') = (1-q^2)(p+p')(p-p') \ ,
\end{equation}
the scattering amplitudes \eqref{bhscatter1} can be rewritten in the following simple form
\begin{equation}\begin{split}\label{bhscatter2}
& \tTv_{\pm\pm}^{\pm\pm} (p,p') =
\frac{p+p'\mp 2 q}{2(p-p')} \big[e (p'\mp q)+e' (p\mp q)\big] +\big(a-\ha\big) \big[e (p'\mp q)-e' (p\mp q)\big] \ ,
\\& \tTv_{\pm\mp}^{\pm\mp}(p,p') =
\frac{p-p'\mp 2 q}{2(p+p')}\big[ e (p'\pm q)+e' (p\mp q)\big] +\big(a-\ha\big) \big[ e (p'\pm q)-e' (p\mp q)\big] \ , %v5 corrected signs
\\& \tTv_{\pm\mp}^{\mp\pm} (p,p') = 0 \ , \qquad \qquad e = \sqrt{p^2 + 1-q^2} \ , \qquad e' = \sqrt{p'{}^2 + 1 - q^2} \ .
\end{split}
\end{equation}
We thus discover that the dependence of the S-matrix on $q$ is remarkably simple:
it can be absorbed
into the shifts of the momentum of the particle $v$ by $-q$ and of the momentum of the
antiparticle $v^*$ by $+q$ (the denominators are invariant under such shifts). %vv4
Note that the shifts apply only to the spatial momenta, not to the energies $e,e'$.

These momentum shifts are precisely the same as those done in eq.~\rf{ze}, which put the action \rf{434}
into the form \rf{fee} with a standard massive dispersion relation. Thus undoing these shifts simply amounts to going back
to the ``unrotated'' basis of fields $y=y_1+iy_2, \ y^*=y_1-i y_2$.
This leads to the following final very simple expression for the S-matrix and dispersion relations
for the $(y,y^*)$ in the $0 \leq q \leq 1$ case:
\begin{equation}\begin{split}\label{bhscatter3}
& \tT_{\pm\pm}^{\pm\pm} (p,p') =
\frac{p+p'}{2(p-p')} (e_\pm p'+e'_\pm p)+\big(a-\ha\big) (e_\pm p'-e'_\pm p) \ ,
\\& \tT_{\pm\mp}^{\pm\mp}(p,p') =
\frac{p-p'}{2(p+p')}(e_\pm p'+e'_\mp p)+\big(a-\ha\big) (e_\pm p'-e'_\mp p) \ ,
\\& \tT_{\pm\mp}^{\mp\pm} (p,p') = 0 \ , \qquad \qquad e_\pm = \sqrt{(p\pm q)^2 + 1-q^2} \ , \qquad e'_\pm = \sqrt{(p' \pm q)^2 + 1-q^2 } \ ,
\end{split}
\end{equation}
where the $\pm$ indices on $e$ indicate
that for states with positive/negative charge we take $p \pm q$ in the dispersion relation.

The form of the S-matrix is thus formally the same as in \rf{bhreduce} but with different $e(p)$: all the 
dependence on $q$ enters through $q$-dependence of the energy $e(p)$ in \rf{bhscatter3}.
To summarize, the S-matrix  admits two  equivalent representations, depending on  whether  we scatter 
the rotated fields  $v$ or   the original   fields  $y$  which 
 are symbolically  (here $e(p,m)\equiv \sqrt{ p^2 + m^2}$ and $p_i=(p,p')$): 
\begin{equation} T^v\big(p_i \pm q , e(p_i,1-q^2)\big) \qquad \textrm{and} \qquad
T\big(p_i, e(p_i\pm q,1-q^2)\big) \ . \nonumber
\end{equation}

This remarkable property of the S-matrix, which is by no means obvious from
the action \rf{434} (having a non-trivial dependence on $q$) should be a consequence
of some symmetry of the underlying integrable model.\foot{Let us note also that the corresponding Pohlmeyer-reduced theory
depends on $q$ only via the rescaling of the mass parameter by $(1-q^2)^{1/2}$. Therefore, its (relativistic) S-matrix takes the same
form as in the $q=0$ case, see Appendix \ref{appd}.}

For the natural gauge choice $a=0$, for which comparison with the standard spin chain is
most direct (length is $J$), we find the following $q$-generalization of \rf{tat}
\be
\tT^{++}_{++}(p,p'){}_{_{a=0}} = { e \, p'^2 + e' \, p^2 \ov p - p'} \ , \ \ \ \ \ \ \ e =\sqrt{ p^2 + 2 qp + 1 }\ , \ \ \
e' =\sqrt{ p'^2 + 2 qp' + 1 } \ . \la{opo} \ee

Let us now comment on
some properties of the tree-level S-matrix \eqref{bhscatter2}. First, it satisfies the following identities
\begin{equation}\begin{split}
& \tTv_{\pm\pm}^{\pm\pm}(p,p') = \tTv_{\pm\mp}^{\pm\mp}(-p',p)\Big|_{e' = -\sqrt{p'^2+1-q^2}} \ ,
\\ & \tTv_{\pm\pm}^{\pm\pm}(p,p') + \tTv_{\pm\pm}^{\pm\pm}(p',p) =0 \ , \qquad \ \ \ \ \tTv_{\pm\mp}^{\pm\mp}(p,p') + \tTv_{\mp\pm}^{\mp\pm}(p',p) = 0 \ , \vphantom{\Big|}
\\ &\big[ \tTv_{\pm\pm}^{\pm\pm}(p,p'{})\big]^* + \tTv_{\pm\pm}^{\pm\pm}(p'{},p) = 0 \ , \qquad \big[
\tTv_{\pm\mp}^{\pm\mp}(p,p'{})\big]^* + \tTv_{\mp\pm}^{\mp\pm}(p'{},p) = 0 \ , \quad p,p' \in \mathbb{R} \ . \vphantom{\Big|}
\end{split}\end{equation}
These are the crossing symmetry, braiding unitarity and hermitian analyticity relations respectively. The latter two combined imply the
expected
QFT unitarity of the S-matrix.
Furthermore, as the T-matrix is diagonal (${\tTv}_{\pm\mp}^{\mp\pm} (p,p')=0 $), it trivially satisfies
the classical Yang-Baxter equation.

In the special case of $q = 1$ when the world-sheet action
is given by the $SU(2)$ WZW model we get massless excitations (see \rf{wes}):
there are left- and right-moving modes, for which $p = -e = p_L$ and $p = e = p_R$ respectively. %v5 - removed factors of -1/2 and 1/2 respectively (typos)
To consider the $q \rightarrow 1$ limit in the S-matrix \eqref{bhscatter2}
let us first multiply it by the Jacobian factor $e'p-ep'$.\footnote{One can easily see that for massless scattering states
the inverse of this factor can be divergent and as such it should be regularized properly. We shall
avoid this issue by working with the standard QFT amplitudes, i.e. coefficients of $\delta^{(2)}(p_1+p_2+p_3+p_4)$.}
We may then compute different possible scattering amplitudes, i.e. left-left (LL), right-right (RR), left-right (LR) and left-right (RL), by simply substituting in the on-shell relations into \rf{fee2} as appropriate.
The LL and RR amplitudes vanish, while for the LR scattering processes we find\footnote{Note that if the in-state consists of a left mode and a right mode, then simple 2-d kinematical considerations show that the out-state must also consist of a left mode and a right mode. Furthermore, the momentum of the ingoing left mode should equal the momentum of the outgoing left mode and similarly for the right mode.}
\begin{equation}\begin{split}\label{bhscattermass1}
{\htTv}_{\pm\pm}^{\pm\pm} (p_{_L},p_{_R}') = & 2 p_{_L} p_{_R}'\big[(1-2a) p_{_L} p_{_R}' - 1 \pm a (p_{_L} + p_{_R}') \big] \ ,
\\ {\htTv}_{\pm\mp}^{\pm\mp}(p_{_L},p_{_R}') = & 2 p_{_L} p_{_R}'\big[(1-2a) p_{_L} p_{_R}' + 1 \mp a (p_{_L} - p_{_R}') \big]\ ,  %v5 removed typo '
\\ {\htTv}_{\pm\mp}^{\mp\pm} (p_{_L},p_{_R}') = & 0 \ ,
\end{split}
\end{equation}
and the RL scattering amplitudes immediately follow from these.
For the gauge choice $a = 0$ this tree-level S-matrix is relativistically invariant as expected (see \rf{ff}).

The simple structure of the S-matrix \rf{bhscatter3} in the $R\times S^3$ sector we have found above
has a direct analog
in the case of strings moving in $AdS_3\times S^1$
(one needs just to reverse sign of the first term in $\tT^{++}_{++}$, etc.).
Using the Lagrangian of Appendix \ref{appa} we have also computed
the full bosonic S-matrix in the $AdS_3 \times S^3$ sector and again
the results for $q=0$ and $0<q\leq 1$ are found to be related by momentum shifts in
the energy as described 
above.\foot{In Appendix \ref{appa}   we comment also  on the  S-matrix including  the massless   $T^4$ modes.}
This pattern also suggests a generalization to the full
\adst superstring with $0\leq q\leq1$, which is discussed in the next section.

%%%%%%%%%%%%%%%%%%%%%%%%%%%%%%%%%%%%%%
\def\hT{\hat T}
\def\nquad{\!\!\!\!\!\!\!\!}
\renewcommand{\theequation}{4.\arabic{equation}}
\setcounter{equation}{0}
\section{Tree-level S-matrix of \adst superstring  with mixed flux\label{sec4}}
%%%%%%%%%%%%%%%%%%%%%%%%%%%%%%%%%%%%%%
The bosonic-sector results of the previous section
suggest a natural generalization to the full tree-level world-sheet S-matrix for the
massive BMN modes of superstring theory on \adst
with a mixed RR-NSNS flux.
%%%%%%%%%%%%%%%%%%%%%%%%%%%%%%%%%%%%%%
\subsection{Vanishing $B$-flux}
%%%%%%%%%%%%%%%%%%%%%%%%%%%%%%%%%%%%%%
Let us start with the $q=0$ (pure RR flux) case. 
The corresponding  massive  tree-level S-matrix   can be found from  its known 
$AdS_5 \times S^5$  counterpart  \cite{km}   by a suitable truncation.
Below we   present
the  resulting     $AdS_3 \times S^3$    S-matrix in the basis where the  massive 
excitations are represented by 
two complex bosonic ($y_\pm, z_\pm)$ \foot{Here $y_+$,$y_-$ are bosonic $S^3$ excitations
denoted by $(y,y^*)=y_1 \pm i y_2$ above. $z_+,z_-$ are there counterparts $(z,z^*) = z_1 \pm i z_2$ in the $AdS_3$ sector
(see Appendix \ref{appa}). A prime on a field  indicates that it has momentum $p'$.}    and two complex fermionic ($\zeta_\pm, \chi_\pm)$
fields:\foot{This ansatz   is only valid at tree level:  at higher loop orders additional scattering processes will appear.
For example, the
$\left|\yp\ym'\right> \rightarrow \left|\zp\zm'\right>$, $\left|\zp\zm'\right> \rightarrow \left|\yp\ym'\right>$,
$\left|\yp\zp'\right> \rightarrow \left|\zp\ym'\right>$, $\left|\zp\yp'\right> \rightarrow \left|\yp\zp'\right>$ 
amplitudes, and similar amplitudes involving the fermions, should all be non-zero at 1-loop.
Their vanishing at the tree level is actually a requirement of symmetry factorization -- see discussion
below.}
\allowdisplaybreaks{
\begin{align} \nonumber
& \text{\textbf{Boson-Boson}}
\\\nonumber & \mathbb{T}\left|\yp\yp'\right> = ( l_1 + c ) \left|\yp\yp'\right>
&& \nquad \nquad \nquad \nquad \nquad \mathbb{T}\left|\yp\ym'\right> = ( l_2 + c ) \left|\yp\ym'\right> + l_4\left|\ip\im'\right> + l_4 \left|\up\um'\right>
\\\nonumber & \mathbb{T}\left|\zp\zp'\right> = (- l_1 + c ) \left|\zp\zp'\right>
&& \nquad \nquad \nquad \nquad \nquad \mathbb{T}\left|\zp\zm'\right> = (- l_2 + c ) \left|\zp\zm'\right> - l_4\left|\up\um'\right> - l_4 \left|\ip\im'\right>
\\\nonumber & \mathbb{T}\left|\yp\zp'\right> = (l_3 + c ) \left|\yp\zp'\right> + l_5\left|\ip\up'\right> - l_5\left|\up\ip'\right>
&& \qquad \quad \ \ \, \mathbb{T}\left|\yp\zm'\right> = (l_3 + c ) \left|\yp\zm'\right>
\\\nonumber & \mathbb{T}\left|\zp\yp'\right> = (- l_3 + c ) \left|\zp\yp'\right> - l_5\left|\up\ip'\right> + l_5\left|\ip\up'\right>
&& \qquad \quad \ \ \, \mathbb{T}\left|\zp\ym'\right> = (- l_3 + c ) \left|\zp\ym'\right>
\\\nonumber &
\\\nonumber & \text{\textbf{Fermion-Fermion}}
\\\nonumber & \mathbb{T}\left|\ip\ip'\right> = c \left|\ip\ip'\right>
&& \nquad \mathbb{T}\left|\ip\im'\right> = l_4 \left|\yp\ym'\right> - l_4 \left|\zp\zm'\right>
\\\nonumber & \mathbb{T}\left|\up\up'\right> = c \left|\up\up'\right>
&& \nquad \mathbb{T}\left|\up\um'\right> = - l_4 \left|\zp\zm'\right> + l_4 \left|\yp\ym'\right>
\\\nonumber & \mathbb{T}\left|\ip\up'\right> = l_5 \left|\yp\zp'\right> + l_5\left|\zp\yp'\right>
&& \nquad \mathbb{T}\left|\ip\um'\right> = c \left|\ip\um'\right>
\\\nonumber & \mathbb{T}\left|\up\ip'\right> = - l_5 \left|\zp\yp'\right> - l_5\left|\yp\zp'\right>
&& \nquad \mathbb{T}\left|\up\im'\right> = c \left|\up\im'\right>
\\\nonumber & 
\\\nonumber & \text{\textbf{Boson-Fermion}}
\\\nonumber & \mathbb{T}\left|\yp\ip'\right> = (l_6 + c)\left|\yp\ip'\right> - l_5\left|\ip\yp'\right>
&& \nquad \mathbb{T}\left|\yp\im'\right> = (l_7 + c)\left|\yp\im'\right> + l_4\left|\up\zm'\right>
\\\nonumber & \mathbb{T}\left|\ip\yp'\right> = (l_8 + c)\left|\ip\yp'\right> - l_5\left|\yp\ip'\right>
&& \nquad \mathbb{T}\left|\ip\ym'\right> = (l_9 + c)\left|\ip\ym'\right> - l_4\left|\zp\um'\right>
\\\nonumber & \mathbb{T}\left|\yp\up'\right> = (l_6 + c)\left|\yp\up'\right> - l_5\left|\up\yp'\right>
&& \nquad \mathbb{T}\left|\yp\um'\right> = (l_7 + c)\left|\yp\um'\right> - l_4\left|\ip\zm'\right>
\\\nonumber & \mathbb{T}\left|\up\yp'\right> = (l_8 + c)\left|\up\yp'\right> - l_5\left|\yp\up'\right>
&& \nquad \mathbb{T}\left|\up\ym'\right> = (l_9 + c)\left|\up\ym'\right> + l_4\left|\zp\im'\right>
\\\nonumber & \mathbb{T}\left|\zp\ip'\right> = (- l_6 + c)\left|\zp\ip'\right> + l_5\left|\ip\zp'\right>
&& \nquad \mathbb{T}\left|\zp\im'\right> = (- l_7 + c)\left|\zp\im'\right> + l_4\left|\up\ym'\right>
\\\nonumber & \mathbb{T}\left|\ip\zp'\right> = (- l_8 + c)\left|\ip\zp'\right> + l_5\left|\zp\ip'\right>
&& \nquad \mathbb{T}\left|\ip\zm'\right> = (- l_9 + c)\left|\ip\zm'\right> - l_4\left|\yp\um'\right>
\\\nonumber & \mathbb{T}\left|\zp\up'\right> = (- l_6 + c)\left|\zp\up'\right> + l_5\left|\up\zp'\right>
&& \nquad \mathbb{T}\left|\zp\um'\right> = (- l_7 + c)\left|\zp\um'\right> - l_4\left|\ip\ym'\right>
\\ & \mathbb{T}\left|\up\zp'\right> = (- l_8 + c)\left|\up\zp'\right> + l_5\left|\zp\up'\right>
&& \nquad \mathbb{T}\left|\up\zm'\right> = (- l_9 + c)\left|\up\zm'\right> + l_4\left|\yp\im'\right> \label{bhsmatans}
\end{align}}
Here the functions $l_i$ and $c$ depending on momenta are given by
\begin{align}
l_1(p,p') = & \frac{(p+p')^2}{2(e'p-ep')}\ , &&
\!\!\!\!\!\!\!\!\!\!\!\!\!\!\!\!\!\!\!\!\!\!\!\!\!\!\!\!\!\!\!\!
\!\!\!\!\!\!\!\!\!\!\!\!\!\!\!\!\!\!\!\!\!\!\!\!\!\!\!\!\!\!\!\!
\!\!\!\!\!\!\!\!\!\!\!\!\!\!\!\!\!\!\!\!\!\!\!\!\!\!\!\!\!\!\!\!
c(p,p') = \big(a-\tfrac12\big)(ep'-e'p)\ , \no \\
l_2(p,p') = &\frac{(p-p')^2}{2(e'p-ep')}\ ,
&&
\!\!\!\!\!\!\!\!\!\!\!\!\!\!\!\!\!\!\!\!\!\!\!\!\!\!\!\!\!\!\!\!
\!\!\!\!\!\!\!\!\!\!\!\!\!\!\!\!\!\!\!\!\!\!\!\!\!\!\!\!\!\!\!\!
\!\!\!\!\!\!\!\!\!\!\!\!\!\!\!\!\!\!\!\!\!\!\!\!\!\!\!\!\!\!\!\!
l_3(p,p') = - \frac{(p-p')(p+p')}{2(e'p-ep')}\ , \no
\\ l_4(p,p') = & - \frac{pp'}{2(e'p-ep')}\big[\sqrt{(e+p)(e'-p')} - \sqrt{(e-p)(e'+p')}\big]\ ,
\no\\ l_5(p,p') = & - \frac{pp'}{2(e'p-ep')}\big[\sqrt{(e+p)(e'-p')} + \sqrt{(e-p)(e'+p')}\big]\ ,
\no \\ l_6(p,p') = & \frac{(p+p')p'}{2(e'p-ep')} \ , &&
\!\!\!\!\!\!\!\!\!\!\!\!\!\!\!\!\!\!\!\!\!\!\!\!\!\!\!\!\!\!\!\!
\!\!\!\!\!\!\!\!\!\!\!\!\!\!\!\!\!\!\!\!\!\!\!\!\!\!\!\!\!\!\!\!
\!\!\!\!\!\!\!\!\!\!\!\!\!\!\!\!\!\!\!\!\!\!\!\!\!\!\!\!\!\!\!\!
l_7(p,p') = - \frac{(p-p')p'}{2(e'p-ep')}
\no \\ l_8(p,p') = & \frac{(p+p')p}{2(e'p-ep')} \ , &&
\!\!\!\!\!\!\!\!\!\!\!\!\!\!\!\!\!\!\!\!\!\!\!\!\!\!\!\!\!\!\!\!
\!\!\!\!\!\!\!\!\!\!\!\!\!\!\!\!\!\!\!\!\!\!\!\!\!\!\!\!\!\!\!\!
\!\!\!\!\!\!\!\!\!\!\!\!\!\!\!\!\!\!\!\!\!\!\!\!\!\!\!\!\!\!\!\!
l_9(p,p') = \frac{(p-p')p}{2(e'p-ep')}\ ,
\no \\ & \!\! e = \sqrt{p^2+1}\ , \qquad \qquad \ \ \ \ \ e' = \sqrt{p'{}^2 + 1} \vphantom{\frac12}\ . \la{gh}
\end{align}
Using the identity \eqref{bhidentity} the functions $l_i$ can be simplified as follows:
\begin{align}\no
l_1(p,p') = & \frac{(p+p')(e'p+ep')}{2(p-p')} \ , &&
\!\!\!\!\!\!\!\!\!\!\!\!\!\!\!\!\!\!\!\!\!\!\!\!\!\!\!\!\!\!\!\!
\!\!\!\!\!\!\!\!\!\!\!\!\!\!\!\!\!\!\!\!\!\!\!\!\!\!\!\!\!\!\!\!
\!\!\!\!\!\!\!\!\!\!\!\!\!\!\!\!\!\!\!\!\!\!\!\!\!\!\!\!\!\!\!\!
l_2(p,p') = \frac{(p-p')(e'p+ep')}{2(p+p')}\ ,
\\ l_3(p,p') = & - \frac{1}{2}(e'p+ep') \ , && \nonumber
\!\!\!\!\!\!\!\!\!\!\!\!\!\!\!\!\!\!\!\!\!\!\!\!\!\!\!\!\!\!\!\!
\!\!\!\!\!\!\!\!\!\!\!\!\!\!\!\!\!\!\!\!\!\!\!\!\!\!\!\!\!\!\!\!
\!\!\!\!\!\!\!\!\!\!\!\!\!\!\!\!\!\!\!\!\!\!\!\!\!\!\!\!\!\!\!\!
\\ l_4(p,p') = & - \frac{pp'}{2(p+p')}\big[\sqrt{(e+p)(e'+p')} - \sqrt{(e-p)(e'-p')}\big] \ , \nonumber
\\ l_5(p,p') = & - \frac{pp'}{2(p-p')}\big[\sqrt{(e+p)(e'+p')} + \sqrt{(e-p)(e'-p')}\big]\ , \nonumber
\\ l_6(p,p') = & \frac{p'(e'p+ep')}{2(p-p')} \ , && \nonumber
\!\!\!\!\!\!\!\!\!\!\!\!\!\!\!\!\!\!\!\!\!\!\!\!\!\!\!\!\!\!\!\!
\!\!\!\!\!\!\!\!\!\!\!\!\!\!\!\!\!\!\!\!\!\!\!\!\!\!\!\!\!\!\!\!
\!\!\!\!\!\!\!\!\!\!\!\!\!\!\!\!\!\!\!\!\!\!\!\!\!\!\!\!\!\!\!\!
l_7(p,p') = - \frac{p'(e'p+ep')}{2(p+p')}\ ,
\\ l_8(p,p') = & \frac{p(e'p+ep')}{2(p-p')} \ , &&
\!\!\!\!\!\!\!\!\!\!\!\!\!\!\!\!\!\!\!\!\!\!\!\!\!\!\!\!\!\!\!\!
\!\!\!\!\!\!\!\!\!\!\!\!\!\!\!\!\!\!\!\!\!\!\!\!\!\!\!\!\!\!\!\!
\!\!\!\!\!\!\!\!\!\!\!\!\!\!\!\!\!\!\!\!\!\!\!\!\!\!\!\!\!\!\!\!
l_9(p,p') = \frac{p(e'p+ep')}{2(p+p')} \ . \label{bhlfunc}
\end{align}
This S-matrix is invariant under a $U(1)^3$ symmetry with the fields
$\{y_\pm,z_\pm,\zeta_\pm,\chi_\pm\}$ charged as follows
\begin{equation}
\pm \{\alpha_1 + \alpha_2,\ \alpha_1 -\alpha_2,\ \alpha_1+\alpha_3,\ \alpha_1-\alpha_3\} \ .\label{bhgbs}
\end{equation}
The index $\pm$ that the fields carry indicates the charge under the $U(1)$ symmetry with parameter $\alpha_1$.
Crucially, with respect to this $U(1)$ there are no reflection processes.
More precisely, the association of the momenta to the $U(1)$ charge of the states is preserved by the scattering process.

\def \adsss  {$AdS_3 \times S^3 \times S^3 \times S^1$\ }

One  may wonder a priori why the above   truncation  of the \adss  S-matrix   should indeed  represent 
the massive sector of the  S-matrix  of the \adst theory   with  pure RR flux. 
This is to large extent fixed by the correspondence  of the spectra and bosonic sectors 
and by the expected supersymmetries.  The   symmetry algebra of the \adss
S-matrix is known to  be 
$\big(\mathfrak{psu}(2|2)\oplus \mathfrak{psu}(2|2)\big) \ltimes \mathbb{R}^3\equiv  \mathfrak{psu}(2|2) ^2 \ltimes \mathbb{R}^3$,
while we find that the corresponding symmetry 
in the \adst theory is  $[\mathfrak{ps}(\mathfrak{u}(1|1)^2) ]^2 \ltimes \mathbb{R}^3$
with the two central elements in the first factor identified
 (see Appendix \ref{appb} for details and notation).\foot{There 
appears to be some confusion in the literature regarding the symmetry preserved by the BMN vacuum  in this theory
(of course, as  in the \adss  case, not all of the symmetry of the S-matrix   may be  visible in the off-shell Lagrangian, see, e.g.,  \ci{afr}).  
In the discussion of the    giant magnons in the  $SU(2)$ sector  of the \adst theory in  \ci{david} 
 the remaining  symmetry was claimed to be 
$\mathfrak{su}(1|1)^2$,  but it was implicitly  assumed that two copies of this algebra  should appear as a  symmetry   when  discussing 
the S-matrix (cf. eq.~(3.18) in the first paper in   \ci{david}). 
In \ci{ahn} the symmetry of the S-matrix  for  the  \adsss  theory was assumed to be $\mathfrak{su}(1|1)$   while 
the corresponding  symmetry for \adst case  was doubled:     $\mathfrak{su}(1|1)^2$. 
In \ci{bor}  the symmetry  of the S-matrix of the \adsss  theory was  taken as   $(\mathfrak{u}(1) \inplus \mathfrak{su}(1|1)^2)\ltimes \mathbb{R}^2$,
which is consistent with the above symmetry algebra 
$[\mathfrak{ps}(\mathfrak{u}(1|1)^2)]^2 \ltimes \mathbb{R}^3$ (with the two central elements in the
first factor identified)
for the \adst case,   taking into account  that  the symmetry should be doubled and the central extensions identified in this limit.}

Indeed, $\mathfrak{psu}(2|2)$ has a $\mathfrak{ps}(\mathfrak{u}(1|1)^2)$   subgroup preserved by the truncation. 
The resulting $[\mathfrak{ps}(\mathfrak{u}(1|1)^2) ]^2 \ltimes \mathbb{R}^3$  symmetry   matches  (modulo the quantum deformation and central elements)
the symmetry of the  S-matrix   of the corresponding Pohlmeyer-reduced theory   \ci{ht2}.
While the limiting  \adst case    was not explicitly discussed in \ci{bor} (which considered  the \adsss theory), we   
 believe  the   massive sector of the corresponding 
string tree-level (strong-coupling)   S-matrix  should   match the above result \rf{bhsmatans},\rf{gh},\rf{bhlfunc}.
It should also be in agreement   with the tree-level part of the   expression in \ci{wu} 
found by direct string-theory computation and claimed to be in agreement with \ci{bor}.\foot{Let 
us mention  also a   remark  in  the first paper in \ci{sax} that the  proposed Bethe 
ansatz agrees in the \adst  limit with the    standard \adss one in the $\mathfrak{psu}(1,1|2)$ sector. 
This is again  consistent with the S-matrix truncation  picture.} 

Indeed, as   we shall explain  in  Appendix \ref{A_2}, the  quadratic fermionic action   that reproduces all of the 
above   amplitudes  involving  two  fermions and two bosons 
is the same   as the \adst limit   of the action  found in \ci{wulf,wu}  directly from the \adsss  superstring action.
This explicitly confirms   that the S-matrix \rf{bhsmatans}  is in agreement with   the world-sheet theory.

%%%%%%%%%%%%%%%%%%%%%%%%%%%%%%%%%%%%%%
\subsection{Non-vanishing $B$-flux }\label{4_2}
%%%%%%%%%%%%%%%%%%%%%%%%%%%%%%%%%%%%%%

The above  observations  combined with the $q\not=0$  bosonic sector results and the requirements of integrability
allow us to conjecture the expression for the tree-level S-matrix for the massive states of the %vv4
superstring theory with a non-zero NSNS flux ($q\not=0$).
From the direct computation of the $yy\to yy$ amplitudes in the $S^3$ sector in the previous section
and of the $yy\to zz$ and $zz\to zz$ amplitudes, which follow from the quartic Lagrangian
in Appendix \ref{appa}, we know that to find the $q\not=0$ generalization of
the functions $l_{1,2,3}$ and $c$ one
should take the $q=0$ expressions given in \eqref{bhlfunc} and modify the dispersion relation
\begin{equation}\label{dis}
e \rightarrow e_\pm = \sqrt{(p\pm q)^2 + 1 -q^2} \ , \qquad \qquad e' \rightarrow e'_\pm = \sqrt{(p'\pm q){}^2 + 1-q^2}\ ,
\end{equation}
where, as in \rf{bhscatter3}, for states with positive/negative charge we use $e_+/e_-$, i.e. we shift the momentum
in the dispersion relation by $ \pm q$.

The functions $l_{6,7,8,9}$ are then constrained by the first requirement of integrability 
-- the {\it symmetry factorization} property of the S-matrix.
In the $q = 0$ case the symmetry algebra of the S-matrix takes the form of a direct sum with central extensions -- see Appendix \ref{appb}.
Combined with integrability this implies the S-matrix should factorize under this structure.
In the theory with mixed RR and NSNS flux, i.e. $q \neq 0$, the global symmetry of the string Lagrangian is unaltered
and the symmetry algebra of the S-matrix should be unchanged -- $q$ should appear in the particular representation used.
Therefore, we expect the S-matrix will still factorize in the same way.

Let us first briefly review the factorization in the $q=0$ case. Formally defining the following tensor products
\begin{equation}\begin{split}\label{bhtens}
\left|y\right> = \left|\phi\right> \otimes \left|\phi\right> \ , \qquad & \left|z\right> = \left|\psi\right> \otimes \left|\psi\right> \ ,
\\
\left|\zeta\right> = \left|\phi\right> \otimes \left|\psi\right> \ , \qquad & \left|\chi\right> = \left|\psi\right> \otimes \left|\phi\right> \ ,
\end{split}\end{equation}
where $\left|\phi\right>$ is bosonic and $\left|\psi\right>$ is fermionic, the factorization then
means that the S-matrix for $\{y,z,\zeta,\chi\}$ can be constructed from an S-matrix for $\{\phi,\psi\}$. For example,
\begin{equation}\begin{split}
\mathbb{T}\left| y_+ \zeta_+'\right> = & (\mathbb{I} \otimes \mathbb{\hT} + \mathbb{\hT} \otimes \mathbb{I}) \, \big(\left|\phi_+\phi_+'\right>\otimes \left|\phi_+ \psi_+'\right>\big) \ ,
\\ \mathbb{T}\left| z_+ \chi_-'\right> = &- (\mathbb{I} \otimes \mathbb{\hT} + \mathbb{\hT} \otimes \mathbb{I}) \, \big(\left|\psi_+\psi_-'\right>\otimes \left|\psi_+ \phi_-'\right>\big) \ ,
\end{split}\end{equation}
where the minus sign in the second case comes from moving two fermions past each other. In the $q=0$ case the factorized tree-level S-matrix for $\{\phi,\psi\}$ is then given by
\begin{align} \nonumber
\mathbb{\hT}\left|\phi_\pm\phi_\pm{}'\right> = & \frac12(l_1 + c)\left|\phi_\pm\phi_\pm{}'\right>\ ,
& \mathbb{\hT}\left|\phi_\pm\phi_\mp{}'\right> = & \frac12(l_2 + c)\left|\phi_\pm\phi_\mp{}'\right> + l_4\left|\psi_\pm\psi_\mp{}'\right>\ ,
\\ \nonumber
\mathbb{\hT}\left|\psi_\pm\psi_\pm{}'\right> = & \frac12(- l_1 + c)\left|\psi_\pm\psi_\pm{}'\right>\ ,
& \mathbb{\hT}\left|\psi_\pm\psi_\mp{}'\right> = & \frac12(- l_2 + c)\left|\psi_\pm\psi_\mp{}'\right> + l_4 \left|\phi_\pm\phi_\mp{}'\right>\ ,
\\ \nonumber
\mathbb{\hT}\left|\phi_\pm\psi_\mp{}'\right> = & \frac12(l_3 + c)\left|\phi_\pm\psi_\mp{}'\right>\ ,
& \mathbb{\hT}\left|\phi_\pm\psi_\pm{}'\right> = & \frac12(l_3 + c)\left|\phi_\pm\psi_\pm{}'\right> - l_5\left|\psi_\pm\phi_\pm{}'\right>\ ,
\\ \label{bhsmatfact}
\mathbb{\hT}\left|\psi_\pm\phi_\mp{}'\right> = & \frac12(-l_3 + c)\left|\psi_\pm\phi_\mp{}'\right>\ ,
& \mathbb{\hT}\left|\psi_\pm\phi_\pm{}'\right> = & \frac12(-l_3 + c)\left|\psi_\pm\phi_\pm{}'\right> - l_5 \left|\phi_\pm\psi_\pm{}'\right>\ .
\end{align}
It is worth noting
that this factorized S-matrix \rf{bhsmatfact} has a $U(1)^2$ symmetry under which $\{\phi,\psi\}$ have charges $\{1,0\}$ and $\{0,1\}$ respectively.

The factorization relies on the following identities
\begin{equation}\label{bh6789}
l_6 = \frac12(l_1 + l_3) \ , \qquad l_8 = \frac12(l_1 - l_3) \ , \qquad l_7 = \frac12(l_2 + l_3) \ , \qquad l_9 = \frac12(l_2 - l_3) \ .
\end{equation}
Observing that the $++\rightarrow ++$ scattering amplitudes in \eqref{bhsmatans} are only built from the $++ \rightarrow ++$ amplitudes in \eqref{bhsmatfact} and the same is true for the $+- \rightarrow +-$, $-+ \rightarrow -+$ and $--\rightarrow --$ scattering processes,
the requirement that the tree-level S-matrix of the $q\not=0$ theory should factorize for any value of $q$ implies that the
functions $l_{6,7,8,9}$ should be given
by the same generalization procedure as was found
for the functions $l_{1,2,3}$ and $c$, i.e. their $q\not=0$ form should be the same as
in \eqref{bhlfunc} with the dispersion relation modified as in \eqref{dis}.

The factorization property does not constrain $l_4$ and $l_5$. To fix $l_4$ and $l_5$ we use a second requirement of integrability -- the classical
{\it Yang-Baxter} equation.\footnote{We have defined the S-matrix ($S$)
as a map from the space $|\Phi_i(p)\Phi_j(p')\ket$ to $|\Phi_k(p)\Phi_l(p')\ket$,
which at leading order is just the identity \eqref{fffi},\eqref{smatdef}. However,
the S-matrix ($\tilde S$) that naturally satisfies the Yang-Baxter equation
(for example, through the use of ZF operators)
\begin{equation}
\tilde S_{12}(p',p'') \tilde S_{23}(p,p'') \tilde S_{12}(p,p') = \tilde S_{23}(p,p') \tilde S_{12}(p,p'') \tilde S_{23}(p',p'') \ , \nonumber 
\end{equation}
is a map from the space $|\Phi_i(p)\Phi_j(p')\ket$ to $|\Phi_l(p')\Phi_k(p)\ket$,
i.e. the momenta are interchanged. This S-matrix is related to ours by composing
with the permutation operator ($\tilde S = {\mathcal P} S$), which flips the two
outgoing states picking up a minus sign when they are both fermions. Therefore, at
leading order it is given by the permutation operator. Using these relations it is
simple to find the appropriate classical Yang-Baxter equation (i.e. with the required
minus signs) that our tree-level S-matrix is required to satisfy.}
The  Yang-Baxter equation  should apply   to the  massive   subsector of the S-matrix
of the \adst theory.\footnote{In a general integrable 
theory the truncated S-matrix for the scattering of all particles 
of a given mass should satisfy the Yang-Baxter equation in its own right. This is a consequence of
the fact that the scattering of particles of different mass should be diagonal in the space of masses
(there can still be a non-trivial S-matrix in flavour space) by the conservation
of higher charges (see also comments in Appendix \ref{appa}).}
We then find that the functions $l_4$ and $l_5$ should
depend on $q$ not only through $e,e'$ but also explicitly and are generalized to $q\not=0$
in slightly different ways depending on the charges of the excitations being scattered:
\begin{equation}\begin{split}\label{param45}
l_4^{\pm\mp\rightarrow\pm\mp}(p,p') = & - \frac{pp'}{2(p+p')}\Big[\sqrt{(e_\pm+p\pm q)(e_\mp'+p'\mp q)} - \sqrt{(e_\pm-p\mp q)(e_\mp'- p'\pm q)}\, \Big] ,
\\ l_5^{\pm\pm\rightarrow\pm\pm}(p,p') = & - \frac{pp'}{2(p-p')}\Big[\sqrt{(e_\pm+p\pm q)(e_\pm'+p'\pm q)} + \sqrt{(e_\pm-p\mp q)(e_\pm'-p'\mp q )}\, \Big].
\end{split}\end{equation}
Here the superscripts label the different scattering processes and $e_\pm$ and $e_\pm'$ are given
by their $q$-dependent expressions in \eqref{dis}.

\

In summary, our result for the tree-level S-matrix for the massive states of the mixed-flux \adst superstring theory is given by
\eqref{bhsmatans} with the parametrizing functions $l_{1,2,3,6,7,8,9}$ and $c$ given by \eqref{bhlfunc}
and the functions $l_{4,5}$ given by \eqref{param45} with the dispersion relation modified as
in \eqref{dis}.
Alternatively, the S-matrix can be represented as a tensor product of two
copies of the factorized S-matrix \eqref{bhsmatfact} with the corresponding
parametrizing functions $l_{1,2,3},\, c$ and $l_{4,5}$ generalized to $q\not=0$
as described above.

In Appendix \ref{A_2}   we   shall   present the  action  quadratic in both fermions 
and bosons  that reproduces the   corresponding amplitudes  of  the  $q\not=0$ 
generalization of the S-matrix   described above  and  explain  how this action  should  
follow from the gauge-fixed superstring action. This  providing a strong indication  that this S-matrix is 
 consistent   with  the \adst  world-sheet theory for $q\not=0$.

\

The factorized S-matrix \eqref{bhsmatfact} has a non-local supersymmetry -- see Appendix \ref{appb} --
(the non-locality appears in the form of a braiding in the coproduct \ci{afrr,km,afr}).
It is therefore natural to expect that the $q\not=0$ generalization of this S-matrix will be invariant under a
modified
supersymmetry algebra with structure constants depending on $q$.
It would be interesting to determine this algebra and relate its central extension
to a $q$-modified dispersion relation. We shall discuss a candidate for this symmetry algebra
in Appendix \ref{appb}.

A natural generalization of the standard magnon dispersion
relation might be the following:
for the positively/negatively charged states we should have
\begin{equation}\la{iii}
e_\pm^2 = 1 -q^2 +   \, (2\hh\sin\frac p2  \pm q)^2 \ ,
\end{equation}
where $\hh$ should be identified with the string tension, $ \hh = {\sql \ov 2 \pi}= { R^2\ov 2\pi \a'}$.\foot{
It is unclear whether  
the expression for $\hh$ is  renormalized for $q \neq 0$.  
In the $q = 0$ case it appears that it is not. 
There is though  a  1-loop shift in $\hh(\l)$  in the case of another 1-parameter deformation  -- 
the $AdS_3\times S^3 \times S^3 \times S^1$  theory 
\ci{ab,bec}.
 To investigate this question  by  a direct 1-loop computation  is an open problem.}
Note   that  for $q=1$    eq. \rf{iii}   reduces to 
$e_\pm =    2\hh\sin\frac p2  \pm 1  $    leading to  a massless    dispersion  relation 
for small $p$. 

 This relation may be  viewed  as a ``lattice''   generalization of the above  string world-sheet 
  dispersion relation \rf{dis}   with ``$\frac p2  \to  \sin\frac p2 $''. 
  Indeed, in  the BMN limit when $\hh$ is large and
the momentum $p$ is small, scaling as $\hh^{-1} $, we find,
after redefining $p \rightarrow \hh^{-1} p$
\begin{equation}
e_\pm^2 = 1 - q^2 + (p \pm q)^2 + \mathcal{O}(\hh^{-2}) \ . \la{get}
\end{equation}
This matches the result \rf{hap},\rf{bhscatter3} we obtained directly
from the string perturbation theory.
In the semiclassical (``giant magnon'' \ci{hma}) limit when $\hh$ is large and
$p$ stays finite we get from \rf{iii}
\begin{equation}\la{gia}
%e_\pm = 2 \hh \, \sin \frac{p}{2} \pm q \cos \frac{p}{2} + \mathcal{O}(\hh^{-1}) \ .
e_\pm = 2 \hh \, \sin \frac{p}{2} \pm q + \mathcal{O}(\hh^{-1}) \ .
\end{equation}
This suggests that the leading classical energy (minus the angular
momentum) of the giant magnon solution
should be unaltered by the $q$-deformation. The expansion \rf{gia}
predicts the presence of a string 1-loop correction to the giant magnon
energy proportional to $q$
(there was no 1-loop correction in the $q = 0$ case \cite{ps,david}).
It would be interesting to derive this correction
by a direct string-theory computation and thus check the
conjectured form of the dispersion relation \rf{iii}.

%%%%%%%%%%%%%%%%%%%%%%%%%%%%%%%%%%%%%%
\section{Concluding remarks\label{sec5}}
\renewcommand{\theequation}{5.\arabic{equation}}
\setcounter{equation}{0}
%%%%%%%%%%%%%%%%%%%%%%%%%%%%%%%%%%%%%%

In this paper we have found the generalization of the tree-level S-matrix for massive BMN-type
excitations of the \adst superstring theory in the case of non-zero NSNS 3-form flux (parametrized by $q \in (0,1)$).
We have directly computed the S-matrix in the bosonic sector discovering
its very simple dependence on $q$
via the modified dispersion relation. Using the requirements of integrability (factorization and Yang-Baxter properties of the S-matrix)
we then suggested its generalization
to the full superstring case.

While we have little doubt 
(on the basis of integrability and arguments in Appendix \ref{A_2})
in the correctness of the proposed fermionic part of the $q\not=0$\   S-matrix,
it would be useful to check it  in full (including 4-fermion part) 
by starting directly with the component form of the corresponding  superstring action. 

One straightforward extension of the computations presented in this paper is to the similar case of
the $AdS_3 \times S^3 \times S^3 \times S^1$ theory with mixed 3-form flux \ci{bsz,cz}, generalizing the
recent discussion of the $q=0$ case in \ci{wu}.
It  would be important also to generalize the semiclassical  1-loop computations   done in 
the \adst theory with pure RR flux  \ci{david,wulf,for,ab,bec,beca,wu}
to the  $q\not=0$ case  in order to  check non-renormalization of the effective string tension $\hh$ 
and determine the  corresponding 1-loop  dressing phases.

The next obvious step is to use the information about the S-matrix to conjecture the
corresponding $q\not=0$ generalization of the asymptotic Bethe ansatz
which for $q=0$ was first conjectured (by analogy with the \adss case) in
\ci{bsz}.\foot{The full structure of the asymptotic Bethe ansatz still remains to be
understood already in the $q=0$ case of the \adst superstring, cf. \ci{sax,ahn,bor,bec,wu}.}
It is natural to expect that as long as $q < 1$ (i.e. away from the WZW point or the case of pure NSNS flux)
the underlying integrable system should be similar to the one describing the pure RR case ($q=0$), i.e. there should be
a ``ferromagnetic'' BPS vacuum with standard massive BMN excitations
(with a candidate dispersion relation suggested above in \rf{iii}).\foot{It should be emphasized again
that the superstring theory case considered here,
where there are no UV divergences and the mass scale is introduced by the choice of a vacuum or a gauge,
is very different from the
case of the quantum bosonic $SU(2)$ principal chiral model with a WZ term \ci{pw,zz,ber}. There,
in the absence of the WZ term (for $q=0$),
there is a dynamical mass generation and thus a massive S-matrix \ci{zzz}. However, for $q\not=0$
there is an RG flow (of $\hh$ and thus of $q$ in \rf{4.1})
between the trivial UV ($q=0$) and non-trivial WZW ($q=1$)
fixed points, so that there is no mass generation and the underlying
S-matrix is massless.}

%%%%%%%%%%%%%%%%%%%%%%%%%%%%%%%%%%%%%%
\section*{Acknowledgments}
%%%%%%%%%%%%%%%%%%%%%%%%%%%%%%%%%%%%%%
We are grateful to A. Babichenko for many useful discussions and collaboration in the initial stages of this project.
We would like to thank T. Klose, M. Kruczenski and R. Roiban for important explanations and comments  and 
also  S. Frolov,  O. Ohlsson Sax, R. Roiban, A. Torrielli, L. Wulff  and K. Zarembo for useful discussions and comments on the draft.
We   thank   A. Sfondrini for pointing out  misprints in eq.~\eqref{bhsmatans}
in the original version of this paper.
BH is supported by the Emmy Noether Programme ``Gauge fields from Strings'' funded by the German Research Foundation (DFG).
The work of AAT is supported by the ERC Advanced Grant No.290456:  ``Gauge theory -- string theory duality'' %vv4
and also by the STFC grant ST/J000353/1.

%%%%%%%%%%%%%%%%%%%%%%%%%%%%%%%%%%%%%%
\appendix
%%%%%%%%%%%%%%%%%%%%%%%%%%%%%%%%%%%%%%

%%%%%%%%%%%%%%%%%%%%%%%%%%%%%%%%%%%%%%
\def\ka{\k}
\def\rS{{\rm S}}
\def\scs{\scriptsize}
\refstepcounter{section}
\def\theequation{A.\arabic{equation}}
\addcontentsline{toc}{section}{Appendices}
\setcounter{equation}{0}
\section*{Appendix A: \\ Expansion   of the action of %bosonic sector of
 \adst string theory with $B$-flux\label{appa}}
\addcontentsline{toc}{section}{A \  Expansion of the action  of \adst string theory with $B$-flux}
%%%%%%%%%%%%%%%%%%%%%%%%%%%%%%%%%%%%%%

Here we shall  discuss  quadratic and quartic terms in the expansion of    \adst string  action   
that are relevant for the derivation of the S-matrix in  \ref{sec4}.

\subsection{Bosonic sector}\label{A_1}

The generalization of the Lagrangian \rf{5.22} to the case when the string moves on $AdS_3 \times S^3\times T^4 $ can be written as
\be \la{2222}
&& L= - \ha\Big[ - \hat G (z) \del^a t \del_a t + \hat F(z) \del^a z_s \del_a z_s
+ 2 \ep^{ab} \hat B_s (z) \del_a z_{s} \del_b t \no \\
&&\ \ \ \ \ \ \ \ \ \ \ \ \ +\, G(y) \del^a \vp \del_a \vp + F(y) \del^a y_s \del_a y_s
+ 2 \ep^{ab} B_s(y) \del_a y_{s} \del_b \vp + \del^a x_k \del_a x_k \Big] \ , \\
&& G= { ( 1 - \fo y^2 )^2 \ov ( 1 + \fo y^2 )^2 }= 1 - y^2 F \ , \ \ \ \ \ \ \ \ \ \ \ \ \ \ \ F = { 1 \ov ( 1 + \fo y^2 )^2 } \ , \la{1225} \\
&&
\hat G = { ( 1 + \fo z^2 )^2 \ov ( 1 - \fo z^2 )^2 }= 1 + z^2 \hat F \ , \ \ \ \ \ \ \ \ \ \ \ \ \ \ \ \hat F = { 1 \ov ( 1 - \fo z^2 )^2 } \ , \la{2225} \\
&& B_s(y) = { q F (y) } \epsilon_{rs} y_{r} \ , \ \ \ \ \ \ \ \
\hat B_s(z) = q \hat F(z) \epsilon_{rs} z_{r} \ .
\ee
Here $x_k$ are the ``massless'' $T^4$ fields.
The $AdS_3$ and $S^3$ parts are related by the formal analytic continuation $t\to \vp, \ z_r \to i y_r$ and then reversing the overall sign of the
Lagrangian.\foot{The sign of coefficient in $\hat B_s(z)$ can be opposite to that
of in $B_s(y)$ but the two cases are related, e.g., by coordinate redefinition $t \to - t$.
The expansion of the action below in the case of $\hat B_s(z) = - q \hat F(z) \epsilon_{rs} z_{r} $
can be found by reversing the sign of $ \epsilon_{rs} z_{r}$ in all terms where it appears. }

Following the discussion in section \ref{sec2}, let us consider the
redefinition of $t$ as in \rf{b} and then T-dualize in the $\vp$ direction. This leads to the following generalization of \rf{le},\rf{523},\rf{p}:
\be
&&\td L = - \sqrt g g^{cd} h_{cd} - P\ep^{cd} \del_d u \Big[
\, b \hat G (\del_c \tv - B_s \del_c y_s ) + G \hat B_s \del_c z_s \Big] \la{lei} \ , \\
&&h_{cd}= - Q \del_c u \del_d u
+ P ( \del_c \tv - B_s \del_c y_s + b \hat B_s \del_c z_s ) ( \del_d \tv - B_s \del_d y_s + b \hat B_s \del_d z_s ) \no\\
&& \ \ \ \ \qquad + F\del_c y_s \del_d y_s + \hat F\del_c z_s \del_d z_s + \del_c x_k \del_d x_k \ , \la{3i} \\
&& Q = G \hat G P \ , \ \ \ \ \ \ \ \ \ \ P = ( G - b^2 \hat G )^{-1} \ . \la{pi} \ee
Fixing the gauge in \rf{gai} (with $\J=1$) we get
the following generalization of \rf{223}
\be
&&\td L = - \sqrt h + \bc P \Big[ b \hat G ( \bc - B_s y'_s) + G \hat B_s z'_{s} \Big] \ , \ \la{2123} \\
&&h = \Big[ \bc^2 Q - P (B_s \dot y_s - b \hat B_s \dot z_s )^2
- F \dot y^2_s - \hat F \dot z^2_s - \dot x^2_k \Big]\no \\
&& \ \ \ \ \ \ \ \ \ \ \ \ \ \ \ \ \ \ \times \Big[ P (\bc - B_r y'_r + b \hat B_r z'_r )^2 + F y'^2_r + \hat F z'^2 _r + x'^2_k \Big]
\no \\
&& \ \ \ \ \ \ \ +
\Big[ P ( B_s \dot y_s - b \hat B_s \dot z_s ) (\bc - B_r y'_r + b \hat B_r z'_r ) - F \dot y_r y'_r - \hat F \dot z_r z'_r -
\dot x_k x'_k \Big]^2\ . \la{kik}
\ee

Expanding $\td L$ in powers of $y_s$ and $z_s$ we get the following generalization of \rf{222},\rf{434}
\be
&&\td L= L_2 + L_4 +... \ , \no \\
&&L_2 = \ha ( \dot z^2_s - z'^2 _s - z_s^2) + q \ep_{sr} z_s z'_r
+ \ha ( \dot y^2_s - y'^2 _s - y_s^2 ) + q \ep_{sr} y_s y'_r + \ha (\dot x^2_k - x'^2 _k ) \ , \la{u22} \\
&& L_4 = \fo\Big[ y^2_s ( 2 y'^2_r + \dot z^2_r + z'^2_r )
- z^2_s ( 2z'^2_r + \dot y^2_r + y'^2_r ) + (y^2_s - z^2_s) ( \dot x^2_k + x'^2_k ) \Big] \no\\
&& \ \qquad +
q \Big[ \ha ( \dot y_r y'_r + \dot z_r z'_r + \dot x_k x'_k ) \ep_{sp} ( y_s \dot y_p - z_s \dot z_p) \no \\
&& \ \ \ \ \ \ \ \ \ - \fo ( \dot y^2_r + y'^2_r + \dot z^2_r + z'^2_r + \dot x^2_k + x'^2_k ) \ep_{sp} ( y_s y'_p - z_s z'_p )
- \fo ( y^2_r - z^2_r) \ep_{sp} ( y_s y'_p + z_s z'_p )
\Big]\no\\
&&
\ \ \ \ + ( a- \ha) \Big\{ \fo ( y^2_s + z^2_s)^2 - \fo (\dot y^2_s + y'^2_s + \dot z^2_s + z'^2_s + \dot x^2_k + x'^2_k )^2
+ ( \dot y_s  y' _s + \dot z_s z' _s + \dot x_k x'_k )^2
\no \\ && \ \ \ \ \ \ \ \ \ \qquad \ \ \
+ q \Big[ - ( \dot y_r y'_r + \dot z_r z'_r + \dot x_k x'_k ) \ep_{sp} ( y_s \dot y_p + z_s \dot z_p) \no \\
&& \ \ \ \ \ \ \ \ \ \ \ \ \ \ \ \ \ \ \ \ \ \
+ \ha ( \dot y^2_r + y'^2_r - y^2_r + \dot z^2_r + z'^2_r - z^2_r + \dot x^2_k + x'^2_k )\ep_{sp} ( y_s y'_p + z_s z'_p ) \Big] \Big\} \ . \la{434app}
\ee
The kinetic terms for $y_s$ and $z_s$ are the same, implying the same massive dispersion relation \rf{hap} while
$x_k$ are massless.

As there are no cubic terms in the bosonic Lagrangian,
and there cannot be a $boson-boson-fermion$ cubic term when fermions are included,
the above quartic Lagrangian (ignoring the massless fields $x_k$) is sufficient to compute
the tree-level S-matrix for the four massive bosons, $y_s$ and $z_s$.
The result of this computation was  summarized in sections \ref{sec3} and \ref{sec4}. 

Furthermore, the  tree-level S-matrix  truncated to  just  the massive mode  sector 
(including all massive  bosons and fermions) should satisfy the classical Yang-Baxter equation 
in its own right as by integrability (i.e. as a consequence of the  existence of higher conserved charges) 
the scattering of particles of different mass should be diagonal in the space of masses
(there can still be a non-trivial S-matrix in flavour space).
It would be of interest to compute the S-matrix for the scattering of both the 
massless excitations with the massive excitations and amongst themselves.
At tree level the allowed scattering  %vv4
processes following from \eqref{434app} are  (here  $x_L$ and $x_R$ 
stand for the left- and right-moving  components   of  $x_k$) 
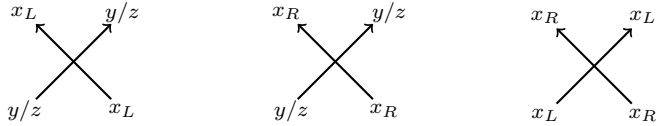
\begin{equation}
\begin{tikzpicture}
\draw[->,thick] (-0.5,-0.5) -- (0.5,0.5);
\draw[->,thick] (0.5,-0.5) -- (-0.5,0.5);
\node at (-0.65,-0.65) {\scs $y/z$};
\node at (0.65,0.65) {\scs $y/z$};
\node at (0.65,-0.65) {\scs $x_L$};
\node at (-0.65,0.65) {\scs $x_L$};
\end{tikzpicture}
\qquad \ \ \ \
\begin{tikzpicture}
\draw[->,thick] (-0.5,-0.5) -- (0.5,0.5);
\draw[->,thick] (0.5,-0.5) -- (-0.5,0.5);
\node at (-0.65,-0.65) {\ \scs $y/z$};
\node at (0.65,0.65) {\ \scs $y/z$};
\node at (0.65,-0.65) {\scs $x_R$};
\node at (-0.65,0.65) {\scs $x_R$};
\end{tikzpicture}
\qquad \ \ \ \
\begin{tikzpicture}
\draw[->,thick] (-0.5,-0.5) -- (0.5,0.5);
\draw[->,thick] (0.5,-0.5) -- (-0.5,0.5);
\node at (-0.65,-0.65) {\scs $x_L$};
\node at (0.65,0.65) {\scs $x_L$};
\node at (0.65,-0.65) {\scs $x_R$};
\node at (-0.65,0.65) {\scs $x_R$};
\end{tikzpicture}
\no \end{equation}
where along each line the spatial momentum is unchanged.

\def\bcp{{\bar \chi}_+}
\def\bcm{{\bar \chi}_-}
\def\cp{{\chi}_+}
\def\cm{{\chi}_-}
\def\by{{\bar y}}
\def\dm{{\del_-}}
\def\dpl{{\del_+}}

\def \hq  {{\hat q}}
\def\bzp{{\zeta_{_R}^*}}
\def\bzm{{\zeta_{_L}^*}}
\def\bzpprime{{\zeta^{\prime\,*}_{_R}}}
\def\bzmprime{{\zeta^{\prime\,*}_{_L}}}
\def\zp{{\zeta_{_R}^{\vphantom{*}}}}
\def\zm{{\zeta_{_L}^{\vphantom{*}}}}
\def\zpm{{\zeta_{_{R,L}}}}
\def\zpprime{{\zeta'_{_R}}}
\def\zmprime{{\zeta'_{_L}}}
\def\bcp{{\chi_{_R}^*}}
\def\bcm{{\chi_{_L}^*}}
\def\bcpprime{{\chi^{\prime\,*}_{_R}}}
\def\bcmprime{{\chi^{\prime\,*}_{_L}}}
\def\cp{{\chi_{_R}^{\vphantom{*}}}}
\def\cm{{\chi_{_L}^{\vphantom{*}}}}
\def\cpm{{\chi_{_{R,L}}}}
\def\cpprime{{\chi'_{_R}}}
\def\cmprime{{\chi'_{_L}}}
\def\by{{y^*}}
\def\byprime{{y^{\prime*}}}
\def\dm{{\del_-}}
\def\dpl{{\del_+}}
\def \hq{\sqrt{1 - q^2} \, }
\def \hatD  {{\rm D}}
\def \G {\Gamma} 

%%%%%%%%%%%%%%%%%%%%%%%%%%%%%%%%%%%%%%%%%%%%%%
\subsection{Fermionic sector} \label{A_2}

A systematic derivation of  the  gauge-fixed superstring action to quartic order in  both bosons and fermions 
 goes beyond the scope of the present paper. 
Here we shall   limit ourselves to   a  discussion of  terms  quadratic in fermions
and quadratic in bosons  that are relevant for the  derivation of the most  non-trivial 
    parts  of the S-matrix \rf{bhsmatans}, 
i.e. scattering processes such as $FF\to BB$,  $BB\to FF$    and $BF \to BF$, 
which,  in particular, involve the   functions $l_4$ and $l_5$ in \rf{param45}.\foot{Note that according 
to  \rf{bhsmatans}  the amplitudes $FF \to FF$ corresponding to 
 quartic fermionic terms in the action  vanish in the light-cone gauge 
$a=\ha$ where $c=0$ (see \rf{gh}).} 

We shall  start with the general form of  the quadratic   term in the type IIB  
superstring action  (thus avoiding  gauge-fixing subtleties of truncating to the supercoset action \ci{bsz,cz}), 
 as was  also done  in the absence of $B$-flux in \ci{wulf,wu}. 
  The fermionic  ``kinetic''  term in  the
 IIB   GS superstring  action in a curved background  is a direct generalization of its  flat-space form:
\be \la{2aa}
&&L_2 =  i (\eta^{ab} \delta_{IJ} -
\epsilon^{ab} \rho_{3IJ}) \del_a x^{ m}  e^{\hat m}_m\ \bar \theta^I \Gamma_{ \hat  m }
( \hatD_b)^{JK} \theta^K \ , \\
  && \hatD_a = \del_a +  {\textstyle{ 1 \ov 4}} \del_a x^k  e_k^{\hat k} \ 
\Big[ (\om_{\hat  m \hat  n\hat   k} - \ha  \rho_3 H_{ \hat  m\hat   n\hat  k}) \G^{ \hat  m\hat   n} -    {\textstyle { 1 \ov
3!}} \rho_1 F_{\hat  m\hat  n\hat  l} \G^{\hat  m\hat  n\hat  l}   \G_{\hat k} \Big]  \ ,  \la{3}\\
&&  \rho_3 = \Big(\begin{array}{cc}  1 & 0 \\ 0  & -1  \end{array}\Big) \ , \ \ \ \ \  \ \ \   
 \rho_1 =  \Big(\begin{array}{cc}  0 & 1 \\ 1 & 0  \end{array}\Big) \ , \ \ \ \ \ 
  \ee
  where  $\hat m, $ etc.,  are     tangent-space indices,  
$\theta^I$ are two  real MW spinors
 and  $\rho_a$  act in the $I,J=1,2$ space.  \  $ \hatD_a$ is
the generalized covariant derivative that appears in the Killing
spinor equation or gravitino transformation law  in type IIB
supergravity 
(we have included only  NSNS   and RR 3-form  background  field  couplings).  
The tangent space components of the fluxes   corresponding to the metric in \rf{2222}  are\foot{Here $ \hat r' ,  \hat s'$ denote   the  spatial 
$AdS_3$ directions $z_s$  and  $ \hat r ,  \hat s$  --  the   $S^3$ transverse directions $y_{s}$.} 
\be \la{hee} 
&&H_{\hat t \hat r'  \hat s'} =    -2 q   \ep_{\hat r'\hat s'}  \ , \ \ \ \ \ \  \ \ \ \ \ \ \ \ \ 
H_{\hat \vp \hat r  \hat s} =  -2 q   \ep_{\hat r\hat s}  \ , \\
&& F_{\hat t \hat r'  \hat s'} =    -2 \hq   \ep_{\hat r'\hat s'}  \ , \ \ \ \ \ \  F_{\hat \vp \hat r \hat s} =  -2 \hq   \ep_{\hat r\hat s}
 \ . \la{fee5} 
\ee
One may   expand   the action  to quadratic order in bosonic fields  using that 
\be 
e^{\hat \vp}_\vp  =  1 - \ha y^2_s + ...\ , \ \ \     e^{\hat s}_s = 1 - \fo  y^2_s + ...\ , \ \ \ \ \ 
\omega_{\hat 3 \hat s} = - y_s  d \vp + ... \ , \ \ \   \omega_{\hat s\hat r} =   y_{[s } d y_{r] } + ... \ ,  \la{1aa}
\ee 
and similar relations  for  the $AdS_3$ part. 
The leading   term in the fermionic action in the light-cone gauge  is found    by   setting 
\be   t= \tau \ , \ \ \ \ \ \ \  \ \ \ \   \vp =\tau \ , 
 \ee
i.e.    choosing    $u\equiv  t + \vp = 2 \tau, \ v\equiv  t-\vp=0$   (cf. \rf{5.1},\rf{b})
and fixing  the  l.c. kappa-symmetry  gauge  $\G^v \theta^I=0$. 
One  then  finds  as  in \ci{rut}  that  the  fermions   split into two groups: 4  massive and 4 massless. The massive  
ones, denoted    by  $\zeta_{_{R,L}}$ and $\chi_{_{R,L}}$  (corresponding to
 $\zeta_{+}$ and $\chi_+$  in  \rf{bhsmatans}), 
  have the following action
 \bea 
\label{eL}
       &&\mathcal{L}_2 = \ \ \  i \bzp(\partial_-  + i q ) \zp +i\bzm (\partial_+ -  i q) \zm -\hq \big(\bzp \zm + \bzm\zp\big) \nonumber
\\ &&\ \qquad  +\  i \bcp(\partial_-  + i q ) \cp +i\bcm (\partial_+ -  i q) \cm -\hq \big(\bcp \cm + \bcm\cp\big)\ .
\eea
Here the $q$-dependent  derivative shifts come   from the  NSNS  flux \rf{hee}  term
  and the   mass terms   --   from the RR  flux   \rf{fee5}  term  in $\hatD$ 
in \rf{3}. 
The equations of motion that follow from \eqref{eL} are 
\begin{equation}\la{lii} 
(\partial_- + i q) \zp + i \hq \zm = 0 \ ,\qquad\qquad  (\partial_+ - i q) \zm  + i \hq \zp = 0 \ ,
\end{equation}
and similar  ones  for $\chi$. Combining them  gives  the following second-order equation 
\begin{equation}
(\partial_+ - i q )(\partial_- + iq) \zpm + (1-q^2)\zpm = 0\ ,
\end{equation}
which is the same as  the  free  equation for the massive fields $(y,y^*)$ and $(z,z^*) $ in the bosonic sector 
 following from  \rf{u22} (note that $ (\partial_+ - i q )(\partial_- + iq) = \del_0^2 - (\del_1 - i q)^2$, cf. \rf{vev}). 
 This  implies   that all massive bosons   and fermions have the same
dispersion relation \rf{hap} or \eqref{dis}.

The  structure of   \rf{2aa} and   the expansions \rf{1aa}  imply  that there are no terms  quadratic in fermions 
and linear in  the ``transverse''  bosons  $(y_s,  z_r)$. To find terms that are quadratic   in both  the    fermions and  the  bosons 
one  may follow the same procedure as used  in the purely bosonic case  in section \ref{2_3} and  the previous   subsection, i.e. 
apply T-duality in the $\vp$ direction and then fix the gauge \rf{gai} where $u= 2 \tau, \ \td \vp =  2 \s$ (here we  choose $a= \ha, \ \J=1$) 
together with $\G^v \theta^I=0$.  An alternative is to follow  the approach of \ci{cal} used for $q=0$ in \ci{wulf,wu}.

 Here  we will not give a systematic derivation  of the resulting Lagrangian  and just 
 indicate which  types of terms one should expect to find.   Let us  focus on the 
terms involving $y$ and $\zeta$ only. 
As  $\del \vp$  terms will   appear  in the $H_3$ and $F_3$ 
 parts of  $\hatD$ in \rf{3}     the corresponding    fermionic  terms linear in $\del_a \vp$  will 
 enter   the  action  in  the same way as  the bosonic WZ term 
 $b_a = \ep_{ab}   \ep_{rs} y_r \del_b y_s$ in  \rf{kik},\rf{434app}, 
i.e. there   will be products of the  connection terms  $    q ( \bzp \zp -  \bzm \zm) $ 
and $  \hq \big(\bzp \zm + \bzm\zp\big)$ with the  bosonic 
second-derivative terms $\del y \del y$.
In addition, there will  be also $q$-dependent terms  where the  fermionic kinetic term is  multiplied 
by the bosonic  $b_a$ term, i.e.  we should get terms like  $ \bzm \zmprime   ( q   \ep_{rs} y_r \dot y_s)$  
 and similar ones   with $ q   \ep_{rs} y_r  y'_s$.\foot{Equivalently, the  T-dualization  procedure with  the fermionic terms 
 included should lead  to $q$-dependent ``cross-terms'' similar 
 to the ones in \rf{434app} (for $a= \ha$)  
 with $\del x_k \del x_k$   terms replaced by the  fermionic  kinetic terms.} 
 To simplify the resulting action one is allowed to use the linearized equations of 
  motion \rf{lii} as this does not change the S-matrix  (in particular, 
   one can always   eliminate terms with time derivatives of the fermions). 
 
 Let us consider  for simplicity   a subset of  quartic terms involving  only 
 $y$ and $ \zp,\zm$. 
 Then the Lagrangian that  reproduces  the $yy\zeta\zeta$   amplitudes in  \rf{bhsmatans} 
(involving  the functions $l_{4,5,6,7,8,9}$ given in \rf{gh} with \rf{dis} and \rf{param45})
 is  found to be\foot{To recall,  in \rf{bhsmatans}   we used the notation $(y_+, y_-) = (y,y^*)$ and $(\zeta_+,\zeta_-) = (\zeta_{_{R,L} }, \zeta^*_{_{R,L}})$. Also note that in \eqref{lop} and \eqref{old} a different normalization of $(y,y^*)$ has been used compared to section \ref{2_3}. To be precise, $(y,y^*)_{\textrm{appendix \ref{A_2}}} = \frac{1}{\sqrt{2}}(y,y^*)_{\textrm{section \ref{2_3}}}$.} %v5
 \bea 
&&\mathcal{L}_4 =  -\tfrac12 \big[ \hq \, \bzm \zp +\tfrac q2 (\bzp   \zp- \bzm   \zm)\big] \dpl \by \dm y \no\\
&&  \ \ \qquad    -\tfrac12 \big[\hq \, \bzp \zm +\tfrac q2 (\bzp   \zp- \bzm   \zm)\big]\dm \by \dpl y \no  
\\ 
&& \ \  \qquad   + \big[\tfrac i4 ( \bzp \zp - \bzm \zm) + \tfrac q2 \bzm \zmprime \big] \by \dpl y 
+  \big[- \tfrac i4 ( \bzp \zp - \bzm \zm)  +  \tfrac q2 \bzmprime \zm \big] \dpl \by  y  \no \\
&& \ \  \qquad 
 - \big[ \tfrac i4 ( \bzp \zp - \bzm \zm) - \tfrac q2 \bzp \zpprime \big] \by \dm y 
 - \big[-\tfrac i4 ( \bzp \zp - \bzm \zm) - \tfrac q2 \bzpprime \zp \big] \dm \by  y \no \\
&&  \ \  \qquad   +\tfrac i4 (\bzp \dpl  \zp - \dpl \bzp   \zp +\bzm \dm  \zm  -\dm \bzm   \zm     ) \by y \ .
\la{lop}
\eea
The structure of other terms, e.g., involving $yy \chi \chi$,     is very similar.
%The terms    are  very similar.  %   (and   as is the 
We   observe    that the $q$-dependent  terms    here indeed have    the    structure expected from the T-duality  based gauge-fixing procedure outlined  above. 

In   the limit of $q = 0$  the Lagrangian  \rf{lop}  reduces to 
\begin{equation}\la{old}
\begin{split}
\mathcal{L}_4 = & -\tfrac12 \bzm \zp  \dpl \by \dm y -\tfrac12 \bzp \zm \dm \by \dpl y \\
& + \tfrac i2 ( \bzp \zp - \bzm \zm) \by y'  - \tfrac i2 ( \bzp \zp - \bzm \zm) \byprime  y 
\\ & +\tfrac i4 (\bzp \dpl  \zp - \dpl \bzp   \zp +\bzm \dm  \zm  -\dm \bzm   \zm     ) \by y \ .
\end{split}
\end{equation}
This   Lagrangian   matches      the     \adst   limit of the   corresponding part of the $AdS_3 \times S^3 \times S^3 \times S^1$  Lagrangian  found  in \ci{wulf,wu}. 
Indeed,   relabelling $y,\zpm  \rightarrow y_2,\chi_1{}_\pm ,
  $  it  is the same as   the $\alpha \to 1$ limit  
of eq.~E.1 in \cite{wu} (keeping  only the fields $y_2, \chi_1$ there), or equivalently,
as the limit $\alpha \to 0$ (keeping  only the fields $y_3, \chi_1$ there).

The above   discussion thus  gives a strong indication that the  expression for 
the $q$-dependent   S-matrix  found  in section \ref{4_2}  using   integrability constraints   
 follows also  directly from the  \adst superstring action.

%%%%%%%%%%%%%%%%%%%%%%%%%%%%%%%%%%%%%%
\def\ka{\m}
\refstepcounter{section}
\def\diag{\textrm{diag}}
\def\theequation{B.\arabic{equation}}
\setcounter{equation}{0}
\section*{Appendix B: \\ Comments on the symmetry algebra of the superstring S-matrix\label{appb}}
\addcontentsline{toc}{section}{B \ Comments on symmetry algebra of the superstring S-matrix}
%%%%%%%%%%%%%%%%%%%%%%%%%%%%%%%%%%%%%%

Below we shall  first  describe 
the supersymmetry of the  tree-level S-matrix of the massive modes in the \adst theory with 
pure RR flux ($q=0$)  which can be found by analyzing the algebra underlying 
the supercoset formulation of the theory. We shall then    comment on possible $q\not=0$ generalization of it. 

Let us  start with    briefly reviewing  the  symmetry  of  the \adss theory  based on the superalgebra $\mathfrak{psu}(2,2|4)$,
which we formulate in terms of $8 \times 8$ (traceless and supertraceless)
supermatrices:
\begin{equation}
\begin{tikzpicture}
\draw[fill,gray!20] (0,4)--(1,4)--(1,3)--(0,3)--(0,4);
\draw[fill,gray!20] (2,4)--(3,4)--(3,3)--(2,3)--(2,4);
\draw[fill,gray!20] (0,2)--(1,2)--(1,1)--(0,1)--(0,2);
\draw[fill,gray!20] (2,2)--(3,2)--(3,1)--(2,1)--(2,2);
\draw[fill,gray!40] (1,3)--(2,3)--(2,2)--(1,2)--(1,3);
\draw[fill,gray!40] (3,3)--(4,3)--(4,2)--(3,2)--(3,3);
\draw[fill,gray!40] (1,1)--(2,1)--(2,0)--(1,0)--(1,1);
\draw[fill,gray!40] (3,1)--(4,1)--(4,0)--(3,0)--(3,1);
\draw (0,0)--(0,4);
\draw (0,4)--(4,4);
\draw (4,4)--(4,0);
\draw (4,0)--(0,0);
\draw (0,2)--(4,2);
\draw (2,4)--(2,0);
\node at (0.25,3.75) {$*$};
\node at (1.25,3.75) {$*$};
\node at (2.25,3.75) {$*$};
\node at (3.25,3.75) {$*$};
\node at (0.25,2.75) {$*$};
\node at (1.25,2.75) {$*$};
\node at (2.25,2.75) {$*$};
\node at (3.25,2.75) {$*$};
\node at (0.25,1.75) {$*$};
\node at (1.25,1.75) {$*$};
\node at (2.25,1.75) {$*$};
\node at (3.25,1.75) {$*$};
\node at (0.25,0.75) {$*$};
\node at (1.25,0.75) {$*$};
\node at (2.25,0.75) {$*$};
\node at (3.25,0.75) {$*$};
\node at (0.75,3.25) {$\bullet$};
\node at (1.75,3.25) {$\bullet$};
\node at (2.75,3.25) {$\bullet$};
\node at (3.75,3.25) {$\bullet$};
\node at (0.75,2.25) {$\bullet$};
\node at (1.75,2.25) {$\bullet$};
\node at (2.75,2.25) {$\bullet$};
\node at (3.75,2.25) {$\bullet$};
\node at (0.75,1.25) {$\bullet$};
\node at (1.75,1.25) {$\bullet$};
\node at (2.75,1.25) {$\bullet$};
\node at (3.75,1.25) {$\bullet$};
\node at (0.75,0.25) {$\bullet$};
\node at (1.75,0.25) {$\bullet$};
\node at (2.75,0.25) {$\bullet$};
\node at (3.75,0.25) {$\bullet$};
\end{tikzpicture}
\label{pic1}
\end{equation}
Here the top left $4 \times 4$ block is the $\mathfrak{su}(2,2)$ bosonic subalgebra with signature $(+,+,-,-)$, while the 
bottom right $4 \times 4$ block is the $\mathfrak{su}(4)$ bosonic subalgebra. The top right and bottom left
$4 \times 4$ blocks contain the Grassmann-odd parts of the algebra. The BMN geodesic can then
be written as a supergroup-valued solution:
\begin{equation}\label{bmn}
f = \exp(\k \tau \, \diag(i,i,-i,-i,i,i,-i,-i)) \ , 
\end{equation}
which preserves the symmetry algebra
$\mathfrak{psu}(2|2)^2$
plus central extensions. The $\mathfrak{psu}(2|2)^2$ algebra is denoted
by the shaded regions in \eqref{pic1}.\foot{That is, this is 
the subalgebra of $\mathfrak{psu}(2,2|4)$ that commutes with the solution.
We use the notation 
$\mathfrak a^2 \equiv  \mathfrak a  \oplus  \mathfrak a $  for an algebra $\mathfrak a$.
Below $\ltimes$ stands for central extension.}

The truncation to the \adst theory is  found  by taking  the elements of $\mathfrak{psu}(2,2|4)$ marked $*$ and $\bullet$ 
in \eqref{pic1}. Each of these sets forms a $\mathfrak{u}(1,1|2)$ superalgebra;  however, combined they give
$\mathfrak{ps}(\mathfrak{u}(1,1|2)^2)$, where the $\mathfrak{ps} $  projections correspond to the vanishing of the overall trace and
supertrace. Eq.~\eqref{bmn} is still a solution  here  and
this allows us to write down the symmetry algebra 
preserved by the BMN geodesic in the \adst theory. It can be found by
taking the ``intersection'' of the shaded areas and the $*$ and $\bullet$ regions in \eqref{pic1}.
The resulting algebra is 
\begin{equation}\label{algebra}
[\mathfrak{ps}(\mathfrak{u}(1|1)^2)]^2 = [\mathfrak{u}(1)\inplus \mathfrak{psu}(1,1)^2]^2\ltimes \mathfrak{u}(1)^2
\rightarrow  [\mathfrak{u}(1)\inplus \mathfrak{psu}(1|1)^2]^2\ltimes \mathfrak{u}(1)\ ,
\end{equation}
where we have dropped one of the $\mathfrak{u}(1)$ central extensions ($* = i$, $\bullet = -i$) as it acts trivially. This is in agreement with the fact that the global bosonic symmetry is $U(1)^3$ (cf. \eqref{bhgbs}).

The massive tree-level S-matrix (in the case of vanishing RR flux, i.e. $q=0$) \eqref{bhsmatans},\eqref{gh} should therefore have
symmetry given by \eqref{algebra} (up to central extensions encoding the energy and momentum),
while the factorized S-matrix \eqref{bhsmatfact},\eqref{gh} 
should only have half this symmetry. 

Explicitly, the generators of the supersymmetry algebra of the factorized S-matrix \rf{bhsmatfact} in the $q=0$ case
are: two $U(1)$ generators $\mathfrak R$ and $\mathfrak L$; four supercharges $\mathfrak Q_{\pm\mp}$ and $\mathfrak S_{\pm\mp}$ ($+$ and $-$ denote the
charges under the $U(1) \times U(1)$ bosonic subalgebra)
and three central extensions $\mathfrak C$, $\mathfrak P$ and $\mathfrak K$. The commutation relations are given by
\begin{align}
&[ \mathfrak R, \, \mathfrak R ] =0 \ , & \nonumber &[ \mathfrak L, \, \mathfrak L ] =0 \ ,
\\
&[ \mathfrak R, \, \mathfrak Q_{\pm\mp} ] = \pm i \mathfrak Q_{\pm\mp} \ , & \nonumber & [ \mathfrak L, \, \mathfrak Q_{\pm\mp} ] = \mp i \mathfrak Q_{\pm\mp} \ ,
\\
&[ \mathfrak R, \, \mathfrak S_{\pm\mp} ] = \pm i \mathfrak S_{\pm\mp} \ , & \nonumber & [ \mathfrak L, \, \mathfrak S_{\pm\mp} ] = \mp i \mathfrak S_{\pm\mp} \ ,
\\
&\{\mathfrak Q_{\pm\mp} , \,\mathfrak S_{\pm\mp}\} = 0 \ , & \nonumber & \{\mathfrak Q_{\pm\mp} , \,\mathfrak S_{\mp\pm}\} =
\mp \tfrac{i
}{2}(\mathfrak R+\mathfrak L) + \mathfrak C \ ,
\\
&\{\mathfrak Q_{\pm\mp} , \,\mathfrak Q_{\pm\mp}\} = 0 \ , & \nonumber & \{\mathfrak Q_{\pm\mp} , \,\mathfrak Q_{\mp\pm}\} = \mathfrak P \ ,
\\&\{\mathfrak S_{\pm\mp} , \,\mathfrak S_{\pm\mp}\} =0 \ , & & \{\mathfrak S_{\pm\mp} , \,\mathfrak S_{\mp\pm}\} = \mathfrak K \ . \label{bcd}
\end{align}
They are consistent with the following set of reality conditions
\begin{equation}\label{bhrc}
\mathfrak R^\dagger = -\mathfrak R \ , \qquad \mathfrak L^\dagger = - \mathfrak L \ , \qquad \mathfrak Q_{\pm\mp}^\dagger = \mathfrak S_{\mp\pm} \ , \qquad \mathfrak P^\dagger = \mathfrak K \ , \qquad \mathfrak C^\dagger = \mathfrak C \ .
\end{equation}
This superalgebra is a centrally-extended semi-direct sum of $\mathfrak u(1)$ (generated by $\mathfrak R - \mathfrak L$) 
with two copies of the superalgebra $\mathfrak{psu}(1|1)$, i.e.
\begin{equation}
[\mathfrak u(1)\inplus \mathfrak{psu}(1|1)^2 ]\ltimes \mathfrak u(1) \ltimes \mathbb{R}^3 \ .
\end{equation}
The central extensions are $\mathfrak R + \mathfrak L$, $\mathfrak C$, $\mathfrak P$ and $\mathfrak K$.

Indeed, we expect $\mathfrak R + \mathfrak L$ to be central as it has the same action on the tensor product states \eqref{bhtens} whether acting on the first or the
second entry. As with the other three central extensions, $\mathfrak C$, $\mathfrak P$ and $\mathfrak K$, there is therefore only a single copy of this $\mathfrak u(1)$ central extension when we consider the symmetry of the full S-matrix
\begin{equation}
[\mathfrak u(1)\inplus \mathfrak{psu}(1|1)^2]^2\ltimes \mathfrak u(1) \ltimes \mathbb{R}^3 \ ,
\end{equation}
in agreement with \eqref{algebra}.

From this discussion it is natural to expect  the symmetry algebra for $q\not=0$ to be the same
as in the $q = 0$ case. The dispersion relation \rf{iii} should then
follow from a modification of the representation.
Indeed, the fact that $\mathfrak R + \mathfrak L$ is central suggests its eigenvalue can be altered
allowing the parameter $q$ to be introduced. 
We leave the detailed study of the relevant 
representation in the $q \not= 0$ case for the future.

\

To conclude this appendix,  we will briefly discuss the details of the invariance of the tree-level S-matrix under supersymmetry in the case $q=0$ following the \adss story \cite{afrr,km,afr,beis1,tor}.
The particular representation of interest to us consists of one complex boson and one complex fermion. The generators have the following action on the one-particle states
\begin{align}
& \mathfrak R\left|\phi_\pm\right>=\pm i\left|\phi_\pm\right> \ , & & \nonumber & & \mathfrak R\left|\psi_\pm\right>=0 \ ,
\\
& \mathfrak L\left|\phi_\pm\right>=0 \ , & & \nonumber & & \mathfrak L\left|\psi_\pm\right>=\pm i \left|\psi_\pm\right> \ ,
\\
& \mathfrak Q_{\pm\mp}\left|\phi_\pm\right>=0 \ , & & \nonumber & & \mathfrak Q_{\pm\mp}\left|\psi_\pm\right>=a \left|\phi_\pm\right> \ ,
\\
& \mathfrak Q_{\pm\mp}\left|\phi_\mp\right>=b\left|\psi_\mp\right> \ , & & \nonumber & & \mathfrak Q_{\pm\mp}\left|\psi_\mp\right>=0 \ ,
\\
& \mathfrak S_{\pm\mp}\left|\phi_\pm\right>=0 \ , & & \nonumber & & \mathfrak S_{\pm\mp}\left|\psi_\pm\right>=c\left|\phi_\pm\right> \ ,
\\
& \mathfrak S_{\pm\mp}\left|\phi_\mp\right>=d\left|\psi_\mp\right> \ , & & \nonumber & & \mathfrak S_{\pm\mp}\left|\psi_\mp\right>=0 \ ,
\\
& \mathfrak C \left|\phi_\pm\right> = C \left|\phi_\pm\right> \ , & & \mathfrak P \left|\phi_\pm\right> = P \left|\phi_\pm\right> \ , & & \mathfrak K \left|\phi_\pm\right> = K \left|\phi_\pm\right> \ , \nonumber
\\
& \mathfrak C \left|\psi_\pm\right> = C \left|\psi_\pm\right> \ , & & \mathfrak P \left|\psi_\pm\right> = P \left|\psi_\pm\right> \ , & & \mathfrak K \left|\psi_\pm\right> = K\left|\psi_\pm\right> \ .\label{bhabcd}
\end{align}
Here $a,b,c,d,C,P$ and $K$ are representation parameters that will eventually be functions of the energy and momentum of the state. For the supersymmetry algebra to close the following constraints should be satisfied
\begin{equation}\begin{split}\label{bhabcdpkc}
a b = P \ , \qquad c d = K \ , \qquad & a d = C + \frac12 \ , \qquad b c = C-\frac12\ .
\end{split}\end{equation}
These can easily be seen to imply that
\begin{equation}\label{bhshort}
C^2 = \frac14 + P K \ ,
\end{equation}
which is just the shortening condition for this atypical representation. Physically, it will be interpreted as the dispersion relation, with $C$ playing the r\^ole of the energy and $P$ and $K$ defined in terms of the momentum. The representation parameters are further constrained by the reality conditions \eqref{bhrc}
\begin{equation}\label{bhrcc}
a^* = d \ , \qquad b^* = c \ , \qquad P^* = K \ , \qquad C^* = C \ .
\end{equation}
We can solve the set of equations \eqref{bhabcdpkc} for $a,b,c,d$ in terms of $C,P$ and $K$:
\begin{equation}\begin{split}\label{bhaabbccdd}
a = \g \sqrt{P}\Big(\frac{2C + 1}{2C - 1}\Big)^\frac14 \ , \qquad &
b = \g^{-1} \sqrt{P}\Big(\frac{2C - 1}{2C + 1}\Big)^\frac14 \ , \\
c = \g \sqrt{K}\Big(\frac{2C - 1}{2C + 1}\Big)^\frac14 \ , \qquad &
d = \g^{-1} \sqrt{K}\Big(\frac{2C + 1}{2C - 1}\Big)^\frac14 \ ,
\end{split}\end{equation}
where $\g$ is a phase parametrizing the normalization of the fermionic states with respect to the bosonic ones and can
therefore be set to one.

To define the action of the symmetry on the two-particle states we need to introduce the coproduct
\begin{align}
\Delta(\mathfrak R) & = \mathfrak R \otimes \mathbb{I} + \mathbb{I} \otimes \mathfrak R \ , & & \nonumber & \Delta(\mathfrak L) & = \mathfrak L \otimes \mathbb{I} + \mathbb{I} \otimes \mathfrak L \ ,
\\
\Delta(\mathfrak Q) & = \mathfrak Q \otimes \mathbb{I} + \mathfrak U \otimes \mathfrak Q \ , & & \nonumber & \Delta(\mathfrak S) & = \mathfrak S \otimes \mathbb{I} + \mathfrak U^{-1} \otimes \mathfrak S \ ,
\\
\Delta(\mathfrak P) & = \mathfrak P \otimes \mathbb{I} + \mathfrak U^2 \otimes \mathfrak P \ , & \Delta(\mathfrak C) = & \mathfrak C \otimes \mathbb{I} + \mathbb{I} \otimes \mathfrak C \ , & \Delta(\mathfrak K) & = \mathfrak K \otimes \mathbb{I} + \mathfrak U^{-2} \otimes \mathfrak K \ ,
\end{align}
and the opposite coproduct, defined as
\begin{equation}
\Delta^{\text{op}}(\mathfrak J) = \mathcal{P}(\Delta(\mathfrak J)) \ ,
\end{equation}
where $\mathfrak J$ is an arbitrary generator and $\mathcal{P}$ defines the graded permutation of the tensor product.

We have deformed the coproduct from the usual one via the introduction of the new abelian generator $\mathfrak U$ ($\Delta(\mathfrak U) = \mathfrak U \otimes \mathfrak U$) \cite{tor}.
This is done according to a $\mathbb{Z}$-grading of the algebra, whereby the charges $-2,-1,1,2$ are associated to the generators $\mathfrak K$, $\mathfrak S$, $\mathfrak Q$, $\mathfrak P$ and the remaining generators are uncharged. The action of $\mathfrak U$ on the single particle states is given by
\begin{equation}
\mathfrak U\left|\phi_\pm\right> = U \left|\phi_\pm\right> \ , \qquad
\mathfrak U\left|\psi_\pm\right> = U \left|\psi_\pm\right> \ .
\end{equation}
This braiding allows for the existence of the non-trivial S-matrix.

The first consequence of this non-trivial braiding is found by requiring that for the central extensions the coproduct should
equal to its opposite, implying
\begin{equation}
\mathfrak P \propto (1 - \mathfrak U^2) \ ,\qquad \qquad \mathfrak K \propto (1 - \mathfrak U^{-2}) \ .
\end{equation}
We fix the normalization of $\mathfrak P$ relative to $\mathfrak K$ by taking both constants of proportionality to be
\begin{equation}
\frac \hh2 = \frac{\sqrt{\lambda}}{4\pi} \ . \la{b15}
\end{equation}
Acting on the single-particle states gives us the relations
\begin{equation}\label{bhpku}
P = \frac \hh2 \, (1 - U^2) \ ,\qquad \qquad K = \frac \hh2 \, (1 - U^{-2}) \ ,
\end{equation}
where $U$ should satisfy, as a consequence of \eqref{bhrcc}, the following reality condition
\begin{equation}\label{bhurccp}
U^* = U^{-1} \ .
\end{equation}
Motivated by the well-known construction in the $AdS_5 \times S^5$ case
(implying a similar one in the \adst case with $q=0$)
we identify $C$ with the energy and define $U$ in terms of the spatial momentum as
\begin{equation}\label{bhcuep}
C = \frac{e}2 \ ,\qquad \qquad U = e^{-\frac i2 p} \ .
\end{equation}
Using \eqref{bhpku} and \eqref{bhcuep} we can substitute in for $C$, $P$ and $K$ in terms of the energy and momentum in the shortening conditions \eqref{bhshort} to find the following dispersion relation
\begin{equation}\la{apiii}
e^2 = 1 + 4 \, \hh^2 \, \sin^2\frac p2 \ .
\end{equation}
In terms of the energy and momentum the representation parameters $a,b,c$ and $d$ \eqref{bhaabbccdd} are
\begin{equation}\begin{split}\label{bhabcdep}
a = \sqrt{\frac \hh2(1-e^{-ip})} \Big(\frac{e + 1}{e - 1}\Big)^\frac14 \ , \qquad &
b = \sqrt{\frac \hh2(1-e^{-ip})} \Big(\frac{e - 1}{e + 1}\Big)^\frac14 \ , \\
c = \sqrt{\frac \hh2(1-e^{ip})} \Big(\frac{e - 1}{e + 1}\Big)^\frac14 \ , \qquad &
d = \sqrt{\frac \hh2(1-e^{ip})} \Big(\frac{e + 1}{e - 1}\Big)^\frac14 \ .
\end{split}\end{equation}
Rescaling $p \rightarrow \hh^{-1} p$ and expanding the various representation parameters to the appropriate order in $\hh^{-1} $ we find
that the factorized tree-level S-matrix \eqref{bhsmatfact} of the theory with pure RR flux ($q=0$) co-commutes with this symmetry.

%%%%%%%%%%%%%%%%%%%%%%%%%%%%%%%%%%%%%%
\refstepcounter{section}
\def\theequation{C.\arabic{equation}}
\setcounter{equation}{0}
\section*{Appendix C: \\ Faddeev-Reshetikhin model for the string on $R\times S^3$ with $B$-flux\label{appc}}
\addcontentsline{toc}{section}{C \ Faddeev-Reshetikhin model for the string on $R \times S^3$ with $B$-flux}
%%%%%%%%%%%%%%%%%%%%%%%%%%%%%%%%%%%%%%

As already discussed in section \ref{sec2}, the motion of the bosonic string on $S^3$ with $B$-flux is described in conformal gauge by the $SU(2)$ principal chiral model with a WZ term
\be
S= - { \ha \hh}\Big[ \int d^2 \s\ \ha \tr ( J_+ J_- ) - { q } \int d^3 \s \ \third \epsilon^{abc} \tr ( J_a J_b J_c)\Big] \ ,
\ \ \ \ \ \ J_a = g^{-1} \del_a g\ . \la{4.19}
\ee
Fixing the residual conformal diffeomorphism symmetry by choosing
$t=\ka \tau$, the conformal gauge (Virasoro) conditions are
\be \tr J^2_\pm = - 2 \ka^2 \ , \la{ver}\ee
while the first-order form of the equations of motion is as in \rf{4.2},\rf{4.21}:
\be
\da J_- + \ha ( 1 + q) [J_+, J_-] =0 \ , \ \ \ \ \ \ \ \ \ \
\db J_+ - \ha ( 1 - q) [J_+, J_-] =0 \ .
\la{4.211} \ee
Note that the $1 \pm q$ factors here can be formally absorbed by a rescaling of either $J_\pm$
or $\s^\pm $.
Let us write down the action that leads to these equations for the currents,
generalizing the $q=0$ case discussed in \ci{fr,kz}.\foot{The Hamiltonian in the case of a
non-zero coefficient of the WZ term was also
discussed in \ci{fr} but our approach will be different.}
For $g \in SU(2)$ we may solve the conditions \rf{ver}
in terms of two unit 3-vector fields $S^k_\pm$ ( $ {\hat \s}^k$ are Pauli matrices and $k=1,2,3$)
\be J_\pm = i \ka S^k_\pm {\hat \s}^k \ , \ \ \ \ \ \ \ \ \ S^k_\pm S^k_\pm = 1 \ . \la{4.3} \ee
The equations of motion \rf{4.211} then become
\be
\da S^i_- - \ka (1 + q) \epsilon^{ijk} S^j_+ S^k_- = 0 \ , \ \ \ \ \ \ \ \ \ %v5
\db S^i_+ + \ka (1 - q) \epsilon^{ijk} S^j_+ S^k_- = 0 \ . \la{4.4} \ee %v5
The equations \rf{4.4} follow from the following action, generalizing
the action in the $q=0$ case given in \ci{kz} (that leads to the FR Hamiltonian \ci{fr})
\be
\rS = \int d^2 \s\ \Big[ \ka (1-q) C_+ (S_-) + \ka (1+q) C_- (S_+) - \ha (1- q^2) \m^2\, S^k_+ S^k_- \Big] \ , \la{4.5} \ee %v5
where\foot{Note that $C_a$ enters the the $SU(2)$ Landau-Lifshitz action, which can be written as $\int d^2 \s \, \big[ C_0 (n) - {1 \ov 4} n'^2_i \big] $
where $n_i$ is a unit vector with equations of motion $\dot n_i = \epsilon_{ijk} n_j n''_k$.}
\be C_\pm ( S) \equiv - \ha \int^1_0 dx\ \epsilon^{ijk} S_i \del_x S_j \del_\pm S_k \ , \ \ \ \ \ \ \
\delta C_\pm = \ha \epsilon^{ijk} \delta S_i S_j \del_\pm S_k\ . \ee
In what follows we shall rescale $\tau,\s$ by $\m$, i.e. effectively set $\m=1$ and assume that
$\s$ is non-compact.

We observe that there is a simple way to relate the actions \rf{4.5}
with $q=0$ and $q\not=0$. Let us make the following conformal transformation
\be\td \s^+ = (1 + q) \s^+ \ , \ \ \ \ \ \ \ \ \td \s^-= (1 - q) \s^- \ , \ \ \ \ \ \ \ \s^\pm = \ha ( \tau \pm \s) \ . \la{ah} \ee
Since the action
\rf{4.5} is not conformally invariant it will change and become formally the same as at $q=0$:
\be
\td \rS = \int d^2 \td \s\ \Big[ \td C_+ (S_-) + \td C_- (S_+) - \ha S^k_+ S^k_- \Big] \ . \la{4.511} \ee
Let us note that the same transformation applied in the Pohlmeyer-reduced (PR) theory
(which is also constructed by starting with first-order equations for the currents and solving the Virasoro conditions)
will also
remove the $q$-dependent factor ($1-q^2$) from the mass term and will therefore relate
the $q=0$ and $q\not=0$ theories (see Appendix \ref{appd}).\foot{$q$ will still enter non-trivially in the relation between the
PR and string sigma model solutions.}

More explicitly, \rf{ah} implies that
\be \la{45} &&
\td \tau= \tau + q \s \ , \ \ \ \ \ \ \ \ \ \ \ \ \ \ \ \ \ \ \ \td \s= \s + q \tau \ , \\
&&
e= \td e + q \td p \ , \ \ \ \ \ p=\td p + q \td e \ , \ \ \ \ \ \
\td e = { e - q p \ov 1- q^2 } \ , \ \ \ \ \ \td p= { p - q e \ov 1- q^2 } \ , \la{466}
\ee
where $p_a=(p_0,p_1)\equiv (e,p)$ is the 2-momentum conjugate to $\s^a= ( \tau, \sigma)$ (i.e. $p_a \s^a = \td p_a\td \s^a$).
Since this is a conformal transformation rather than a Lorentz boost the mass
gets rescaled: $p_a^2 = (1-q^2) \td p^2_a$.

Let us now explicitly
solve the unit-vector constraints in \rf{4.3} by introducing two independent
complex scalar fields as
\begin{equation}\label{4.52}
{S_\pm^1+iS_\pm^2} = 2 \sqrt{1 - |\p_\pm|^2} \ \phi _\pm
\ , \ \ \ \ \
\qquad S^3_\pm=1-2|\phi _\pm|^2 \;.
\end{equation}
Substituting into \rf{4.5} we find the following first-order action for $\phi _+$, $\phi _-$ (generalizing the
corresponding action \ci{kz} in the $q=0$ case)
\begin{eqnarray}\label{453}
\rS&=&\int_{}^{}d^2\s\,\Big\{
{ {i} } (1-q)\ \phi _-^*\partial _+\phi _-
+ {{i} } (1+q)\ \phi _+^*\partial _-\phi _+
\nonumber \\ &&\ \ \ \ \ \ \ \ \ \ \ \ \ \ \
- (1-q^2) \Big[ \sqrt{(1-|\phi _+|^2) (1-|\phi _-|^2)}
(\phi _+^*\phi _-+\phi _-^*\phi _+) \no \\
&& \ \ \ \ \ \ \ \ \ \ \ \ \ \ \ \ \ \ \ \ \ \ \ \ \ \qquad - |\phi _+|^2- |\phi _-|^2
+ 2 |\phi _+|^2|\phi _-|^2 \Big]
\Big\} \;.
\end{eqnarray}
If we rescale $\phi_\pm$ to have canonically normalized kinetic terms then $q$ will enter the potential terms in a
complicated way.
However, the coordinate transformation \rf{ah},\rf{45} provides a short-cut to determine the dependence on $q$
by starting with the $q=0$ expression.

Let us first look at quadratic terms in \rf{453} near the trivial vacuum $\phi_\pm=0$:\foot{It is easy to see
that $\phi_\pm=0$ is the only choice for a vacuum state, modulo $SO(3)$ rotation.}
\be
L_2 = i (1-q) \phi _-^*\partial _+\phi _-
+ i (1+q) \phi _+^*\partial _-\phi _+ + (1-q^2) \ (\phi _+ - \p_-) ( \p_+^* - \p_-^*) \ .
\la{454}\ee
For $q=0$ the dispersion relation is \ci{kz} \
\be (e + 1)^2 - p^2 = 1
\ , \la{455}\ee
so that there is a particle (magnon or BMN) state which is light at small $p$ and an antiparticle state that decouples at low momenta.
For $q\not=0$ we find\foot{The same result is found using \rf{466}
to get the generalization of the dispersion relation at $q=0$:
$(\td e + 1)^2 - \td p^2 = 1 $.}
\be
( e + 1)^2 - (p + q )^2 = 1-q^2 \ ,
\la{jk}
\ee
which has a solution
\be
e = \sqrt{ (p + q )^2 + 1-q^2 } - 1
\ . \la{555}
\ee
This is the same dispersion relation (up to an overall energy shift)
as was found from the $S^3$ string sigma model in \rf{did}.
If we allow for the overall shifts of the energy and the momentum then the dispersion relation
becomes the standard massive one with $q$-dependent mass
\be e^2 - p^2 = 1-q^2 \ . \la{fj} \ee
Equivalently, starting with the $q\not=0$ action in \rf{453} and doing the $U(1)$ redefinition
\be \la{red}
\p_\pm \to e^{ i \td \tau } \p_\pm\ ,\ \ \ \ \ \ \ \ \ \ \ \ \ \ \ \td \tau =\tau + q \s\ ,
\ee
one removes the $|\p_+|^2 +|\p_-|^2$ term from the quadratic part
of \rf{453} (while the remaining terms stay invariant) thus ending up with the dispersion relation \rf{fj}
with unshifted momentum and energy.

Next, let us discuss the S-matrix starting with the $q=0$ case.
Following \ci{kz} we redefine $\p_\pm\to e^{ i \tau} \p_\pm$ in the $q=0$ analog of \rf{453} so that
the $|\p_+|^2 + |\p_-|^2$ quadratic terms are eliminated, or equivalently, shift $e$ in \rf{455} to get the standard
relativistic dispersion relation $e^2 - p^2 =1$.
The interaction vertices and thus the S-matrix will still be non-relativistic.\foot{Recall 
that the origin of this non-invariance is in the Virasoro plus temporal gauge conditions that are solved
by \rf{4.3}: originally the currents $S_\pm$ should transform as vectors but the invariance is then broken
by the unit-norm condition.
One may formally ask for $\p_\pm$ in \rf{453} to
transform under Lorentz boosts as 2d Weyl fermions making their kinetic terms invariant but then
the interaction terms in \rf{453} will still fail to be invariant.}

After the field redefinition making the
$\p_+,\p_-$ propagator the standard Lorentz-invariant massive one (so that we have both positive and negative energy states)
one is still to decide how to quantize the theory.\foot{In \ci{kz} the ``wrong'' vacuum was chosen
in which the negative energy states are all empty
with a hidden motivation of getting a standard ``ferromagnetic'' type S-matrix.
This amounts to the use of the retarded rather than the causal propagator (just as was the case in the LL model).
Then it is straightforward to compute the corresponding two-particle S-matrix as it will be given simply
by summing bubble graphs \ci{kz}.}
This will not be important at the tree level we are interested in here
as the tree-level S-matrix is given simply by the quartic vertices in the action.

The prescription of \ci{kz} gave the following quantum S-matrix\foot{Here the energies are $e= \sqrt{p^2 + 1} $ and $e'=\sqrt{p'^2 + 1} $. Recall that we set the
scale parameter or effective coupling $\m$ equal to 1. Below we also inserted as formal coupling constant $\kappa$ as in \rf{FRe}.} %vv4
\be \la{ss}
\SS_{\rm FR} (p,p') = \frac{x-x' - 2i\kappa }{x -x' +2i \kappa } = 1 + { 4 i \kappa \ov x'-x} +...
\ ,
\ee
where
$x=x(p)$ (and similarly $x'=x(p')$) is related to the momentum $p$ as
\be\label{ra}
x= {1 \ov p} ( \sqrt{ p^2 +1 } +1) \ , \ \ \ \ \ \ \ \ \ \ \ \ \ p = \frac{2x}{x^2-1} \ , \ \ \ \ \ e = \frac{x^2+1}{x^2-1} \ .
\ee
To find the $q\not=0$ generalization of this S-matrix
we may use the coordinate or momentum transformation trick \rf{45},\rf{466}:
we first put tildes on $p,p'$ in \rf{ss},\rf{ra} and then use \rf{466} where now $e^2 - p^2 = 1- q^2$, i.e.\foot{Note
that compared to \rf{jk} here there are no shifts of $p_0$ and $p_1$ as we assumed
that the original $q=0$ kinetic term took a standard massive form due to the $\p_\pm$ redefinition -- see \rf{gh} and the surrounding discussion.}
\be
\td x = x(\td p) = {1 \ov \td p} ( \sqrt{ \td p^2 +1 } + 1 )\ , \ \ \ \ \ \ \ \ \
\td p = { p - q \sqrt{ p^2 + 1- q^2}\ov 1-q^2} \ , \ \ \ \ \ \td e= \sqrt{ \td p^2 + 1 } \ , \la{qqq}
\ee
with the corresponding $S$-matrix now being given by (cf. \rf{ss},\rf{ffr})
\be \la{sss} && \SS(p,p') = \frac{\td x -\td x' -2 i\kappa}{ \td x -\td x' +2i\kappa } = 1 + \frac{4 i\kappa }{\td x' -\td x }+....
= 1 + \frac{4 i \kappa \td p \td p' }{ \td p ( \td e' + 1) - \td p' ( \td e + 1 ) } +...
\; ,
\ee
where we have explicitly shown the tree-level part.

In contrast to what we observed in the sigma model case in section \ref{sec3}, here
the $q$-dependence of the tree-level S-matrix cannot be found by just generalizing the dispersion relation as in %vv4
\rf{bhscatter3}. This may not be surprising
given that the FR S-matrix \eqref{ffr} did not agree with the string sigma model S-matrix in \rf{tat} already in the $q=0$ case.

Indeed, despite sharing the same classical integrable structure the $R\times S^3$ gauge-fixed string
sigma model, its Faddeev-Reshetikhin formulation
and its Pohlmeyer reduction all have different tree-level S-matrices. This may be attributed to the fact that these S-matrices %vv4
are computed for different objects (and also are effectively gauge-dependent quantities).

%%%%%%%%%%%%%%%%%%%%%%%%%%%%%%%%%%%%%%
\refstepcounter{section}
\def\theequation{D.\arabic{equation}}
\setcounter{equation}{0}
\section*{Appendix D:\\ Pohlmeyer-reduced theory for superstring on \adst \\ with mixed flux\label{appd}}
\addcontentsline{toc}{section}{D \ Pohlmeyer-reduced theory for superstring on \adst with mixed flux}
%%%%%%%%%%%%%%%%%%%%%%%%%%%%%%%%%%%%%%

Pohlmeyer reduction (PR) for classical string theory on $AdS_3 \times S^3$ leads to a combination of
complex sine-Gordon and complex sinh-Gordon models which admits a natural superstring
generalization \ci{gt1,gt2}. The corresponding S-matrix near trivial (BMN-type) vacuum
is relativistic and was studied in \ci{ht1,ht2}. Given that string supercoset sigma model and its PR counterpart are closely related
(at least at the classical and 1-loop level) sharing, in particular, the same integrable structure,
it is of interest to study how the PR model gets modified upon switching on non-zero NSNS 3-form flux, i.e. for $q\not=0$. As we will show below, somewhat unexpectedly, the modification
is remarkably simple: one just needs to replace the mass parameter $\m$ of the PR theory
(which is essentially the same as the BMN parameter $\J$ in the string sigma model context
setting the mass scale via the Virasoro condition) as
\be \mu \ \to \ \sqrt{ { 1- q^2}} \, \mu \ . \la{muq} \ee
As a result, the PR theory depends on $q$ only through the mass $\sqrt{ 1- q^2} \, \mu $ of the elementary excitations
which will thus have the same dispersion relation as ``rotated'' string excitations in \rf{vev}
(i.e. with unshifted momentum $p=\hat p$ in \rf{hap}).

We shall first consider the bosonic theory in the conformal gauge where the starting equations are the same as in the FR case -- eqs.\rf{4.19},\rf{ver}.
Before turning to the group-theoretic formulation that naturally generalizes to the superstring (supercoset) case
it is useful to describe the construction of the bosonic PR theory using explicit embedding coordinate parametrization
of the sigma model.

%%%%%%%%%%%%%%%%%%%%%%%%%%%%%%%%%%%%%%
\subsection{Pohlmeyer reduction of bosonic $AdS_3 \times S^3$ string in embedding coordinates\label{secapp1}}
%%%%%%%%%%%%%%%%%%%%%%%%%%%%%%%%%%%%%%

%%%%%%%%%%%%%%%%%%%%%%%%%%%%%%%%%%%%%%
\subsubsection{PR model for string on $R \times S^3$}
%%%%%%%%%%%%%%%%%%%%%%%%%%%%%%%%%%%%%%

We shall start with the string action \rf{4.1} in the conformal gauge written
in terms of $\mathbb{R}^4$ embedding coordinates
\begin{equation}\label{actionembed}
\rS= \frac{\sqrt\lambda}{2\pi}\Big( \ha \int d^2 \s \; \big[ \partial_+ X \cdot \partial_- X + \Lambda (X^2 - 1) \big]
+ \third q\, \int d^3 \s \; \epsilon^{abc} \epsilon_{mnpq} X_m \partial_a X_n \partial_b X_p \partial_c X_q \Big)\ .
\end{equation}
The resulting equations of motion and the Virasoro constraints are\footnote{Note that when varying the WZ term the four vectors $\delta X$ and $\partial_a X$ are orthogonal to $X$ (as can be seen by varying or differentiating the sphere constraint, $X^2 = 1$). Therefore they only span a three-dimensional subspace of $\mathbb{R}^4$ so that $\epsilon_{mnpq}\delta X_m \partial_a X_n \partial_b X_p \partial_c X_q = 0$.}
\be\label{eom1}
&& \partial_+ \partial_- X_m - \Lambda X_m + q K_m = 0 \ , \qquad X^2_m = 1 \ , \ \ \ \ \ \ K_m = \epsilon_{mnpl} X_n \partial_+ X_p \partial_- X_l \ , \\
&& (\partial_\pm X_m)^2 = \mu^2 \ .
\ee
Note that here
\be X \cdot \partial_\pm X = 0 \ , \ \ \ \ \ \ \ \ \ \
\Lambda = - \partial_+ X \cdot \partial_- X \ . \la{lam}
\ee
Next, we introduce the fields $(\varphi, \chi)$ of the reduced theory\footnote{Note that as the Virasoro constraints imply that $\partial_+ X$ and $\partial_- X$ are vectors with norm $\mu$, $\varphi$ is half the angle between them.}
\begin{equation}
\partial_+ X \cdot \partial_- X = \mu^2 \cos 2\varphi \ ,\qquad \qquad K \cdot \partial_\pm^2 X = f_\pm(\varphi) \partial_\pm \chi \ ,
\end{equation}
where $f_\pm(\vp) $ are to be determined.
As the set of vectors $\{X, \partial_+ X, \partial_- X, K$\} form a basis of $\mathbb{R}^4$ (assuming $\varphi$ is non-zero) we can write $\partial_\pm^2 X$ as linear combinations of them
\begin{equation}
\partial_\pm^2 X = - \mu^2 X + 2 \partial_\pm \varphi \cot 2\varphi \, \partial_\pm X - 2 \partial_\pm \varphi
\operatorname{csc} 2\varphi \, \partial_\mp X + \frac{f_\pm \partial_\pm \chi}{\mu^4 \sin^2 2\varphi} K \ .
\end{equation}
Taking the inner product we find the following equation of motion for $\varphi$
\begin{equation}\label{eomphi1}
\partial_+\partial_- \varphi + \frac{f_+f_- \partial_+\chi\partial_-\chi}{2\mu^6\sin^3 2\varphi} + \frac{\mu^2(1-q^2)}2\sin2\varphi = 0\ .
\end{equation}
If we assume that $f_+ = -f_-$ then
\begin{equation}\label{fpm1}
f_+ = - f_- = A \sin^2\varphi \ , \ \ \ \ \ \ \ \
\end{equation}
is a solution of
\begin{equation}\label{equationf}
\partial_-\Big(\frac{\partial_+^2 X \cdot K}{f_+}\Big) - \partial_+\Big(\frac{\partial_-^2 X \cdot K}{f_-}\Big) = 0 \ .
\end{equation}
Another solution is
\begin{equation}\label{fpm2}
f_+ = f_- = B \cos^2\varphi \ .
\end{equation}
Note that in solving \eqref{equationf} the $q$-dependence drops out and therefore these solutions are exactly the same as in the $q=0$ case.
Taking $f_\pm$ to be given by \eqref{fpm1}
we get for $\chi$
\begin{equation}\label{eomchi}
\partial_-(\tan^2\varphi \, \partial_+\chi) + \partial_+(\tan^2\varphi \, \partial_-\chi) = 0 \ .
\end{equation}
Choosing $A = 4\mu^3$ the equation of motion for $\varphi$ \eqref{eomphi1} is then given by
\begin{equation}\label{eomphi2}
\partial_+\partial_- \varphi - \operatorname{sec}^2\varphi\tan\varphi \, \partial_+\chi\partial_-\chi + \ha (1-q^2) \mu^2 \sin2\varphi = 0\ .
\end{equation}
The equations \eqref{eomchi} and \eqref{eomphi2} are those of the complex sine-Gordon model with mass-squared $(1-q^2)\mu^2$, i.e. they can be found
from the following Lagrangian
\begin{equation}\label{lagpr}
L= \partial_+\varphi\partial_-\varphi + \tan^2\varphi \, \partial_-\chi \partial_+\chi + \ha (1-q^2)\mu^2 \cos 2\varphi \ .
\end{equation}
Thus the only effect of $q$ in the PR theory is to modify the mass parameter. In particular, at the WZW points $q = \pm1$ we find,
as might be expected, a massless theory
(representing the $SU(2)/U(1)$ gauged WZW model). In the above derivation the r\^oles of the two solutions \eqref{fpm1},\eqref{fpm2} can be interchanged. This modifies the reduced theory Lagrangian \eqref{lagpr} by the replacement $\tan^2\varphi \rightarrow \cot^2\varphi$.

%%%%%%%%%%%%%%%%%%%%%%%%%%%%%%%%%%%%%%
\subsubsection{PR models for strings on $AdS_3 \times S^1$ and $AdS_3$\la{secads}}
%%%%%%%%%%%%%%%%%%%%%%%%%%%%%%%%%%%%%%

The reduction for strings on $AdS_3 \times S^1$ works in much the same way as that for strings on ${R}_t \times S^3$.
Here we start with the action
\begin{equation}\label{actionembedads}
\rS=\frac{\sqrt\lambda}{2\pi}\Big(\ha \int d^2 \s \; \big[\partial_+ Y \cdot \partial_- Y + \tilde \Lambda (Y^2 + 1)\big] + {\textstyle {1 \ov 3} }\, q \,
\int d^3 \s \; \epsilon^{abc} \epsilon_{\mu\nu\rho\sigma} Y^\mu \partial_a Y^\nu \partial_b Y^\rho \partial_c Y^\sigma \Big)\ ,
\end{equation}
where $Y^\mu$ are the coordinates on $\mathbb{R}^{2,2}$ with signature $(-,-,+,+)$ and $\tilde \Lambda$ is a Lagrange multiplier imposing the $AdS_3$
constraint.
We fix the conformal gauge as well as the $S^1$ angle
$\varphi= \mu \tau$. The resulting equations of motion and the Virasoro constraints are then
\be\label{eom1ads}
&& \partial_+ \partial_- Y_\m - \tilde \Lambda Y_\m + q \tilde K_\m= 0 \ , \qquad Y^2 = - 1 \ ,\qquad
\tilde K^\mu = \epsilon^{\mu}{}_{\nu\rho\sigma} Y^\nu \partial_+ Y^\rho \partial_- X^\sigma \ , \\
&& (\partial_\pm Y)^2 = - \mu^2 \ . \la{virads}
\ee
Introducing the reduced theory fields $\phi$ and $\vartheta$
\begin{equation}
\partial_+ Y \cdot \partial_- Y = -\mu^2 \cosh 2\phi \ , \qquad \tilde K \cdot \partial_\pm^2 Y = \pm 4\mu^3\sinh^2\phi \, \partial_\pm \vartheta \ ,
\end{equation}
we find that they satisfy the following second-order equations
\begin{equation}\begin{split}\label{eomads}
\partial_+\partial_- \phi &- \operatorname{sech}^2\phi\tanh\phi \, \partial_+\vartheta\partial_-\vartheta +\ha (1-q^2) \mu^2 \sinh2\phi = 0\ ,
\\
& \partial_-(\tanh^2\phi\,\partial_+\vartheta) + \partial_+(\tanh^2\phi\,\partial_-\vartheta) = 0 \ .
\end{split}\end{equation}
These equations are those of the complex sinh-Gordon model
that follow from the Lagrangian\footnote{An alternative Lagrangian with $\tanh^2\phi \rightarrow \coth^2\phi$ is found by taking instead
$ \tilde K \cdot \partial_\pm^2 Y = 4\mu^3\cosh^2\phi \, \partial_\pm \vartheta \ .$}
\begin{equation}\label{lagprads}
L= \partial_+\phi\partial_-\phi + \tanh^2\phi \, \partial_-\vartheta \partial_+\vartheta - \ha (1-q^2) \mu^2 \cosh 2\phi \ .
\end{equation}
Again, the only effect of $q$ is to modify the mass parameter and $q = \pm1$
corresponds to a massless theory ($SL(2,R)/U(1)$ gauged WZW model).

It is of interest to consider the case when the string moves just on $AdS_3$ \ci{gt2,ht3}
that corresponds to the limit $\mu \to 0$.
To take the $\mu \rightarrow 0$ limit of the Lagrangian \rf{lagprads} we should first generalize it by introducing the auxiliary field $a_\pm$
\begin{equation}\begin{split}\label{lagpradsgauge}
L= &\partial_+\phi\partial_-\phi + \sinh^2\phi \, \partial_-\vartheta \partial_+\vartheta
\\ & - a_- \, \sinh^2\phi \, \partial_+ \vartheta - a_+ \, \sinh^2\phi \, \partial_-\vartheta + a_+ a_- \cosh^2\phi - \ha (1-q^2) \mu^2 \cosh 2\phi \ .
\end{split}\end{equation}
Integrating out $a_\pm$ gives back \eqref{lagprads}. To get a finite and non-trivial $\mu \rightarrow 0$ limit we
shift $\phi$ and rescale $\vartheta$ and $a_\pm$ as
\begin{equation}
\{\phi,\vartheta,a_\pm\} \rightarrow \{\phi - \log \mu , \ \mu \, \vartheta , \ \mu \, a_\pm\} \ .
\end{equation}
The resulting Lagrangian is then given by
\begin{equation}
L=\partial_+\phi \partial_- \phi + \fo e^{2\phi} (\partial_+ \vartheta - a_+)(\partial_- \vartheta - a_-) - \fo (1-q^2) e^{2\phi}\ .
\end{equation}
This can be written as
\be\label{laglag2}
&& L= \partial_+\phi \partial_- \phi + \fo e^{2\phi} \partial_+\xi_{\boldsymbol{+}}\partial_-\xi_{\boldsymbol{-}} - \fo ({1-q^2}) e^{2\phi}\ , \\
&& \ \ \ \ \ \ \ a_\pm \equiv \partial_\pm\tilde \xi_{\boldsymbol{\pm}}\ , \ \ \ \ \ \ \ \ \ \ \ \ \ \xi_{\boldsymbol{\pm}} \equiv
\vartheta - \tilde \xi_{\boldsymbol{\pm}} \ .
\ee
One can alternatively get this Lagrangian through the reduction procedure for $\mu=0$, starting with the following definitions of the reduced-theory fields
\begin{equation}
\partial_+ Y \cdot \partial_- Y = -\ha e^{2\phi} \ , \qquad \qquad \tilde K \cdot \partial_\pm^2 Y = \pm \fo e^{4\phi} \partial_\pm \xi_{\boldsymbol{\pm}} \ .
\end{equation}
Note that
in the course of taking the $\mu \rightarrow 0$ part of the diffeomorphism symmetry has been restored.
This is a consequence of the fact that for $\mu\rightarrow 0$ the conformal-gauge constraints \eqref{virads} are invariant under conformal reparametrizations. The conformal reparametrizations acting on the reduced-theory fields are given by
\be \label{cp} \s^\pm \rightarrow f_{_\pm}(\s^\pm) \ , \ \ \ \ \ \ \ \
\partial_\pm \rightarrow f' _{_\pm} \partial_\pm \ , \ \ \
\qquad e^{2\phi} \rightarrow f'_{_+ } f'_{_-} e^{2\phi} \ ,
\qquad \xi_{\boldsymbol{\pm}} \rightarrow f_{_\mp}^{-1} \xi_{\boldsymbol{\pm}} \ .
\ee
To describe the physical degrees of freedom of the string this symmetry should be fixed. One way of doing this is to observe that the
classical equations for $\xi_{\boldsymbol{\pm}}$ imply
\begin{equation}
\partial_\pm U_\mp = 0 \ , \ \ \ \ \ \ \ \ \ \ \ \ U_\pm \equiv e^{2\phi}\partial_\pm\xi_{\boldsymbol{\pm}} \ ,
\end{equation}
where $U_\pm$ transform under the conformal reparametrizations \eqref{cp} as
$U_\pm \rightarrow f'^2_{_\pm} U_\pm $.
Therefore, these fields can be fixed to be equal to $\g= \{+1,0,-1\}$ depending on their sign. Then the Lagrangian \eqref{laglag2} becomes
\begin{equation}\label{laglag}
L= \partial_+\phi \partial_- \phi + \fo
\g\,
e^{-2\phi} - \fo ({1-q^2}) e^{2\phi}\ .
\end{equation}
As long as $q \neq \pm 1$ we can shift $\phi$ to find
that this Lagrangian is equivalent to either the sinh-Gordon, Liouville or cosh-Gordon Lagrangian respectively.
In the case of $q = \pm1$ we have either the Liouville Lagrangian, a free boson or the Liouville Lagrangian with the ``wrong'' sign of the potential.

%%%%%%%%%%%%%%%%%%%%%%%%%%%%%%%%%%%%%%
\subsubsection{Comments on relation between classical solutions of string and PR models}
%%%%%%%%%%%%%%%%%%%%%%%%%%%%%%%%%%%%%%

Solutions of the reduced theory with $q\not=0$ are formally related to solutions of the reduced theory with $q=0$ through the following conformal rescaling of the 2d coordinates
\begin{equation}\label{trans} \s^\pm \rightarrow (1 \pm q) ^{-1} \s^\pm \ , \ \ \ \ \ \ \ \ \ \ \ \
\partial_\pm \rightarrow (1 \pm q) \partial_\pm \ .
\end{equation}
A similar observation was made in the discussion of the FR model (cf. \rf{4.2},\rf{ah}).\foot{Let us note that this transformation relating solutions to solutions is a symmetry for generic $q$ only if $\s$ is decompactified.}
Indeed, if we formally consider the transformation \eqref{trans} in \rf{4.2} without letting $J_\pm$ transform
then we recover the equations of the principal chiral model with $q=0$.
Since the fields of the PR theory are essentially the currents of the string sigma model, this
explains why the transformation \eqref{trans} maps between $q=0$ and $q\not=0$ cases of the reduced theory.

The standard prescription (for $q=0$) of how to reconstruct string sigma model
solutions from the solutions of the PR theory
is to take a solution of the complex sine-Gordon equations ($\varphi_0, \chi_0$) and solve the second-order linear equation
\begin{equation}
\partial_+\partial_- X_m + \mu^2\cos 2\phi_0\, X_m = 0 \ .
\end{equation}
For non-zero $q$ this equation is modified, implying that, while the solutions of the reduced theories with $q=0$ and $q\not=0$
are related
simply by \eqref{muq}, the corresponding solutions of the string theory will in general have a more
non-trivial relation.

Another general conclusion is that since the 1-loop string partition function should be equal to the 1-loop partition function of the PR theory \ci{hit}
and the latter should depend on $q$ only via $\mu^2 \to (1-q^2) \mu^2$
the same should apply to the string partition function.
Let us now consider some simple examples of solutions, introducing the following explicit coordinates on $S^3$
\be X_1 + i X_2 = & \sin \theta \, e^{i\phi_1} \ ,
\ \ \ \ \ \ \ \ \ \ \ \ \ \ \ \ X_3 + i X_4 = & \cos \theta \, e^{i\phi_2} \ ,
\ee
in terms of which the Lagrangian in \eqref{actionembed} is given by \rf{1}, i.e.
\begin{equation}\label{actioncoord}
L=
\partial_+ \theta \partial_-\theta +\sin^2\theta\, \partial_+\phi_1\partial_-\phi_1 + \cos^2\theta\, \partial_+\phi_2\partial_-\phi_2 + q \sin^2\theta\, (\partial_+\phi_1 \partial_- \phi_2 - \partial_+\phi_2 \partial_- \phi_1)
\ .
\end{equation}

Let us first consider
the analogue of the BMN solution
\be \label{bmnsols}
X_1 + i X_2 = 0 \ , \qquad & X_3 + i X_4 = e^{i \mu \tau}\ ,
\ \ \ \ \ \ \ \ \ \ \
\theta = \phi_1 = 0 \ , \qquad & \phi_2 = \mu \tau \ ,
\ee
for which the corresponding reduced theory solution is the vacuum one
\begin{equation}\label{bmnsolr}
\varphi = 0 \ , \qquad \qquad \chi = 0 \ .
\end{equation}
Expanding \eqref{actioncoord} around \eqref{bmnsols} to the leading order gives
\begin{equation}
L= \partial_+\theta\partial_- \theta + \theta^2 \partial_+\phi_1\partial_+\phi_2- \mu^2 \theta^2 + q \mu \theta^2(\partial_+\phi_1-\partial_-\phi_1) + \partial_+ \phi_2 \partial_-\phi_2 \ .
\end{equation}
Transforming ($\theta$, $\phi_1$) to cartesian coordinates we find the following spectrum of fluctuation frequencies
\begin{equation}\la{mmm}
\pm \sqrt{(p-q \mu)^2 + (1-q^2)\mu^2 } \ ,\ \ \ \pm \sqrt{(p+q \mu)^2 + (1-q^2)\mu^2 } \ ,\ \ \ \pm p \ ,
\end{equation}
where $p$ is spatial momentum (which is integer if $\s$ is $2\pi$ periodic).
Expanding \eqref{lagpr} around \eqref{bmnsolr} in a similar way
gives
\begin{equation}
L= \partial_+ \varphi \partial_-\varphi + \varphi^2 \partial_+ \chi \partial_- \chi - (1-q^2)\mu^2 \varphi^2 \ .
\end{equation}
Transforming to cartesian coordinates we find the following spectrum of fluctuation frequencies
\begin{equation}\begin{split}
& 2 \ \times \ \pm \sqrt{p^2 + (1-q^2)\mu^2} \ .
\end{split}\end{equation}
This is the same as the massive part of the string spectrum in \rf{mmm} up to $q$-shifts in the spatial momentum.

Another simple explicit solution is the circular spinning string
\begin{equation}\begin{split}\label{circstring}
& X_1 + i X_2 = \sqrt{\frac{\nu + q}{2\nu}}e^{i(\nu-q)\tau + i \sigma} \ , \qquad X_3 +i X_4 = \sqrt{\frac{\nu-q}{2\nu}}e^{i(\nu+q)\tau-i\sigma} \ ,
\\
& \sin^2\theta = \frac{\nu + q}{2\nu} \ , \qquad \phi_1 = (\nu-q)\tau + \sigma \ , \qquad \phi_2 = (\nu+q)\tau - \sigma \ ,
\end{split}\end{equation}
where
\begin{equation}
\mu = \sqrt{\nu^2+1 -q^2} \ .
\end{equation}
Translated into the reduced theory this solution becomes\foot{
One may wonder how the form of this solution
is consistent with the claim that the PR solutions should depend on $q$ only via
$\bar \mu =\m \sqrt{ 1- q^2} $.
The PR equations are invariant under the formal transformation:
$\sigma^\pm \to (1\mp q)^{-1} \sigma^\pm $ and $\m \to ({1-q^2})^{1/2} \mu$ (see also the related transformation \eqref{trans}).
Performing this transformation on the $(\phi,\theta)$ solution above we find
\be \no
\cos 2 \phi = 1 - 2{ \bar \m}^{-2}\ , \ \ \ \ \ \ \
\theta = (\bar \m-{\bar \mu} ^{-1}) \tau\ , \ \ \ \ \ \ \
\bar \mu = \m \sqrt{1-q^2} \ . \ee
That is the solution can be put into a form such that it depends on $q$ only via $\bar \m$ and thus
satisfies the PR equations. This transformation does not respect $\sigma$-periodicity.
However, if and when the reduction procedure preserves the periodicity of a classical string solution
is a subtle issue even for $q=0$ \cite{hit,i}, which we will not address here.}
\begin{equation}\label{circrt}
\cos 2\phi = \frac{\nu^2-1-q ^2}{\nu^2+1-q ^2}\ , \qquad\qquad
\theta = \frac{\left(\nu^2- q ^2\right) (\tau - q \sigma)}{\sqrt{\nu^2+1-q ^2}} \ .
\end{equation}
Expanding \eqref{actioncoord} around \eqref{circstring} and \eqref{lagpr} around \eqref{circrt} to quadratic order,
we find the following characteristic equations
respectively
\begin{equation}\begin{split}
(\omega^2 - p^2)\big[ &(\omega^2-p^2)\left(\omega^2-p^2+4(1-q^2)\right)-4 \mu^2 (\omega+q p)^2\big] = 0\ ,
\\ & (\omega^2-p^2)\left(\omega^2-p^2+4(1-q^2)\right)-4 \mu^2 (\omega+q p)^2 = 0\ .
\end{split}\end{equation}
Ignoring the trivial (longitudinal) string massless mode, we get
the same characteristic equation, i.e. the same spectrum of fluctuation frequencies
and the same 1-loop partition function.

Note that for $q = 0$ we get of course the same frequencies
\begin{equation}
\pm \sqrt{p^2 + 2(\mu^2-1) \pm 2 \sqrt{(\mu^2-1)^2 + \mu^2 p^2}}
\end{equation}
as found for the same solution in the ${AdS}_{5} \times S^{5}$ case.
At the WZW point, $q =1$, we get
\begin{equation}
\{ -p \, , \ -p \, ,\ \ p \pm 2 \mu \} \ ,
\end{equation}
i.e. the spectrum is massless
up to a shift in the momentum.

%%%%%%%%%%%%%%%%%%%%%%%%%%%%%%%%%%%%%%
\def\third{ {\textstyle{ 1\ov 3}}}
\subsection{Pohlmeyer reduction in group-theoretic approach\label{secapp2}}
%%%%%%%%%%%%%%%%%%%%%%%%%%%%%%%%%%%%%%

To extend the Pohlmeyer reduction to include fermions we need first to formulate it in terms of
group variable parametrization based on describing the principal chiral model for group $G$
as the $ G \times G \ov G$ coset sigma model corresponding the symmetric coset space
\begin{equation}
\frac{G_{_L} \times G_{_R}}{G_{_0}} \ , \la{coo}
\end{equation}
where $G_{_{L,R}}$ are two copies of the group $G = SU(2)$ and $G_{_0}$ is the diagonal subgroup isomorphic to $G$. Taking a group-valued field
\begin{equation}
f = \left( \begin{array}{cc} g_{_L} & 0 \\ 0 & g_{_R} \end{array}\right) \in G_{_L} \times G_{_R} \ ,
\end{equation}
we construct the left-invariant current
\begin{equation}\label{lic}
\EuScript{J} = f^{-1} d f = \left(\begin{array}{cc} \mathcal{J}_{_L} = g_{_L}^{-1} dg_{_L} & 0 \\ 0 & \mathcal{J}_{_R} = g_{_R}^{-1} dg_{_R} \end{array} \right) \in \mathfrak g_{_L} \oplus \mathfrak g_{_R}\ .
\end{equation}
The algebra $\mathfrak g_{_L} \oplus \mathfrak g_{_R}$ admits an $\mathbb{Z}_2$ automorphism
\begin{equation}
\Omega\left(\begin{array}{cc} a_{_L} & 0 \\ 0 & a_{_R} \end{array} \right) = \left(\begin{array}{cc} a_{_R} & 0 \\ 0 & a_{_L} \end{array} \right) \
\end{equation}
with the invariant subspace given by the diagonal subalgebra $\mathfrak g_{_0}$.
The trace is clearly invariant under this automorphism and hence we have the following orthogonal decomposition of the algebra
\begin{equation}\label{od}
\mathfrak g_{_L} \oplus \mathfrak g_{_R} = \mathfrak g_{_0} \oplus \mathfrak p \ .
\end{equation}
Decomposing the left-invariant current \eqref{lic} under the orthogonal decomposition \eqref{od}
\begin{equation}\begin{split}
\EuScript{J} = \EuScript{A} + \EuScript{P} \ , \qquad & \EuScript{A} = \left(\begin{array}{cc} \mathcal{A} & 0 \\ 0 & \mathcal{A} \end{array}\right) \ , \qquad \ \ \mathcal{A} = \frac12(\mathcal{J}_{_L} + \mathcal{J}_{_R}) \ ,
\\ & \EuScript{P} = \left(\begin{array}{cc} \mathcal{P} & 0 \\ 0 & -\mathcal{P} \end{array}\right) \ , \qquad \mathcal{P} = \frac12(\mathcal{J}_{_L} - \mathcal{J}_{_R}) \ ,
\end{split}\end{equation}
the action \eqref{4.1} can be written as
\begin{equation}\label{action2}
\rS=-\frac{\sqrt\lambda}{2\pi}\Big[\int d^2x \; \ha{\operatorname{Tr}} (\EuScript{P}_+\EuScript{P}_- ) - q
\int d^3 x \; \textstyle{\frac23} \epsilon^{abc} \, \widetilde{\operatorname{Tr}} (\EuScript{P}_a \EuScript{P}_b \EuScript{P}_c )\Big]\ .
\end{equation}
Here $\widetilde {\operatorname{Tr}}$ is defined as
\begin{equation}
\widetilde {\operatorname{Tr}} \left(\begin{array}{cc} a_{_L} & 0 \\ 0 & a_{_R} \end{array}\right) = {\operatorname{Tr}}(a_{_L}) - {\operatorname{Tr}}(a_{_R}) \ .
\end{equation}
$\Tr$ is normalized to -1 compared to $\tr$ which is normalized to -2.
If the usual trace is used in the WZ term it vanishes as a consequence of
the $\mathbb{Z}_2$ automorphism of the algebra.\foot{In particular, this action written in terms of $\mathcal{P}$ agrees
with the action in \ci{cz}.}
To recover the action \eqref{4.1} from \eqref{action2} we notice that the latter admits the following gauge symmetry
\be &&
f \rightarrow f g_{_0} \ , \qquad g_{_0} \in G_{_0} \ ,\ \ \ \ \ \
\EuScript{A} \rightarrow g_{_0}^{-1} \EuScript{A} g_{_0} + g_{_0}^{-1} d g_{_0} \ , \qquad \EuScript{P} \rightarrow g_{_0}^{-1} \EuScript{P} g_{_0} \ ,
\ee
which follows from the cyclicity of both ${\operatorname{Tr}}$ and $\widetilde {\operatorname{Tr}}$. Using this symmetry to fix $g_{_R} = \boldsymbol{1}$ we find that $\mathcal{A} = \mathcal{P}$. Defining
\begin{equation}
J = 2\mathcal{A} = 2\mathcal{P}
\end{equation}
and substituting into \eqref{action2} we recover \eqref{4.1}.

The equations of motion following from \eqref{action2} can be projected onto $\mathfrak g_{_L}$\footnote{Or alternatively onto $\mathfrak g_{_R}$ -- by construction, the equations are equivalent.}
\begin{equation}\label{eomp}
\mathcal{D}_- \mathcal{P}_+ + \mathcal{D}_+ \mathcal{P}_ - - 2 q [\mathcal{P}_-,\mathcal{P}_+] = 0\ , \qquad \mathcal{D}_\pm = \partial_\pm + [\mathcal{A}_\pm,] \ ,
\end{equation}
while the conformal gauge Virasoro constraints are given by
\begin{equation}
{\operatorname{Tr}}(\EuScript{P}_\pm^2) = -\mu^2\ .
\end{equation}
Another condition is the flatness condition for the current $\EuScript{J}$
\begin{equation}
d\EuScript{J} + \EuScript{J} \wedge \EuScript{J} = 0 \ ,
\end{equation}
which can be decomposed under the orthogonal decomposition \eqref{od} and projected onto $\mathfrak g_{_L}$ to give
\begin{equation}\begin{split}\label{mc}
d\mathcal{P} + \mathcal{A} \wedge \mathcal{P} + \mathcal{P} \wedge \mathcal{A} = 0 \ , \ \ \ \ \ \ \ \ \ \
d\mathcal{A} + \mathcal{A} \wedge \mathcal{A} + \mathcal{P} \wedge \mathcal{P} = 0 \ .
\end{split}\end{equation}
The equation of motion \eqref{eomp} and the first equation of \eqref{mc} can be rewritten as
\begin{equation}\label{eomhalf}
\mathcal{D}_+ \mathcal{P}_- + q [\mathcal{P}_+,\mathcal{P}_-] = 0\ , \qquad \mathcal{D}_- \mathcal{P}_+ - q [\mathcal{P}_-,\mathcal{P}_+] = 0 \ .
\end{equation}
The Pohlmeyer reduction starts by introducing a constant matrix $\tilde T = \left(\begin{array}{cc} T & 0 \\ 0 & -T \end{array}\right) \in \mathfrak p$ normalized as ${\operatorname{Tr}} \, \tilde T^2 = -1$. We then solve the Virasoro conditions fixing the $G_{_0}$ gauge symmetry as
\begin{equation}\label{ppm}
\mathcal{P}_+ = \mu T \ , \qquad \mathcal{P}_- = \mu g^{-1} T g \ , \qquad g \in G = SU(2) \ .
\end{equation}
Substituting into \eqref{eomhalf} we find that these equations are solved by parametrizing
\begin{equation}\label{apm}
\mathcal{A}_+ = g^{-1}\partial_+ g + g^{-1} A_+ g - q \mu T \ , \qquad\qquad \mathcal{A}_- = A_- + q \mu g^{-1} T g \ ,
\end{equation}
where $A_\pm$ take values in the subalgebra of $\mathfrak g=\mathfrak {su}(2)$ that commutes with $T$
(we shall denote this subalgebra as $\mathfrak h$). This is a $\mathfrak u(1)$ subalgebra that is spanned by $T$ itself.

Finally, to find the equation of motion of the reduced theory we substitute \eqref{ppm} and \eqref{apm} into the second equation of \eqref{mc} to give
\begin{equation}
\partial_-(g^{-1}\partial_+g + g^{-1} A_+ g) - \partial_+ A_- + [ A_- , g^{-1} \partial_+ g + g^{-1} A_+ g ] =(1-q^2) \mu^2 [T, g^{-1} T g] \ .
\end{equation}
Again, as in the embedding coordinate parametrization of the previous section, we see that
the only effect of the WZ term is to rescale the mass-squared parameter $\mu^2$ by $(1-q^2)$. We can then
follow the final steps of the usual PR approach \ci{gt1,gt2}
to find a gauged WZW model for the coset $G/H = SU(2)/U(1)$ plus an integrable potential with coefficient
$(1-q^2)\mu^2 $.
Choosing a particular parametrization of $g$ and integrating out the gauge field $A_\pm$ one recovers the complex sine-Gordon model in agreement with the embedding coordinate reduction approach.\foot{Let us note again that while in the reduced theory solutions for zero and non-zero $q$ are formally
related by the simple transformation \eqref{trans}, the corresponding string solutions will have a more non-trivial relation. This can be seen from the change of variables \eqref{ppm} and \eqref{apm}, where the $q$ does not enter via a rescaling of $\mu$ by $\sqrt{1-q^2}$. }
In particular as the gauge group is abelian the WZW model can be either vector or axially gauged. In the former case we find the complex sine-Gordon Lagrangian with $\cot^2\varphi$ and in the latter case with $\tan^2\varphi$ -- see the comment below eq.~\eqref{lagpr}.
Note also that the reduced theory for $q=1$, i.e. for
the $SU(2)$ WZW model, is given by the standard $SU(2)/U(1)$ gauged WZW model.

A similar construction in
the case of $G=SL(2,R)$ will lead to the reduced theory given by the gauged WZW model for the coset $G/H = SL(2,R)/U(1)$ plus an integrable potential
with coefficient $(1-q^2)\mu^2$, equivalent after gauge fixing
to the complex sinh-Gordon model (in agreement with the discussion in section \ref{secads}).

In general, the above reduction procedure will work
for any sigma model with a target space of the form \rf{coo} (times $R_t$)
but for generic $G$ there will be additional $\text{\rm rank} \, G -1$ massless modes
(for any value of $q$).

Finally, let us give also the expression for the Lax connection corresponding to reduced theory equations.
The set of sigma model equations \eqref{mc} and \eqref{eomhalf} follow from a Lax connection with spectral parameter $z$
(cf.\rf{laxx})
\begin{equation}
\mathcal{L}_\pm = \mathcal{A}_\pm \pm q \mathcal{P}_\pm + z^{\pm 1}\sqrt{1-q^2} \, \mathcal{P}_\pm \ . \la{ax}
\end{equation}
Substituting the change of variables \eqref{ppm}, \eqref{apm} we find the Lax connection of
the reduced theory:
\be
\mathcal{L}_+ = g^{-1} \partial_+ g + g^{-1} A_+ g + z \sqrt{1-q^2} \, \mu\, T \ ,
\qquad
\mathcal{L}_- = A_- + z^{-1} \sqrt{1-q^2} \, \m\, g^{-1} T g \ . \la{axx}
\ee
Again, the dependence on $q$ is only via the rescaling of $\m$ by $\sqrt{1-q^2}$.

%%%%%%%%%%%%%%%%%%%%%%%%%%%%%%%%%%%%%%
\def\rr{{\sqrt{\frac{q\m}{2\zeta}}}}
\def\htheta{{\hat \theta}}
\subsection{Pohlmeyer reduction for superstring on $AdS_3 \times S^3$ with mixed flux \label{secapp3}}
%%%%%%%%%%%%%%%%%%%%%%%%%%%%%%%%%%%%%%

Here we follow \ci{gt1,gt2} and start from the coset superspace
\begin{equation}
\frac{PSU(1,1|2)_{_L} \times PSU(1,1|2)_{_R}}{SU(1,1) \times SU(2)}\ ,
\end{equation}
where denominator is the diagonal subgroup of the bosonic subgroup of the numerator.
The algebra has $\mathbb{Z}_4$ orthogonal decomposition, which schematically takes the form
\footnotesize
\begin{equation*} \nonumber \begin{split} %v5
&
\left(\begin{array}{cccc} \tfrac12(a+b) & 0 & 0 & 0
\\ 0 & \tfrac12(c+d) & 0 & 0
\\ 0 & 0 & \tfrac12(a+b) & 0
\\ 0 & 0 & 0 & \tfrac12(c+d) \end{array}\right)_0
+
\left(\begin{array}{cccc} 0 & \tfrac12(\alpha + i\beta) & 0 & 0
\\ \tfrac12(\nu + i \delta) & 0 & 0 & 0
\\ 0 & 0 & 0 & \tfrac12(\beta - i \alpha)
\\ 0 & 0 & \tfrac12(\delta - i\nu) & 0 \end{array}\right)_1
\\
& +
\left(\begin{array}{cccc} \tfrac12(a-b) & 0 & 0 & 0
\\ 0 & \tfrac12(c-d) & 0 & 0
\\ 0 & 0 & \tfrac12(b-a) & 0
\\ 0 & 0 & 0 & \tfrac12(d-c) \end{array}\right)_2
+
\left(\begin{array}{cccc} 0 & \tfrac12(\alpha - i\beta) & 0 & 0
\\ \tfrac12(\nu - i \delta) & 0 & 0 & 0
\\ 0 & 0 & 0 & \tfrac12(\beta + i \alpha)
\\ 0 & 0 & \tfrac12(\delta + i \nu ) & 0 \end{array}\right)_3
\end{split}\end{equation*} %v5
\normalsize
We decompose the left-invariant Maurer-Cartan one-form as
\begin{equation}\begin{split}
\EuScript{J} = g^{-1} d g = \EuScript{J}_0 + \EuScript{J}_1 + \EuScript{J}_2 + \EuScript{J}_3
\ , \ \ \ \ \ \ \ \ \ \ d \EuScript{J} + \EuScript{J} \wedge \EuScript{J} = 0\ .
\end{split}\end{equation}
The resulting equations of motion can be written in terms of $\EuScript{J}$ projected onto one copy of $PSU(1,1|2)$
\begin{equation}
\EuScript{J}\Big|_{PSU(1,1|2)_{_L}} = \mathcal{J}_0 + \mathcal{J}_1 + \mathcal{J}_2 + \mathcal{J}_3\ ,
\end{equation}
where $\mathcal{J}_{0,2}/\mathcal{J}_{1,3}$ are elements of the Grassmann-even/odd subalgebra of $PSU(1,1|2)$.

The superstring action is \ci{cz}\footnote{This is equivalent to the action in \ci{cz} up to sign conventions.
The action has $\kappa$-symmetry, is integrable, and reduces to the standard supercoset GS
action in the limit $q \rightarrow 0$.}
\be\label{actionsuperstring} &&\rS=
\frac{\sqrt\lambda}{2\pi}\Big[\int d^2x \; \ha {\operatorname{STr}}\Big[\EuScript{J}_{2+}\EuScript{J}_{2-} + \ha {\sqrt{1-q^2}}(\EuScript{J}_{1+}\EuScript{J}_{3-} - \EuScript{J}_{1-}\EuScript{J}_{3+})\Big] \no
\\ &&\ \ \ \ \ \ \ \ \ \ \ \ \ \ \ \ -\, q \int d^3 x \; \epsilon^{abc} \, \widetilde{\operatorname{STr}}\Big[ {\textstyle{ 2\ov 3}} \EuScript{J}_{2a} \EuScript{J}_{2b} \EuScript{J}_{2c} + \EuScript{J}_{1a}\EuScript{J}_{3b}\EuScript{J}_{2c} + \EuScript{J}_{3a}\EuScript{J}_{1b}\EuScript{J}_{2c}\ \Big]\ .
\ee
The equations of motion are then given by ($\mathcal{D}_\pm = \partial_\pm + [\mathcal{J}_{0\pm},]$)
\begin{align}
\mathcal{D}_- \mathcal{J}_{2+} + \mathcal{D}_+ \mathcal{J}_{2-} &+ \sqrt{1-q^2} \big( [\mathcal{J}_{1-},\mathcal{J}_{1+}] - [\mathcal{J}_{3-},\mathcal{J}_{3+}]\big) \nonumber %v5
\\ &- q\big( 2 [\mathcal{J}_{2-},\mathcal{J}_{2+}] + [\mathcal{J}_{1-},\mathcal{J}_{3+}] +[\mathcal{J}_{3-},\mathcal{J}_{1+}]\big) = 0 \ , \label{eq1} %v5
\\ [\mathcal{J}_{2+},\mathcal{J}_{3-} - {\textstyle \frac{1 + \sqrt{1-q^2}}{q} } \mathcal{J}_{1-}]
& - [\mathcal{J}_{2-},\mathcal{J}_{3+} + {\textstyle \frac{1 - \sqrt{1-q^2}}{q} } \mathcal{J}_{1+}] = 0 \ , \label{eq2}
\\ [\mathcal{J}_{2+},\mathcal{J}_{1-} - {\textstyle \frac{1 - \sqrt{1-q^2}}{q} } \mathcal{J}_{3-}] &
- [\mathcal{J}_{2-},\mathcal{J}_{1+} + {\textstyle \frac{1 + \sqrt{1-q^2}}{q} } \mathcal{J}_{3+}] = 0 \ , \label{eq3}
\end{align}
to be supplemented by the Maurer-Cartan equations
\begin{align}
\partial_- \mathcal{J}_{0+} - & \partial_+ \mathcal{J}_{0-} + [\mathcal{J}_{0-},\mathcal{J}_{0+}] + [\mathcal{J}_{2-},\mathcal{J}_{2+}] + [\mathcal{J}_{1-},\mathcal{J}_{3+}] + [\mathcal{J}_{3-},\, \mathcal{J}_{1+}] = 0 \ , \label{eq4}
\\ & \mathcal{D}_- \mathcal{J}_{2+} - \mathcal{D}_+ \mathcal{J}_{2-} + [\mathcal{J}_{1-},\, \mathcal{J}_{1+}] + [\mathcal{J}_{3-},\mathcal{J}_{3+}] = 0 \ , \label{eq5}
\\ & \mathcal{D}_- \mathcal{J}_{1+} - \mathcal{D}_+ \mathcal{J}_{1-} + [\mathcal{J}_{2-},\, \mathcal{J}_{3+}] + [\mathcal{J}_{3-},\mathcal{J}_{2+}] = 0 \ , \label{eq6}
\\ & \mathcal{D}_- \mathcal{J}_{3+} - \mathcal{D}_+ \mathcal{J}_{3-} + [\mathcal{J}_{2-},\, \mathcal{J}_{1+}] + [\mathcal{J}_{1-},\mathcal{J}_{2+}] = 0 \ , \label{eq7}
\end{align}
and the Virasoro constraints
\begin{equation}
{\operatorname{STr}}(\mathcal{J}_{2\, \pm}^2) = 0 \ . \label{eq8}
\end{equation}
Let us introduce the parameter\foot{Note that for the three special points $q = \{0,\pm1\}$, we have $\zeta = q$.}
\begin{equation}
\zeta = \frac{1 - \sqrt{1-q^2}}{q} = \frac{q}{1 + \sqrt{1-q^2}} \ . \la{88}
\end{equation}
Taking linear combinations of \eqref{eq2} and \eqref{eq3}, they can be written as
\begin{align}
& [\mathcal{J}_{2+},\mathcal{J}_{1-} - \zeta \mathcal{J}_{3-}] = [\mathcal{J}_{2-},\mathcal{J}_{3+} + \zeta \mathcal{J}_{1+}] = 0 \ , \label{eq9}
\end{align}
while combining \eqref{eq1} and \eqref{eq5} we find the following first-order equations for $\mathcal{J}_{2\pm}$
\begin{align}
& \mathcal{D}_+ \mathcal{J}_{2-} + q\, [\mathcal{J}_{2+},\mathcal{J}_{2-}] -\ha q\,
[\mathcal{J}_{1-} + \zeta^{-1} \mathcal{J}_{3-},\ \mathcal{J}_{3+} +\zeta \mathcal{J}_{1+}] = 0 \ , \label{eq10}
\\ & \mathcal{D}_- \mathcal{J}_{2+} - q\, [\mathcal{J}_{2-},\mathcal{J}_{2+}] - \ha q\,
[\mathcal{J}_{1-} - \zeta \mathcal{J}_{3-},\ \mathcal{J}_{3+} - \zeta^{-1}\mathcal{J}_{1+}] = 0 \ . \label{eq11}
\end{align}
Note that in the $q \rightarrow 0$ limit we have $\zeta \rightarrow 0$ and $q \zeta^{-1} \rightarrow 2$ and hence these equations agree with the usual ones at $q = 0$.

On the equations of motion (i.e. using \eqref{eq9}) we can choose the following $\kappa$-symmetry gauge
\begin{equation}\label{ksymfix}
\mathcal{J}_{1-} = \zeta \mathcal{J}_{3-} \equiv \zeta \mathcal{Q}_- \ , \qquad \mathcal{J}_{3+} = -\zeta \mathcal{J}_{1+} \equiv -\zeta \mathcal{Q}_+ \ .
\end{equation}
Then \eqref{eq10} and \eqref{eq11} simplify to the form which is the same as in the bosonic case \eqref{eomhalf}. Therefore, the Pohlmeyer reduction can be carried out in the same way as before, i.e. we first
solve the Virasoro constraints \rf{eq8} using the $G$-gauge symmetry as
\begin{equation}\label{sol2}
\mathcal{J}_{2+} = \mu T \ ,\qquad \qquad \mathcal{J}_{2-} = \mu g^{-1} T g \ ,
\end{equation}
where $T \in \mathfrak {su}(1,1) \oplus \mathfrak {su}(2) \subset \mathfrak{psu}(1,1|2)$ satisfies ${\operatorname{STr}}(T^2) = 0$, is non-zero in both the $\mathfrak {su}(1,1)$ and $\mathfrak {su}(2)$ sectors of the algebra.
Explicitly, we can write $T $ as $T = \mu T_1 + \mu T_2$
where $T_1 \in \mathfrak {su}(1,1)$ and $T_2 \in \mathfrak {su}(2)$ and ${\operatorname{Tr}}(T_1^2) = {\operatorname{Tr}}(T_2^2)$.
The matrix $T$ defines an additional $\mathbb{Z}_2$ orthogonal decomposition of the algebra $\mathfrak{psu}(1,1|2)$:
\begin{equation}\label{perppar}
\mathfrak{psu}(1,1|2) = \mathfrak{psu}(1,1|2)^\perp \oplus \mathfrak{psu}(1,1|2)^\parallel \ , \qquad [T,\mathfrak{psu}(1,1|2)^\perp] = 0 \ , \quad \{T,\mathfrak{psu}(1,1|2)^\parallel\} = 0 \ .
\end{equation}
We can then solve the first-order equations \eqref{eq10} and \eqref{eq11} in the $\kappa$-symmetry gauge \eqref{ksymfix}:
\begin{equation}\label{sol0}
\mathcal{J}_{0+} = g^{-1}\partial_+ g + g^{-1} A_+ g - q \mu T \ , \qquad \mathcal{J}_{0-} = A_- + q \mu g^{-1} T g \ .
\end{equation}
Here $A_\pm \in \mathfrak{psu}(1,1|2)^\perp_{_{\text{even}}} \equiv \mathfrak h$, i.e.
$\mathfrak h$ is the subalgebra of $\mathfrak {su}(1,1) \oplus \mathfrak {su}(2)$ that commutes with $T$, i.e. it is $\mathfrak u(1)\oplus \mathfrak u(1)$ generated by $T_1$ and $T_2$.
The Pohlmeyer reduction proceeds
by substituting the $\kappa$-symmetry fixing \eqref{ksymfix} in equations \eqref{eq6} and \eqref{eq7}
\begin{equation}\begin{split}
& \mathcal{D}_- \mathcal{Q}_+ - \zeta \, \mathcal{D}_+ \mathcal{Q}_- - \zeta \, [\mathcal{J}_{2-},\mathcal{Q}_+] - [\mathcal{J}_{2+}, \mathcal{Q}_-] = 0 \ ,
\\ & \zeta \, \mathcal{D}_- \mathcal{Q}_+ + \mathcal{D}_+ \mathcal{Q}_- - [\mathcal{J}_{2-},\mathcal{Q}_+] + \zeta \, [\mathcal{J}_{2+}, \mathcal{Q}_-] = 0 \ .
\end{split}\end{equation}
Taking linear combinations, and substituting in for $\mathcal{J}_{2\pm}$ from \eqref{sol2} and $\mathcal{J}_{0\pm}$
from \eqref{sol0} we find the following equations for $\mathcal{Q}_\pm$ ($D_\pm =\partial_\pm + [A_\pm,]$)
\begin{equation}\label{fof}
D_- \mathcal{Q}_+ - \mu \sqrt{1-q^2} [T,g^{-1}(g\mathcal{Q}_- g^{-1}) g] = 0 \ , \qquad D_+(g \mathcal{Q}_- g^{-1}) - \mu\sqrt{1-q^2}[T,g \mathcal{Q}_+ g^{-1}] = 0 \ .
\end{equation}
Defining
\begin{equation}
\label{frel}
\mathcal{Q}_+^\parallel = c \, \Psi_R \ , \quad (g \mathcal{Q}_- g^{-1})^\parallel = c \, \Psi_L \ , \quad
\mathcal{Q}_+^\perp = c \, \tilde\Psi_R \ , \quad (g \mathcal{Q}_- g^{-1})^\perp = c \, \tilde\Psi_L \ , \ \ \ \ c = \text{\rr} \ ,
\end{equation}
and projecting onto the parallel and perpendicular subspaces \eqref{perppar} we find that $\tilde \Psi_{L,R}$ satisfy
\begin{equation}
D_- \tilde \Psi_R = D_+ \tilde \Psi_L = 0 \ ,\la{d88}
\end{equation}
and the residual $\kappa$-symmetry can be used to fix them to zero.
The final system of equations describing the reduced theory is given by \eqref{eq4} and \eqref{fof} after substituting in for the new set of variables $\{g,A_\pm,\Psi_{R,L}\}$
\be &&
\partial_-(g^{-1}\partial_+g + g^{-1} A_+ g) - \partial_+ A_- + [ A_- , g^{-1} \partial_+ g + g^{-1} A_+ g ] \nonumber
\\ &&\ \ \ \ \ \ \ \ \ = (1-q^2) \mu^2 [T, g^{-1} T g] + \ \sqrt{1-q^2}\,\m\ [\Psi_R,g^{-1}\Psi_L g] \ ,\la{d89}
\\
&&D_- \Psi_R - \sqrt{1-q^2}\, \mu\, [\, T,g^{-1}\Psi_Lg] = 0 \ , \quad D_+ \Psi_L - \sqrt{1-q^2}\,
\mu\, [\, T,g \Psi_R\, g^{-1}] = 0 \la{d90}
\ee
Therefore, we find the same reduced system of equations as in the $q=0$ theory, but with
$\mu \to \sqrt{1-q^2}\, \m$. We can then follow the final steps of the usual PR approach \cite{gt1,gt2}
to find the corresponding action of the PR model as that of
the gauged WZW model for the coset $\frac{SU(1,1) \times SU(2)}{U(1)\times U(1)}$ plus a potential and
fermionic terms ($k$ is the coupling of the PR model)
\be
&& \rS
= \ \frac{k}{4\pi} \operatorname{STr} \Big[ \tfrac{1}{2}\,\int d^2x \; %v5
\ g^{-1}\partial_+ g\ g^{-1}\partial_- g\
- \tfrac{1}{3}\,\int d^3x \; %v5
\ \epsilon^{mnl} \ g^{-1} \partial_m g\ g^{-1}\partial_n g \ g^{-1}\partial_l g \no
\\ && \qquad + \,\int d^2x \;
\ \big[ A_+\partial_- gg^{-1} - A_-g^{-1}\partial_+ g
- g^{-1} A_+g A_- + A_+A_-
+ (1-q^2)\mu^2\,g^{-1} T g T \big]
\no \\ && \label{gwzw} \qquad + \int d^2x\; \big( \Psi_L T D_+ \Psi_L + \Psi_R T D_- \Psi_R
+ \sqrt{1-q^2}\mu\;g^{-1}\Psi_L g \Psi_R\big)\Big]\,.
\ee

Let us note that as in the bosonic models of sections \ref{secapp1} and \ref{secapp2}, the reduced theory solutions for zero and non-zero $q$ are
formally related by the simple transformation \eqref{trans} along with
\begin{equation}
\Psi_R \rightarrow \sqrt{ 1+q}\, \Psi_R \ , \qquad \qquad \Psi_L \rightarrow \sqrt{ 1-q} \, \Psi_L \ .
\end{equation}
However, the corresponding string solutions will have a more non-trivial relation. This can be seen from the change of variables used above: \eqref{sol2},\eqref{sol0},\eqref{ksymfix},\eqref{frel}, which is not just given by rescaling $\mu$ by $\sqrt{1-q^2}$.
The same is then true also for the corresponding reduced theory action \eqref{gwzw}.

The set of supercoset sigma model equations \eqref{eq1}--\eqref{eq7} follow from the flatness condition for the following Lax connection (with spectral parameter $z$)\footnote{Up to sign conventions this is equivalent to the Lax connection in \ci{cz}
written in light-cone coordinates.}
\begin{equation}\begin{split}
\mathcal{L}_+ = & z^{-1} \tilde q (\mathcal{J}_{3+} + \zeta \mathcal{J}_{1+}) + \mathcal{J}_{0+} + q \mathcal{J}_{2+} + z \, \tilde q (\mathcal{J}_{1+} - \zeta \mathcal{J}_{3+}) + z^2 \sqrt{1-q^2} \, \mathcal{J}_{2+} \ ,
\\ \mathcal{L}_- = & z \, \tilde q (\mathcal{J}_{1-} - \zeta \mathcal{J}_{3-}) + \mathcal{J}_{0-} - q \mathcal{J}_{2-} + z^{-1} \tilde q (\mathcal{J}_{3-} + \zeta \mathcal{J}_{1-}) + z^{-2} \sqrt{1-q^2} \, \mathcal{J}_{2-} \ ,\label{d94}
\end{split}\end{equation}
with
\begin{equation}
\tilde q^2 = \frac{q \sqrt{1-q^2}}{2\z} \ .
\end{equation}
On substituting here the change of variables used in the Pohlmeyer reduction we find
\begin{equation}\begin{split}
\mathcal{L}_+ = & g^{-1} \partial_+ g + g^{-1} A_+ g + z (\mu \sqrt{1-q^2})^{1/2} \, \Psi_R + z^2 \mu \sqrt{1-q^2} \, T \ ,
\\
\mathcal{L}_- = & A_- + z^{-1} (\mu \sqrt{1-q^2})^{1/2} \, g^{-1} \Psi_L g + z^{-2} \mu \sqrt{1-q^2} \, g^{-1} T g \ .\la{d99}
\end{split}\end{equation}
This gives indeed the Lax connection of the PR theory.
Note again that $q$ enters here only via a simple rescaling of the mass parameter $\mu \to \sqrt {1-q^2}\, \mu $.

\

In the $q=0$ case the PR model expanded near the trivial vacuum has the same massive spectrum (with mass $\mu$) as the
BMN-type spectrum of small fluctuations in the string sigma model.
The corresponding massive tree-level S-matrix of the Pohlmeyer reduction of the superstring on $AdS_3 \times S^3$
is {\it relativistically invariant} \cite{ht1,ht2}. It
formally has the same structure as the superstring S-matrix
in \eqref{fffi},\eqref{bhsmatans} with $\sqrt{\lambda} \rightarrow k$ being the
coupling of the reduced theory and with the functions of momenta  $l_{1,2,3,4,5} $ given by (here the function $c=0$)
\be 
{\textstyle {l_1 = \coth \frac \htheta 2\ , \qquad l_2 = - \tanh \frac\htheta 2 \ , \qquad  l_3 = 0 \ ,\qquad
\ l_4= - \frac12 \operatorname{sech} \frac \htheta 2 \ ,  \qquad 
l_5 = \frac12 \operatorname{csch} \frac \htheta 2\ . } }\la{d97}
\ee
 Here $\hat\theta$ is the difference of the two rapidities
\be \htheta=\theta-\theta' \ , \qquad p =\m \sinh \theta \ ,\quad e =\m \cosh \theta \ , \qquad
p' = \m\sinh \theta' \ ,\quad e' =\m \cosh \theta' \ .
\ee
The functions $l_{6,7,8,9}$ are then defined as in \eqref{bh6789}.
The 1-loop result for the PR S-matrix and a conjecture for its all-order expression based on supersymmetry was given in \cite{ht2}.

Since the generalization to the $q\not=0$ case is found simply by replacing $\mu \to \sqrt {1-q^2}\, \mu $
in the relativistic PR Lagrangian \rf{gwzw}, the corresponding S-matrix thus remains the same as in the $q=0$ case.

\bigskip

%%%%%%%%%%%%%%%%%%%%%%%%%%%%%%%%%%%%%%

%%%%%%%%%%%%%%%%%%%%%%%%%%%%%%%%%%%%%%

%%%%%%%%%%%%%%%%%%%%%%%%%%%%%%%%%%%%%%
\end{document}
%%%%%%%%%%%%%%%%%%%%%%%%%%%%%%%%%%%%%%